\documentclass[preprint]{elsarticle}

\usepackage{graphicx}
\usepackage{amssymb,amsmath}
\usepackage{bm}

\newcommand{\be}{\begin{eqnarray}}
\newcommand{\ee}{\end{eqnarray}}
\newcommand{\nn}{\nonumber\\}
\newcommand{\nin}{\noindent}
\newcommand{\la}{\langle}
\newcommand{\ra}{\rangle}

\renewcommand{\theequation}{\arabic{section}.\arabic{equation}}

\begin{document}

\title{Z(2) Gauge Neural Network and its Phase Structure
}

\date{\today}

\author{Yusuke Takafuji} 
\author{Yuki Nakano} 
\author{Tetsuo Matsui\corref{cor1}}
\address{%
Department of Physics, Kinki University, 
Higashi-Osaka, 577-8502 Japan
}%
\cortext[cor1]{Corresponding author}


\begin{abstract}
We study general phase structures of neural-network models 
that have Z(2) local gauge symmetry.
The Z(2) spin variable $S_i =\pm1$ on the $i$-th site describes
a neuron state as in the Hopfield model, and the Z(2) gauge variable
$J_{ij}=\pm1$ describes a state of the synaptic connection between  
$j$-th and $i$-th neurons. The gauge symmetry allows for 
a self-coupling energy among $J_{ij}$'s such as $J_{ij}J_{jk}J_{ki}$, 
which describes reverberation of signals.  Explicitly, we consider
the three models; (I)  
annealed model with full and partial  connections of $J_{ij}$,
(II) quenched model with full connections where $J_{ij}$ is
 treated as a slow quenched variable, and  (III) quenched  
three-dimensional lattice model with the nearest-neighbor connections.
By numerical simulations, we examine their phase structures
paying attention to the effect of reverberation term,  and 
compare them each other and with the
annealed 3D lattice model which has been studied beforehand.
By noting the dependence of thermodynamic quantities upon the total number
of sites and the connectivity among sites, we obtain a coherent 
interpretation to understand these results.  
Among other things, we find that 
the Higgs phase of the annealed model is separated
into two stable spin-glass phases in the quenched cases (II) and (III). 
\\

\nin

\end{abstract}

\maketitle

\section{Introduction}
\setcounter{equation}{0} 

Although it is just a collection of neurons and other biological cells, 
the human brain exhibits quite various functions such as 
learning patterns and recalling them.
We are still in a  way to obtain a  physical understanding of 
 these functions in a comprehensive manner.
As a well-known and convincing step on this way, one may refer to
the Hopfield model\cite{hopfield} for explanation of associative memory.
Here, the state of $i$-th neuron is described simply by 
a Z(2) variable as  $S_i=1$(excited), -1(unexcited).
This model is a good example of a simple physical model 
that describes essential mechanism of some functions of the human brain.

Concerning to the learning process, many models have been proposed and 
studied\cite{learning}.
In Ref.\cite{matsui,kemukemu}, the Z(2) gauge model is introduced as a 
model of learning, where the state of synaptic connections from
the $j$-th neuron to the $i$-th neuron
is described by the gauge variable $J_{ij}=1$(excitatory connection)
and  -1(inhibitory connection).
The plasticity of $J_{ij}$ is described 
by the equation of motion which basically decreases
the gauge-invariant energy subject to fluctuations caused by
random noises. The reason why the local gauge symmetry is implemented 
there is two fold; 

(i) There is a freedom to
assign two physical neuron states(excited and unexcited)
to a two-valued local variable $S_i=\pm1$
(See Ref.\cite{kemukemu} for more details).

(ii) In the real human brain, electric signals 
are transfered according to the rule of electromagnetism which
is based on the U(1) local gauge symmetry. 
Because Z(2) is a subgroup of U(1) group, the 
Z(2) gauge symmetry may be viewed as a remnant of this U(1) 
symmetry.

A by-product of the gauge symmetry is that
the time evolution of $J_{ij}$ automatically involves a
term suggested by Hebb's law\cite{hebb}, $J_{ij}(t+\Delta t)-J_{ij}(t) 
\propto
S_iS_j +\cdots$\cite{kemukemu}. 

In Ref.\cite{kemukemu} we introduced a simple Z(2) gauge model
defined on the three-dimensional (3D) lattice and studied
its  various aspects such as the phase structure, ability of learning 
and recalling patterns, etc. We obtained some interesting results
such as an existence of confinement phase in which
neither learning and recalling is possible.

Although these results warrant that the Z(2) gauge neural network
is worth further studies, the 3D lattice model
has some flaws as a model of the human-brain functions. For example,
(i) its nearest-neighbor connections of the 3D 
lattice model are too scarce compared with those of the human brain,
and (ii) treating $S_i$ and $J_{ij}$ on an equal footing
in the time evolution is not realistic because, in the human brain,   
the rate of time variation
of synaptic strength is much slower than that of neuron state.
One certainly needs to incorporate ample synaptic connections
and different scales in time evolution of variables.

In this paper we consider various models of Z(2) gauge neural network
respecting these two  points,
and study their properties numerically. Comparison of these results with
those of the above 3D lattice model\cite{matsui,kemukemu} 
provides us with more understanding of the general properties
of the Z(2) gauge network
and relevance of Z(2) gauge symmetry to modeling 
the human brain. In particular, we consider certain quenched models
in which the synaptic variables are treated as slow-varying
quenched variables, and find that there appear spin-glass (SG) states.

The paper is organized as follows:
In Sect.2 we explains the explicit models (I-III) in detail.
In Sect.3-5 we present the results of Monte Carlo (MC) simulations for
the phase structure of each model.
In Sect.6 we present conclusions and discussions.
In Appendices A-E some technical topics are studied.

\section{Z(2) Gauge Models}
\setcounter{equation}{0} 

In this section we introduce the Z(2) gauge models, Model I-III, 
that we shall study.
They  are classified by the following two points;

(a) magnitude of connectivity of synaptic connections; 
full and  partial connections or
the nearest-neighbor connections on the 3D lattice,

(b) nature of synaptic variables; annealed one ($J_{ij}$ varies
in a similar time-scale as $S_i$) or quenched one
($J_{ij}$ varies much slower than $S_i$). 

These models are listed in Table 1 
together with the 3D annealed lattice model which we call Model 0.
We note here that Models 0 and III are put on the 3D lattice
where neurons reside on lattice sites, and therefore the distance between 
a pair of neurons can be defined. In contrast,  Models II and III 
have full (or partial) connections and there are no concepts
of distance between neurons as long as one does not introduce 
explicitly a metric space in which neurons reside. 
  
{\small
\begin{center}
\begin{tabular}{|p{0.5cm}|p{3cm}|p{2.5cm}|p{1.5cm}|} 
\hline
& \  model    &\ connections& \ \shortstack{synaptic \\ variables}  
\\ \hline
\ 0&\ 3D lattice/anneal &\ 3D lattice &\ anneal   
\\ \hline
\ I&\ full/anneal       &\ full and partial  &\ anneal  
\\ \hline
\ II&\ full/quench     &\ full  &\ quench  
\\ \hline
\ III&\ 3D lattice/quench &\ 3D lattice &\ quench   
\\
\hline
\end{tabular}\\
\vspace{0.3cm}
\end{center}
Table 1. List of Z(2) gauge models classified by connections and
treatment of synaptic variables: Model 0 studied in Ref.\cite{kemukemu}
and Models I-III studied in the present paper.
}

\subsection{Model 0: Annealed 3D Lattice Model}

Before going to Models I-III, let us review Model 0 and 
its results\cite{kemukemu} briefly.
The energy $E_0$ of Model 0 is given by\cite{c3}
\begin{eqnarray}
 E_0 &=& -c_1 \sum_x \sum_{\mu} S_{x+\mu} J_{x\mu} S_x
  -c_2 \sum_{x} \sum_{\mu > \nu} J_{x\mu} J_{x+\mu,\nu} 
J_{x+\nu ,\mu} J_{x\nu}.
\label{energy0}
\end{eqnarray}
$S_x (=\pm 1)$ 
is the neuron variable ($S_i$) put on the site $x$ of the 3D lattice and 
 $J_{x\mu}  (=\pm 1)$ is the synaptic variable ($J_{ij}$) 
 put on the link $(x,x+\mu)$ 
connecting nearest-neighbor sites, where $\mu=1,2,3$ is the direction index
as well as the unit vector in that direction.
$c_1$ and $c_2$ are real parameters.
The $c_1$-term, the Hopfield energy\cite{hopfield}, 
describes the process of signals transferring
from the neuron at $x$ to the neuron at $x+\mu$ (and vice versa), and
the $c_2$-term describes the process of signals running around
the contour $(x,x+\mu,x+\mu+\nu,x+\nu)$ (and the reversed one) as 
a reverberating circuit\cite{hebb,reverberating}.
This $c_2$-term also corresponds to the magnetic energy in the 
U(1) gauge theory\cite{wilson}.
 
$E_0$ is invariant under the Z(2) gauge transformation,
\begin{eqnarray}
S_x \rightarrow S_{x}^{'} \equiv V_x S_x, \ \
 J_{x\mu} \rightarrow J_{x\mu}^{'} \equiv V_{x+\mu} 
 J_{x\mu} V_{x},\ \
 V_x &=& \pm 1,
\label{z2gaugetrsf}
\end{eqnarray}
due to $V_x^2=1$.

We introduce the fictitious temperature $T$ as a parameter to
control the fluctuations of $S_x$ and $J_{x\mu}$.
The partition function at $T$ is given by
\be
Z_0 &=& \sum_S\sum_J \exp(- E_0),\nn
\sum_S&\equiv& \prod_x\sum_{S_x=\pm1},\
\sum_J \equiv \prod_{x}\prod_{\mu}\sum_{J_{x\mu}=\pm1},
\label{model0}
\ee
where we have included the inverse  temperature $\beta \equiv 1/T$ 
into the coefficients $c_1$ and $c_2$ ($c_i$ is proportional to $\beta$).
The average 
$\la O(S,J) \ra$ of a function $O(S,J)$ w.r.t. $Z_0$ is given by
\be 
\langle O(S,J) \rangle &=& \frac{1}{Z_0} \sum_{S}\sum_J\ 
O(S,J) \exp(- E_0).
\label{thermalaverage}
\end{eqnarray}

\begin{figure}[h]
\begin{center}
\hspace{-1.6cm}
\vspace{-0.5cm}
\includegraphics[width=6.5cm]{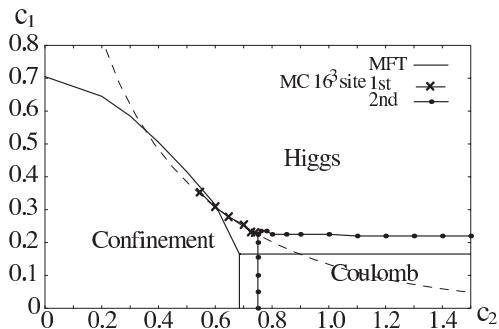}
\end{center}
\caption{
Phase structure of the 3D Z(2) lattice gauge model, Model 0 of 
(\ref{model0}), in the $c_2$-$c_1$ plane (taken from Ref.\cite{kemukemu}). 
The crosses and filled squares 
denote first and second-order transitions respectively.
The real curves are the result of MFT, and 
the dashed curve is the self-dual curve on which
a part of phase transition points may locate. 
}
\label{phase3dlgt}
\end{figure}

The phase diagram in the $c_2$-$c_1$ plane is given in Fig.\ref{phase3dlgt}.
There are three phases as listed in Table.2,
where the order parameters of the mean-field theory (MFT) and  
the ability of learning and recalling patterns are also given.
The order of transition is of first-order for the confinement-Higgs 
transition, while it is of second-order for the confinement-Coulomb
transition and for the Coulomb-Higgs transition\cite{mftmodel0}.

\vspace{0.2cm}
{\small
\begin{center}
\begin{tabular}{|p{2cm}|p{1cm}|p{1cm}|p{3.5cm}|} 
\hline
 \  phase    &\ $\langle J_{x\mu} \rangle $ &\ $\langle S_x \rangle $  
   &\  ability 
\\ \hline
\ Higgs       &\ $\neq 0$  &\ $\neq 0$  &\ learning and recalling 
\\ \hline
\ Coulomb     &\ $ \neq 0$  &\ \quad  $0$  &\ learning 
\\ \hline
\ confinement &\ \quad  $0$   &\ \quad  $0$     &\  N.A.
\\
\hline
\end{tabular}
\end{center}

Table2. Phases, order parameters 
of the  mean-field theory\cite{drouffe}, and ability 
of learning and recalling patterns 
in the 3D Z(2) lattice gauge model, Model 0 of  (\ref{model0})
(See Ref.\cite{kemukemu}). The names for three phases are those
used in lattice gauge theory\cite{wilson}.
}\\

\subsection{Model I: Annealed Model with full and partial connections}

The annealed  model with full and partial connections involves $N$ neurons.
The state of the neuron at the $i$-th site
($i=1,\cdots, N$) is described by the 
neuron variable $S_i (=\pm1)$,   and the state of synaptic connection
connecting $j$-th neuron and $i$-th neuron is described by
the synaptic variable $J_{ij} (=\pm 1)$.
In this paper we consider the case of symmetric coupling, 
$J_{ij}=J_{ji}$,  and only 
$J_{ij}$ with $i < j$ are independent\cite{symmetric}. 
The total number of independent variables of $J_{ij}$ is  
$N_{\rm l}\equiv {}_NC_2=N(N-1)/2$. 
We also have the connection parameter $\epsilon_{ij}$, 
\be
\epsilon_{ij} =
\left\{
\begin{array}{ll}
1 & {\rm connected},\\
0 & {\rm disconnected}.
\end{array}
\right.
\ee

The energy $E_{\rm I}(\epsilon)$ for a fixed configuration 
of connections,  $\epsilon \equiv\{\epsilon_{ij}\}$, 
is given by
\begin{eqnarray}
 E_{\rm I}(\epsilon)&=&-c_{1}\sum_{i<j}\epsilon_{ij}S_{i}J_{ij}S_{j}
 -\frac{c_{2}}{N}
\sum_{i<j<k}\epsilon_{ij}\epsilon_{jk}\epsilon_{ki}J_{ij}J_{jk}J_{ki}.
\label{energy1}
\ee
Each term is depicted in Fig.\ref{fig2}.
It may be viewed as a direct extension of the energy (\ref{energy0}) 
of Model 0.  
We note that the  $c_2$-term for reverberation consists of the product of
{\it three} $J$'s  in contrast to the product of
four $J$'s in the 3D model, Model 0, reflecting 
the difference of the minimum number of $J$'s to
construct  nontrivial (not a constant) gauge-invariant term. 
We have introduced the factor $N^{-1}$ in the coefficient of 
the $c_2$-term for later convenience.
One may include other gauge-invariant  terms to the energy, such as 
$c_4 J_{ij}J_{jk}J_{kl}J_{li}$, but the properties of 
the ``minimum" form (\ref{energy1}) should be studied first.
$ E_{\rm I}(\epsilon)$ is invariant under Z(2) gauge transformation
similar to (\ref{z2gaugetrsf}),
\begin{eqnarray}
S_i \rightarrow S_{i}^{'} \equiv V_i S_i,\ \
 J_{ij} \rightarrow J_{ij}^{'} \equiv V_{i} 
 J_{ij} V_{j},\ \
 V_i = \pm 1.
\label{z2gaugetrsf2}
\ee

\begin{figure}[b]
\begin{center}
\hspace{-1.6cm}
\includegraphics[width=7cm]{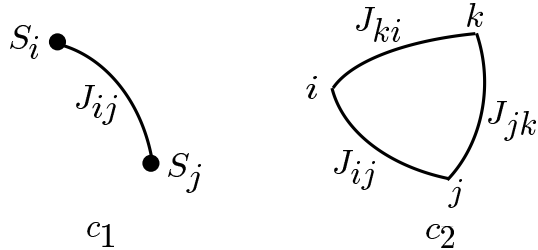}
\end{center}
\caption{
Each term of the energy $E_{\rm I}(\epsilon)$ of (\ref{energy1})
of Model I for the case $\epsilon_{ij}=\epsilon_{jk}=\epsilon_{ki}=1$. 
The filled circle denotes $S_i$ and the segment denotes $J_{ij}$.   
}
\label{fig2}
\end{figure}

The partition function $Z_{\rm I}(\epsilon)$ for a fixed configuration 
$\epsilon$ and the average $\langle O\rangle_\epsilon$
over $Z_{\rm I}(\epsilon)$ are given by
\be
Z_{\rm I}(\epsilon)&=&\sum_S\sum_J\exp\left(-E_{\rm I}(\epsilon)\right),\nn
\langle O\rangle_\epsilon&=&\frac{1}{Z_{\rm I}(\epsilon)}\sum_S\sum_J O(S,J)
\exp\left(-E_{\rm I}(\epsilon)\right),\nn
\sum_S&\equiv&\prod_i\sum_{S_i=\pm1},\
\sum_J\equiv\prod_{i < j}\sum_{J_{ij}=\pm1}.
\label{averageo1}
\end{eqnarray}

The connectivity $p(\epsilon)$ for a fixed set of $\epsilon_{ij}$ is defined by
\be
p(\epsilon)= \frac{1}{{}_NC_2}\sum_{i < j}\epsilon_{ij}.
\ee
The average $\la O(S,J)\ra_p$ for a fixed value $p$ of connectivity 
is defined by
\be
\hspace{-0.5cm}\la O(S,J) \ra_p &=& \frac{1}{N_\epsilon}
\sum_{\epsilon}\delta_{p(\epsilon),p}\la O(S,J) 
\ra_\epsilon,\nn
{N_\epsilon}&=& \sum_{\epsilon}\delta_{p(\epsilon),p},\
\sum_\epsilon \equiv \prod_{i < j}\sum_{\epsilon_{ij}=0,1}.
\label{averageo12}
\ee
Namely, we sum over different ``samples" with the same value $p$ of 
connectivity, where 
each sample has different configurations of $\epsilon_{ij}$. 
To judge phase boundaries, 
we measure the internal energy $U$ and the
specific heat $C$ defined by
\begin{eqnarray}
U&\equiv&\langle E_{\rm I}\rangle_p,\ 
C\equiv\langle E_{\rm I}^2\rangle_p-\langle E_{\rm I}\rangle_p^2.
\end{eqnarray}
We note that, for the case of full connections $p=1$,
$\epsilon_{ij}=1$ and so the summation over $\epsilon_{ij}$ is
unnecessary.

\subsection{Model II: Quenched model with full connections}

In Model II, the synaptic variables $J_{ij}$ are treated as slowly varying 
quenched variables.
Then a suitable way to take average may be to (i) consider a configuration
of $J_{ij}$, which we call a sample, generated by certain probability
$P(J)$ and take an average over first variables $S_i$, and then (ii)
take average over different samples of $J_{ij}$. Explicitly, as $P(J)$ 
we take the Boltzmann factor of the reverberation term ($c_2$-term) of energy,
and write the average over $P(J)$ by 
\be 
\la f(J)\ra_J \equiv \sum_J f(J)P(J).
\ee
  Then we have the final average $\la O(S,J) \ra$ of $O(S,J)$ as
\begin{eqnarray}
E_1(S,J)&=&-c_{1}\sum_{i<j}S_{i}J_{ij}S_{j},\ \ 
Z_1(J)=\sum_{S}\exp\left(-E_1(S,J)\right),\nn
\langle O(S,J)\rangle_{S}&\equiv&\frac{1}{Z_1(J)}\sum_{S}O(S,J)\exp\left
(-E_1(S,J)\right),\nn
E_2(J)&=&-\frac{c_{2}}{N}\sum_{i<j<k}J_{ij}J_{jk}J_{ki},\ \ 
Z_{2}=\sum_{J}\exp\left(-E_2(J)\right),\nn
P(J)&=&\frac{1}{Z_2}\exp\left(-E_2\right),\ \ 
\la f(J) \ra_J\equiv \sum_{J}f(J)P(J),\nn
\langle O(S,J)\rangle&=&\la\la O(S,J)\ra_S\ra_J.
\label{averageo2}
\end{eqnarray}
Similar treatment has been adopted
  in the theory of SG\cite{sg, skmodel}.
However, the distribution $P_{\rm SG}(J)$ of $J_{ij} \in (-\infty,
\infty)$ is taken there as a Gaussian form,
\be
P_{\rm SG}(J) \propto \exp\left(-a\sum_{i<j}(J_{ij}-J_0)^2\right),
\label{psgj}
\ee
which has no correlations among $J_{ij}$ in strong contrast with
$P(J)$ of (\ref{averageo2}). 
Also we note that $P_{\rm SG}(J)$
of (\ref{psgj}) loses Z(2) gauge symmetry for $J_0 \neq 0$\cite{thoulouse}.

As the thermodynamic quantities, we consider
\begin{eqnarray}
U&\equiv&\langle E_1(S,J)\ra,\nn
C&\equiv&\la E_1(S,J)^2\ra-\la\langle E_1(S,J)\rangle_S^2\ra_J.
\label{ucmodel2}
\ee
We also measure the following order parameters $m$ and $q$,
\be
m&\equiv&\frac{1}{N}\sum_{i}\langle S_{i} \rangle
=\frac{1}{N}\sum_{i}\la\langle S_{i}\rangle_{S}\ra_J,\nn
q&\equiv&
\frac{1}{N}\sum_{i}\la\langle S_{i} \rangle_{S}^2\rangle_{J}.
\label{mqmodel2}
\end{eqnarray}
$m$ and $q$ are the generalization of the order parameters
of SG\cite{sg,skmodel} to the present model. Namely,
if $m=0$ and $q\neq 0$, then we call this the SG phase. 
Here one may guess that $m$ is  the average of a {\it gauge-variant} quantity
$S_i$ and should vanish according to Elitzur's theorem\cite{elitzur}.
In fact, we show in Appendix A 
that this theorem holds also for quenched systems, so $m=0$ always.
We shall see that our simulation confirms this point. In contrast, $q$
is gauge-invariant and free from Elitzur's theorem, 
and develops nonvanishing values 
in some regions.  

One may expect that the similar set of averages
\be
m_J&\equiv&\frac{1}{N_l}\sum_{i<j}\langle J_{ij} \rangle
=\frac{1}{N_l}\sum_{i<j}\la\langle J_{ij}\rangle_{S}\ra_J,\nn
q_J&\equiv&
\frac{1}{N_l}\sum_{i<j}\la\langle J_{ij} \rangle_{S}^2\rangle_{J},
\end{eqnarray}
are able to serve as order parameters for the ``gauge-glass" phase.
However, they give rise to trivial values
\be
m_J=0,\ q_J=1,
\end{eqnarray}
and don't work as order parameters.

\subsection{Model III: Quenched lattice model}

The quenched lattice model is defined in a similar manner as Model II
but on the 3D lattice.
Its energies $E_1(S,J)$ and $E_2(J)$ are given by 
\begin{eqnarray}
E_1(S,J)&=&-c_{1}\sum_{x}\sum_{\mu=1}^3S_{x+\mu}J_{x\mu}S_{x},\nn
E_2(J)&=&-c_{2}\sum_{x}\sum_{\mu<\nu}
J_{x\nu}J_{x+\nu,\mu}J_{x+\mu,\nu}J_{x\mu}.
\ee
Each term has the same form as in $E_0$ of Eq.(\ref{energy0}).
We first take the average over $S_x$ as
\be
\hspace{-0.8cm}
Z_1(J)&=&\sum_{S}\exp\left(-E_1(S,J)\right),\nn
\hspace{-0.8cm}
\langle O(S,J)\rangle_{S}&=&\frac{1}{Z_1(J)}
\sum_{S}O(S,J)\exp\left(-E_1(S,J)\right),
\ee
where $\sum_S$ and $\sum_J$ are defined in Eq.(\ref{model0}).

Then quenched averages are taken w.r.t. $P(J)$ as
in Model II,
\be 
P(J)&=&\frac{\exp\left(-E_2(J)\right)}{Z_2},\ \
Z_2=\sum_{J}\exp\left(-E_2(J)\right),\nn
\langle O\rangle&=&\langle \la O\ra_S\rangle_J\equiv
\sum_{J}\langle O(S,J)\rangle_{S}P(J).
\label{annealpj}
\end{eqnarray}

As observables we measure $U,C,q$ which are defined
by the same expressions (\ref{ucmodel2}) and (\ref{mqmodel2})
as in Model II.

Before going to MC results of next section, 
we account here for some details of our MC method.
We first use Metropolis 
algorithm\cite{metropolis} for update of variables.
For some cases of large hysteresis (such as 
Fig.\ref{uc1confinementcoulomb}a below), 
we adopt the multicanonical method\cite{multicanonical}.
For Model I, the typical number of sweeps for single run is 5000, and
we estimate errors using data of 20 runs.
For Model II, typical sweep number for a fixed configuration
of quenched variable $J_{ij}$ is 5000$\times 20$, and  we repeat it 
for  typically 200 samples(configurations) of $J_{ij}$.
For Model III, we use the periodic boundary condition, and the 
typical number of sweeps is either 5000$\times 20$ over 200
samples or $500\times20$ over  1000 samples.

\clearpage

\section{Model I:
Annealed Model with full and partial connections}
\label{sec2}
\setcounter{equation}{0} 

In this section we present the results of MC simulations of 
Model I, the annealed model with full and partial connections.
 We study the case of $p=1$ in Sec.\ref{sec2}.1 and
 $p < 1$ in Sec.\ref{sec2}.2.
 
\subsection{full connections ($p=1.0$)}

\begin{figure}[h]
\begin{center}
\begin{minipage}{0.49\hsize}
\includegraphics[width=5.6cm]{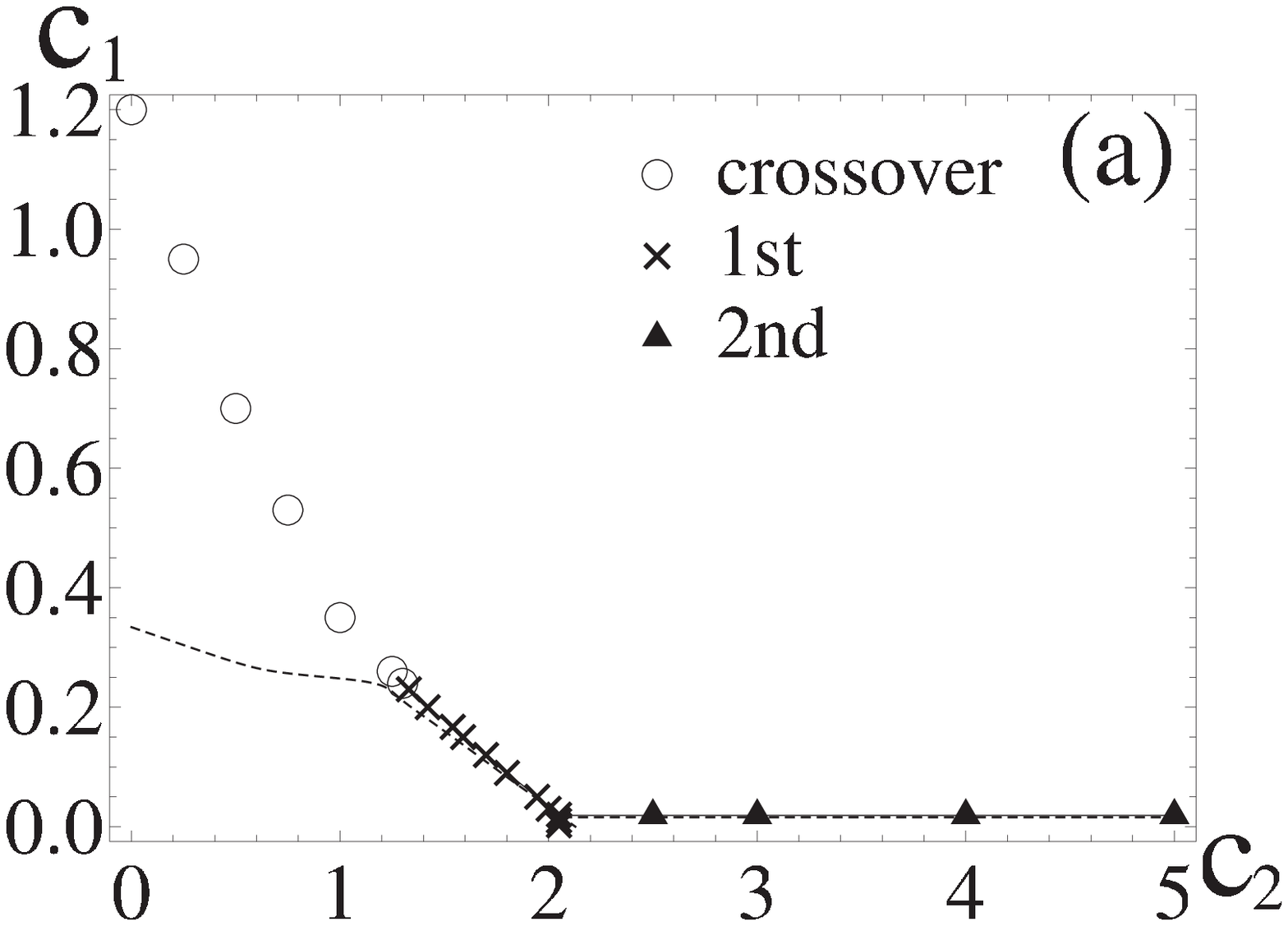}
\end{minipage}
\begin{minipage}{0.49\hsize}
\includegraphics[width=5.6cm]{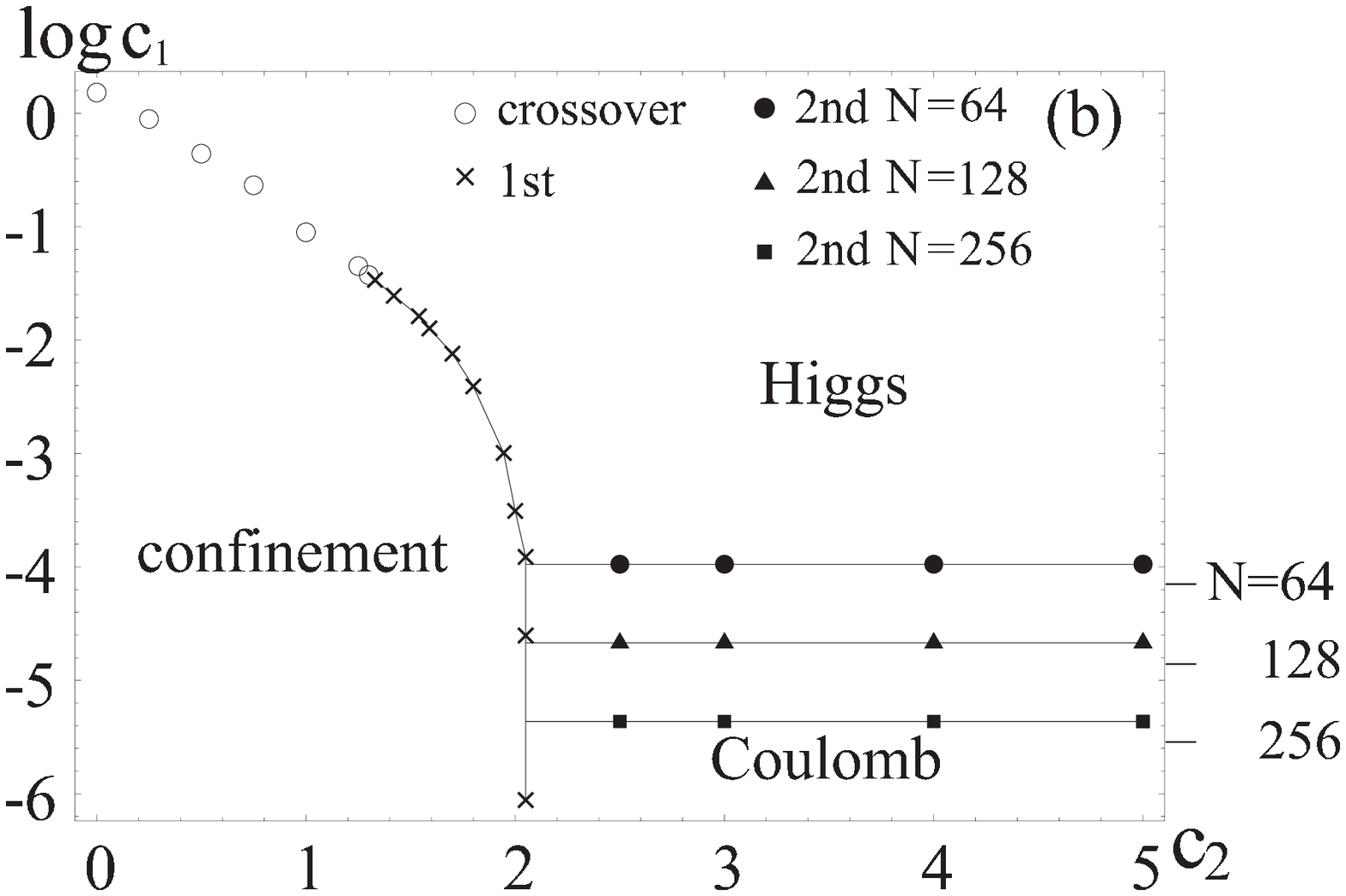}
\end{minipage}
\end{center}
\vspace{-0.2cm}
\caption{
Phase structure  in the $c_2$-$c_1$ plane of Model I
with full connections $p=1$. (a) $N=64$, (b) Several $N$. 
As indicated, there are first-order transitions, 
second-order ones and crossovers. The critical value $c_{1c}$ of $c_1$ 
 for $c_2 \lesssim 2.0$ 
has almost no $N$ dependence, whereas  $c_{1c}$ 
for $c_2 \gtrsim 2.0$ 
behaves as $O(1/N)$.
The latter $c_{1c}$ should approach to 
the critical value  of the infinite-range Ising model
$c_{1c}=1/N$ of Appendix C in the limit of $c_2\to\infty$ and large $N$.
It gives Log(1/N)=-4.16(N=64), -4.85(N=128),
-5.55(N=256), as marked on the vertical axis in (b).
In (a) the dashed curves are results of MFT of Appendix D.
They are almost same as MC results for $c_2 \gtrsim 1.3$ but
give first-order transitions instead of crossovers for 
$0 \leq c_2 \lesssim 1.3$.
}
\label{phase1}
\end{figure}

In Fig.\ref{phase1} we present the phase diagram for $p=1.0$ in 
the $c_2-c_1$ plane.
There is one crossover curve and two curves of phase boundaries: \\
(i) $0 \le c_2 \lesssim 1.3$ Crossover;\\
(ii)$1.3 \lesssim c_2 \lesssim 2.0$; First-order transitions;\\
(iii)$2.0 \lesssim c_2 $; Second-order transitions.\\
They are determined by  the peak of $C$ and possible  discontinuity of $U$.
Before going into the details of the analysis of each transition,
let us present some analytic arguments (a-c) related to our MC results.\\

(a) The case $c_2=0$ can be analyzed exactly by the single-link sum, 
because  $J_{ij}$ is factorized  in the $c_1$-term. For example,
for $p=1$, the partition function is calculated as
\be
Z_{\rm I}(c_2=0)&=&\sum_{S}\sum_J\prod_{i<j}\exp(c_1S_iJ_{ij}S_j)=
\sum_S\prod_{i< j}\sum_{J_{ij}}\exp(c_1S_iJ_{ij}S_j)\nn
&=&\sum_S\left(2\cosh c_1\right)^{N_{\rm l}}=2^{N}\left(2\cosh c_1\right)^{N_{\rm l}},
\ee
where we used $\cosh(c_1S_iS_j)=\cosh c_1$ due to $S_i^2=1$.

\clearpage

\nin
Because $Z_{\rm I}$ has no singularity in $c_1$, there are no phase transitions
at $c_2=0$. This is consistent with Fig.\ref{phase1} where we have a crossover
at $c_2 \lesssim 1.3$.
In Appendix B, we study the case $p < 1$ where
$U$ and $C$ are shown to have a form $U(p)=p\; U(p=0),\ C(p)=p\; C(p=0)$.
\\

(b) In the region of large $c_2$ (explicitly speaking,
$c_2 \gtrsim 2.0$), fluctuations of $J_{ij}$ are
small and so the $c_2$-term of $E_{\rm I}$ becomes almost constant.
We note that the (fully-connected)  $c_2$-term  of the energy has
the lowest value $U=-c_2 {}_NC_3/N$ 
at $J_{ij}J_{jk}J_{ki}=1$ for all the triangle $ijk$.
This is achieved by the trivial configuration $J_{ij}=1$ and 
its gauge transformed ones\cite{nodegeneracy}.

To estimate the critical value $c_{1c}$ in this region,
one may set $J_{ij}=1$.
Then the  behavior of the system at $p=1$ is controlled by the 
$c_1$-term with $J_{ij}=1$.
That is, the system reduces to the so-called 
infinite-range Ising (IRI) spin  model,
the energy $E_{\rm IRI}$ of which is given by
\be
E_{\rm IRI} = -c_1\sum_{i < j}S_i S_j.
\label{irm}
\ee   
In Appendix C we analyze the IRI model 
by the saddle-point method, which gives rise to the exact result
for $N \to \infty$.
We see there that the nontrivial phase structure is obtained
for small $c_1$ such that $c_1 \propto N^{-1}$.
In fact, there is a second-order phase transition at $c_1 N=1$.
Then, if we consider the internal energy $U_{1,2}$ and
the specific heat $C_{1,2}$ of the $c_{1,2}$-term of the energy
separately as
\be
E_{\rm I}&=&E_1+E_2,\nn
U_a&=&\la E_a\ra,\ C_a=\la E_a^2\ra-\la E_a\ra^2,\ (a=1,2),
\label{separateuc}
\ee 
we expect $U_1, C_1=O(N),\ U_2, C_2=O(N^2)$ for $c_2 \gtrsim 2.0$.
That is, the magnitudes of $U_a, C_a$ are of different order for $a=1,2$
for the choice $c_1=O(N^{-1})$ and $c_2=O(N^0)$.
This consideration is supported by Fig.\ref{phase1}b, which shows
that the exact value of $c_{1c}$ for $c_2 \gtrsim 2.0 $ 
is very near to the value of IRI model, i.e., $c_{1c}=1/N$.
The discrepancy is attributed to the corrections
of $O(N^{-2})$ and $O(c_2^{-1})$.

At first, it may sound strange to have a phase transition
for $c_2 \gtrsim 2.0$ because $U_1$ and $U_2$ are unbalanced there.
However, that transition is {\it not} due to the competition between these
two terms, but the competition between the energy and entropy of
the $c_1$-term itself (with fixed $J_{ij}$'s) as explained above. 
Therefore, that unbalance does not matter.  

On the contrary, as we shall see, in the region $c_2 \lesssim 2.0$,
the critical value $c_{\rm 1c}$ of $c_1$ is  $O(N^0)$,
which is almost independent of $N$. Then, both $U_1$ and $U_2$
is of $O(N^2)$ with $c_1, c_2 =O(N^0)$.
In summary, we always take $c_2 =O(N^0)$, and therefore $U_2=O(N^2)$, 
whereas
we allow $c_1$ to vary from $O(N^{-1})$ to $O(N^0)$, so
$U_1$ varies from $O(N)$ to $O(N^2)$ accordingly. \\

(c) Let us comment on the phase structure obtained by 
MFT based on a variational principle\cite{feynman}, 
which is summarized in Appendix D. 
As shown in Fig.\ref{phase1}a, it predicts that the first-order 
confinement-Higgs transition
continues down to $c_2=0$ instead of the MC results
which has an end point $c_2\simeq 1.3$ at which the first-order
terminates and becomes crossover. We note here that
this MFT does not necessarily predict the correct results
even in the limit $N\to \infty$ in contrast with the Sherrington-Kirkpatrick
model\cite{skmodel}. This is due to the $c_2$-term which has mutual couplings
among $J_{ij}$.\\

Let us see the details of each phase transition(crossover) 
in Fig.\ref{phase1}.
We consider the following four cases (i)-(iv) in order.\\

\begin{figure}[t]
\begin{center}
\hspace{-0.6cm}
\includegraphics[width=4.8cm]{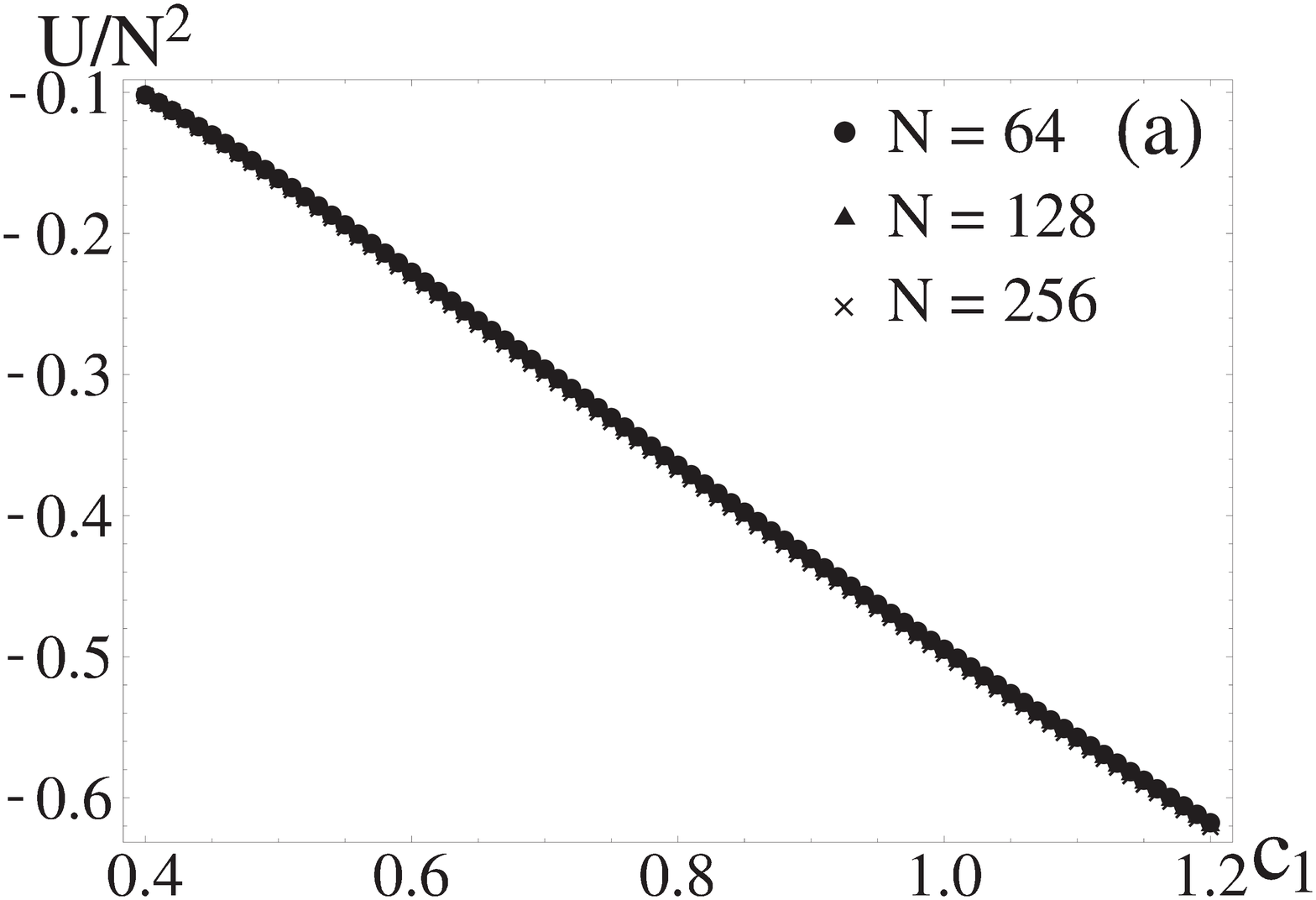}
\includegraphics[width=5cm]{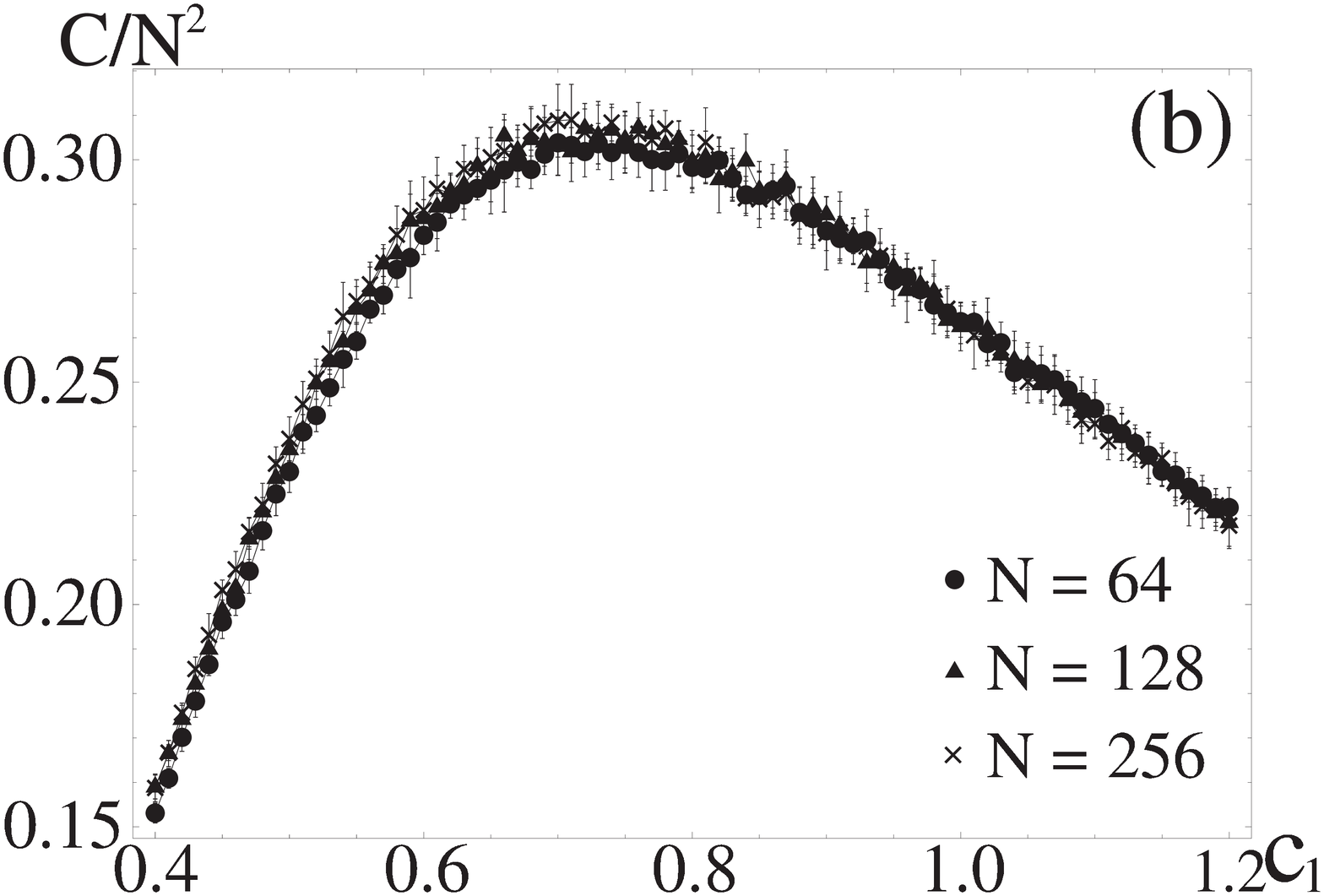}\\
\end{center}
\caption{
(a) $U/N^2$ and (b) $C/N^2$ of Model I at $c_2=0.5$ and $p=1.0$.
There is no systematic development of the peak of $C$, so there are 
no transitions but a crossover. 
}
\label{uc1confinementhiggs}
\end{figure}

\nin
(i) Between confinement and Higgs phases($0 \leq c_2 \lesssim 1.3$)

In Fig.\ref{uc1confinementhiggs} we present $C$ and $U$
between the confinement and Higgs phases at $c_2=0.5$.
The round peak of $C$ has no development as $N$ increases.
So we conclude that there is only a crossover between these
two phases. This is consistent with the above argument (a) for $c_2=0$
that there is no phase transition along $c_2=0$.\\

\nin
(ii) Between confinement and Higgs phases($ 1.3 \lesssim c_2 \lesssim 2.0$)

\begin{figure}[t]
\begin{center}
\includegraphics[width=5.3cm]{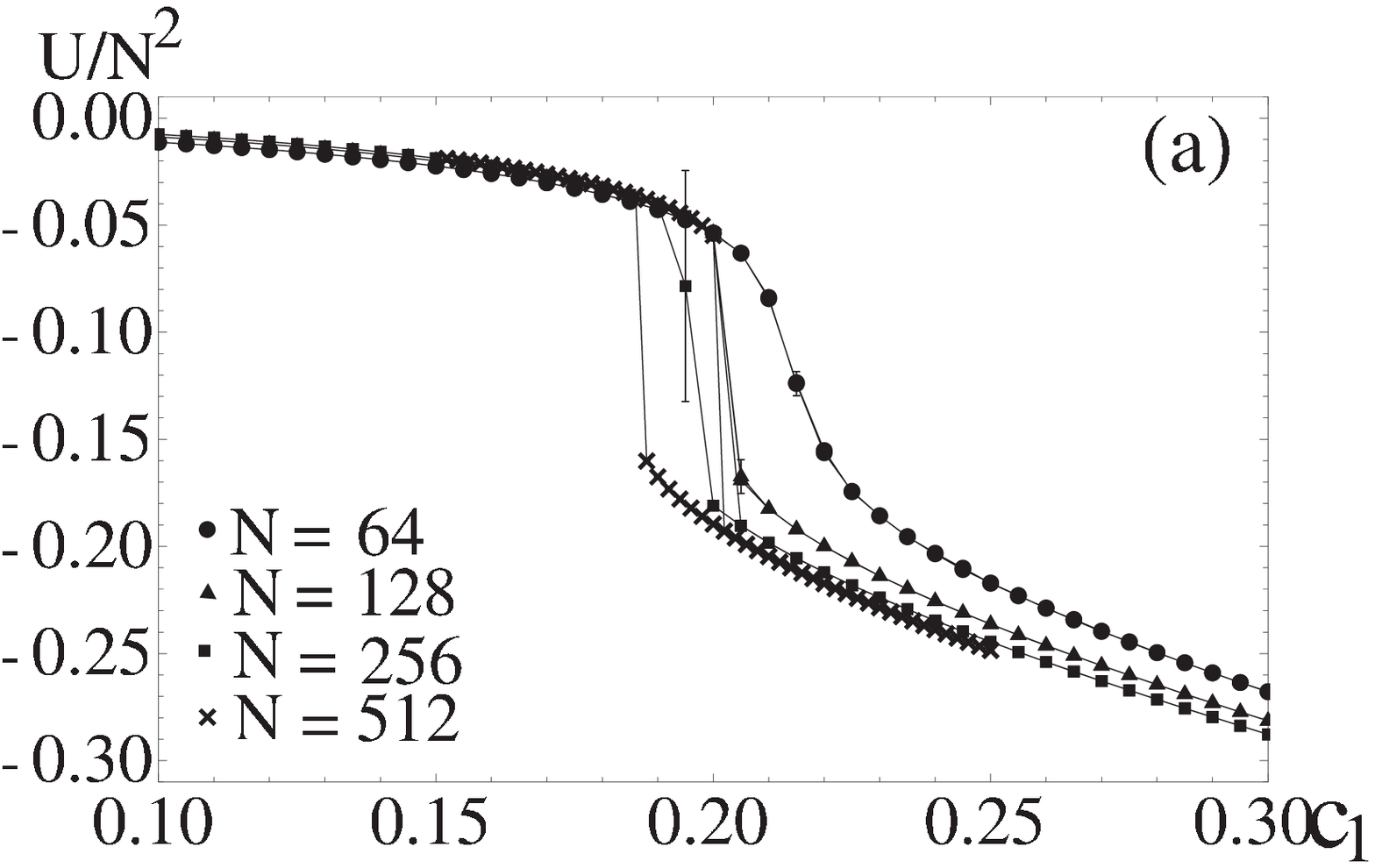}
\includegraphics[width=5.3cm]{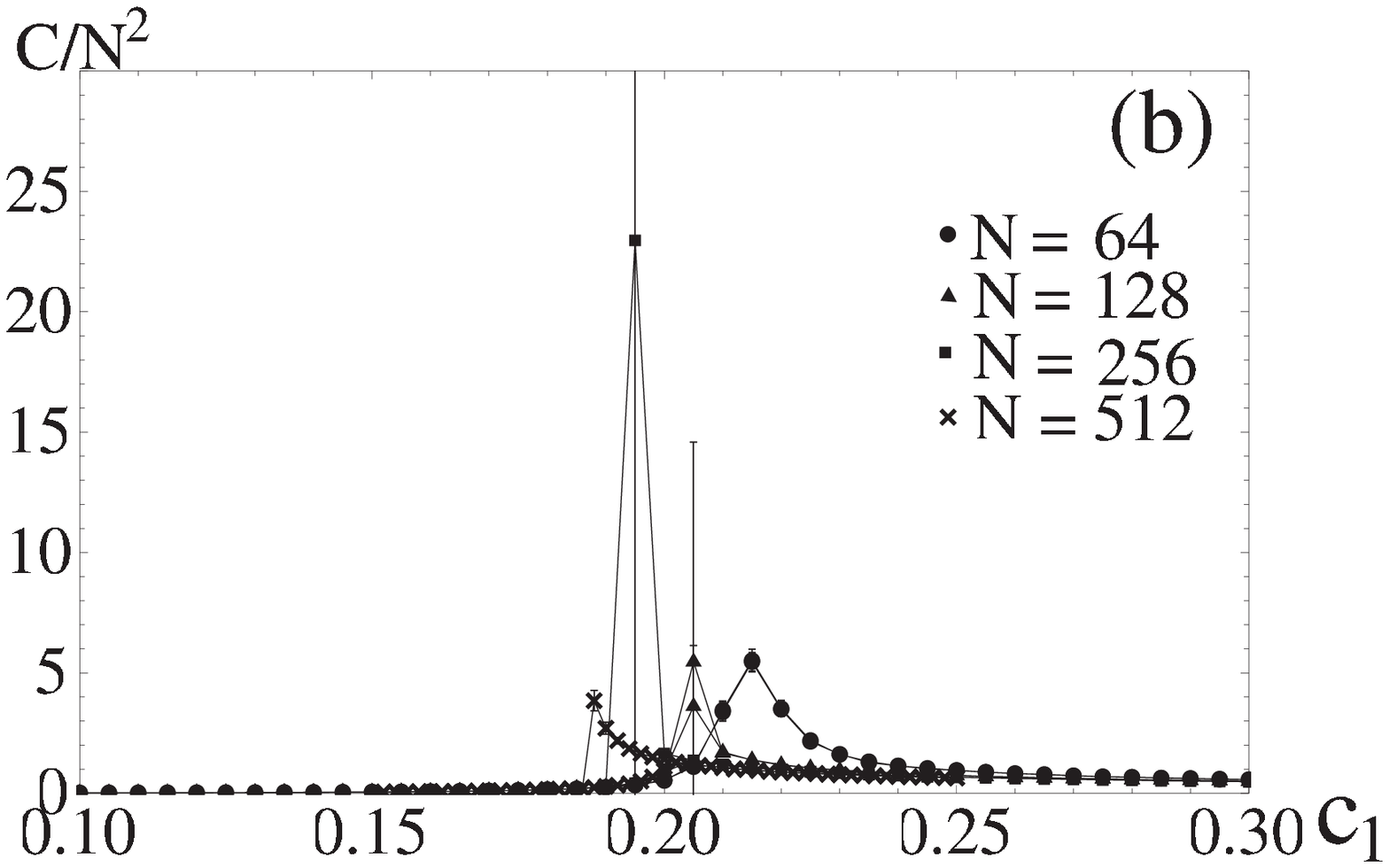}\\
\end{center}
\caption{
(a) $U/N^2$, (b) $C/N^2$  of Model I vs. $c_1$ at $c_2=1.4$ and $p=1.0$.
There is a hysteresis in $U$, 
and the peak of $C$ develops rapidly as $N$ increases.
So there is a first-order  transition at $c_1 \simeq 0.2$. 
}
\label{uc1confinementhiggs2}
\end{figure}

In Fig.\ref{uc1confinementhiggs2} we present $C/N^2$ and $U/N^2$
for $c_2=1.4$.
The peak of $C$ is sharp and develops rapidly as $N$ increases, and 
also $U$ exhibit a jump (small hysteresis).
So we conclude that there is a first-order transition
between these two phases in this region of $c_2$.\\

\nin
(iii) Between Higgs and Coulomb phases($2.0 \lesssim c_2$)

\begin{figure}[h]
\begin{center}
\includegraphics[width=5cm]{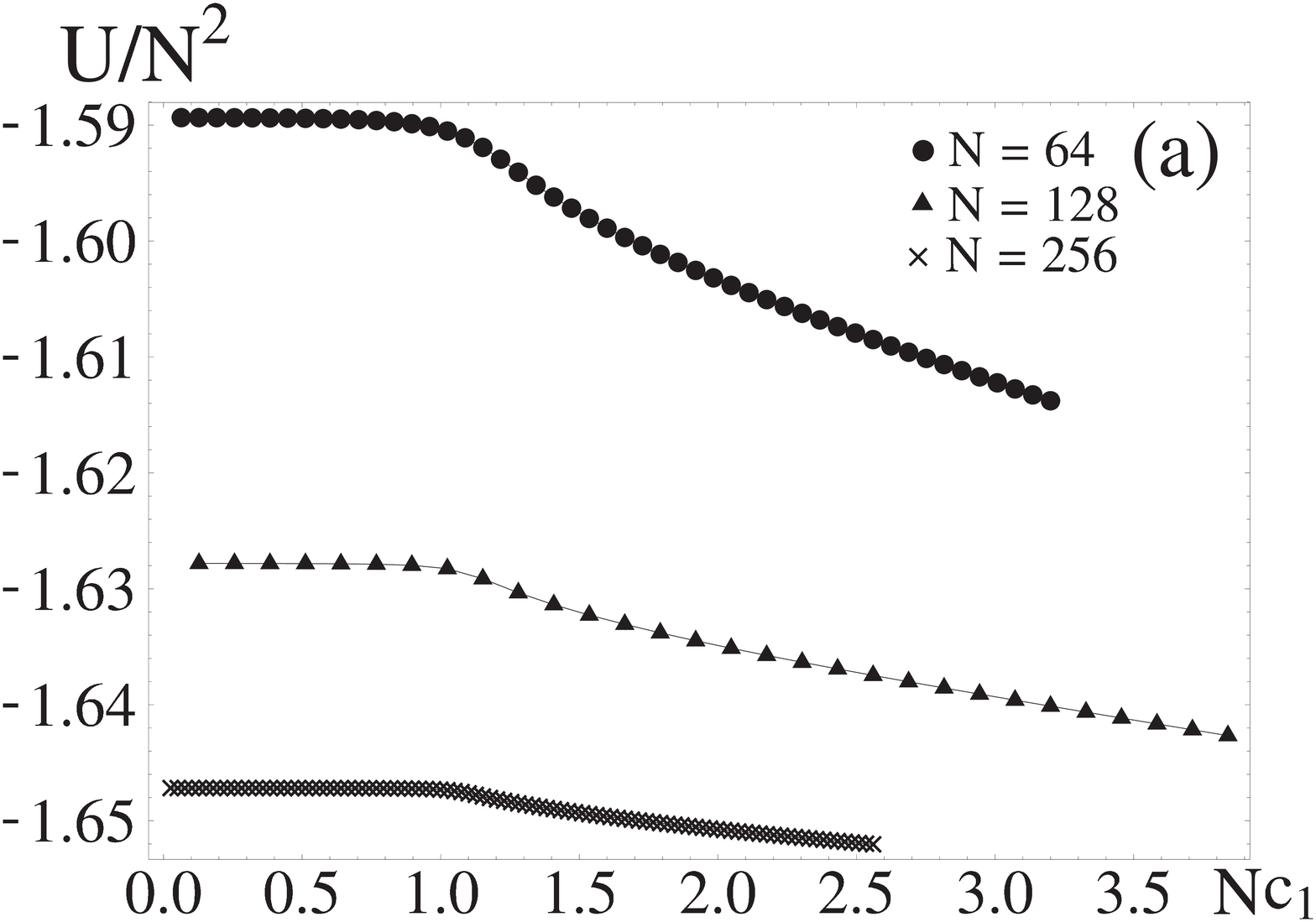}
\includegraphics[width=5.2cm]{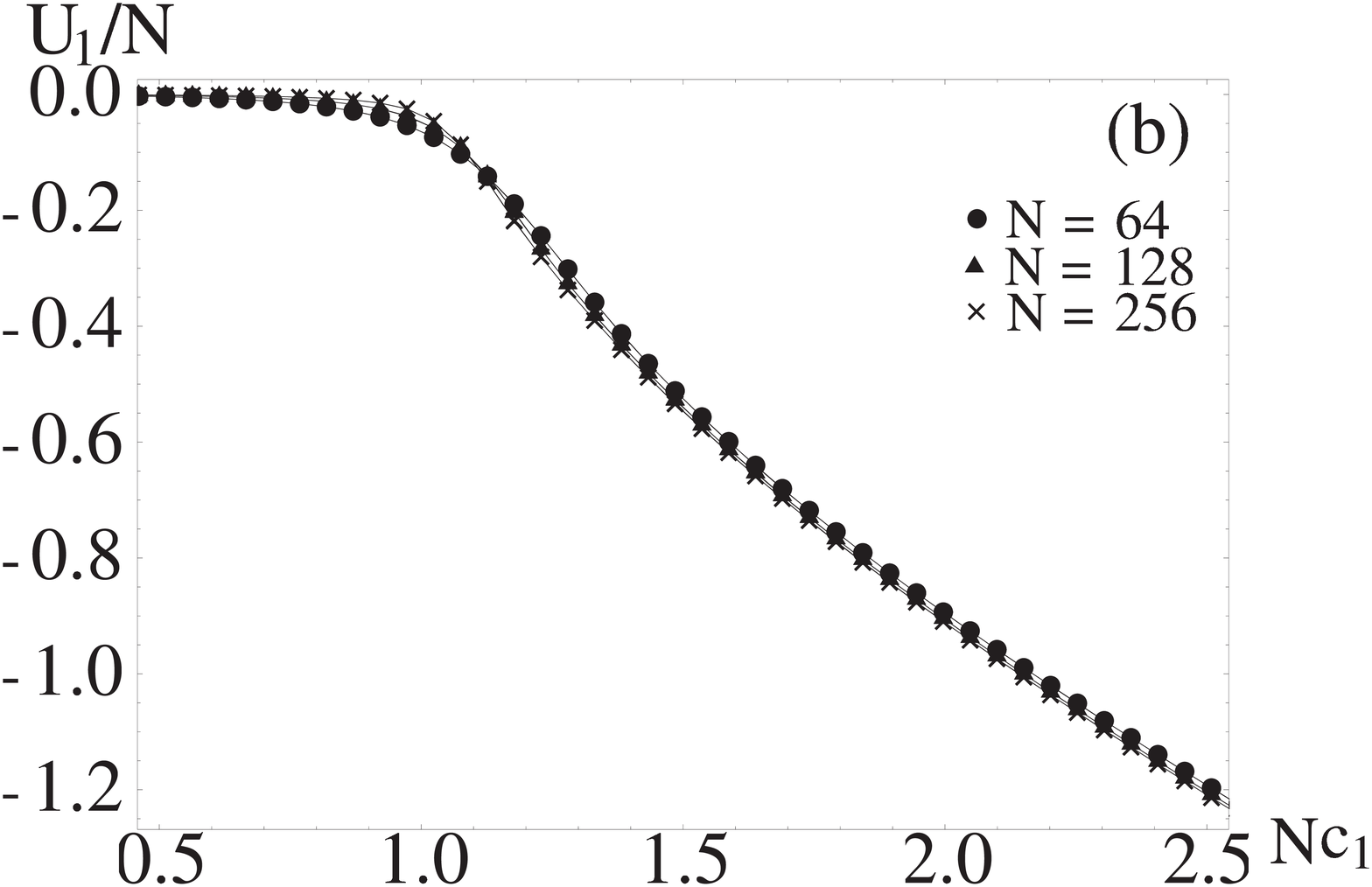}\\
\vspace{0.3cm}
\includegraphics[width=5.2cm]{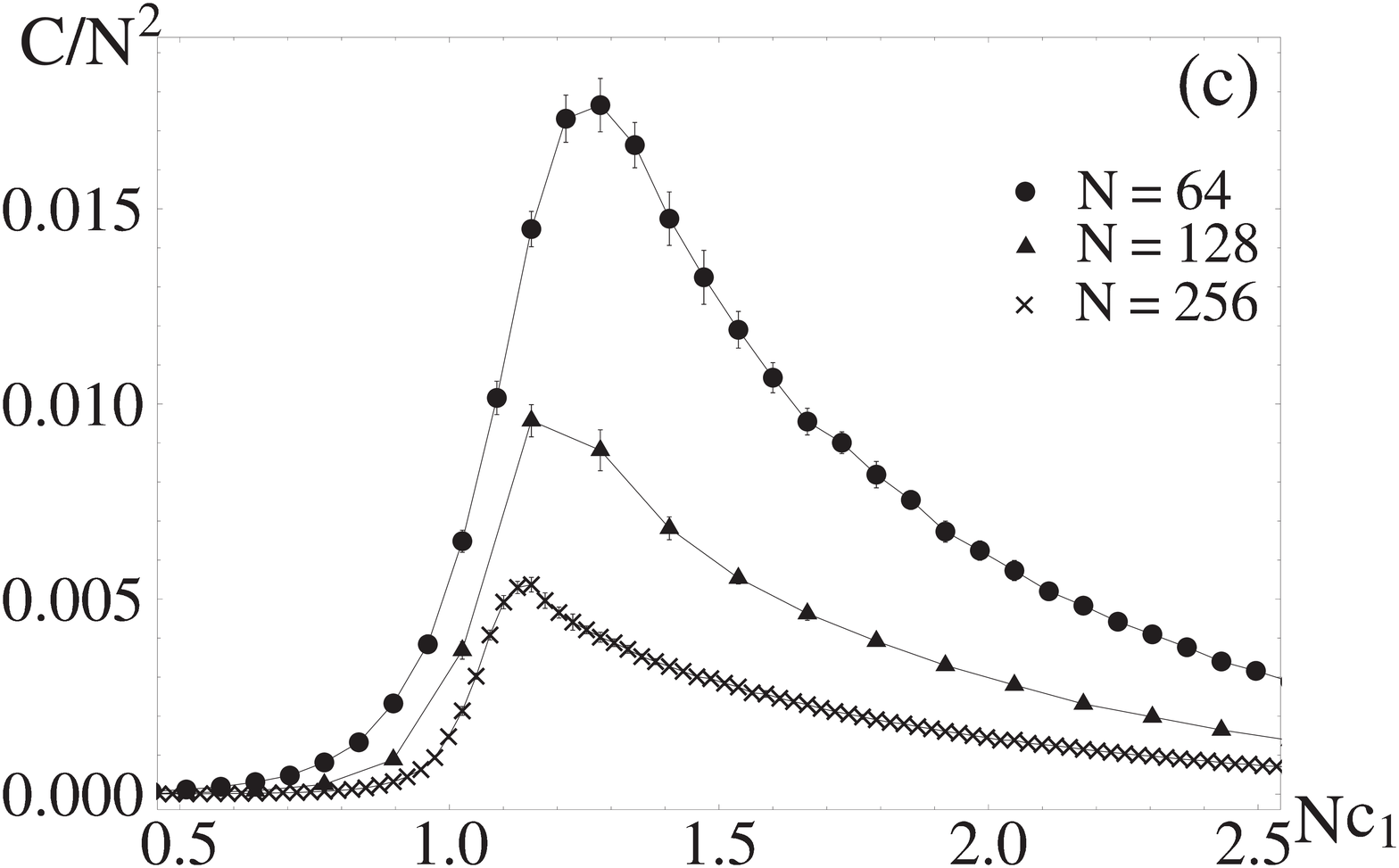}
\includegraphics[width=4.9cm]{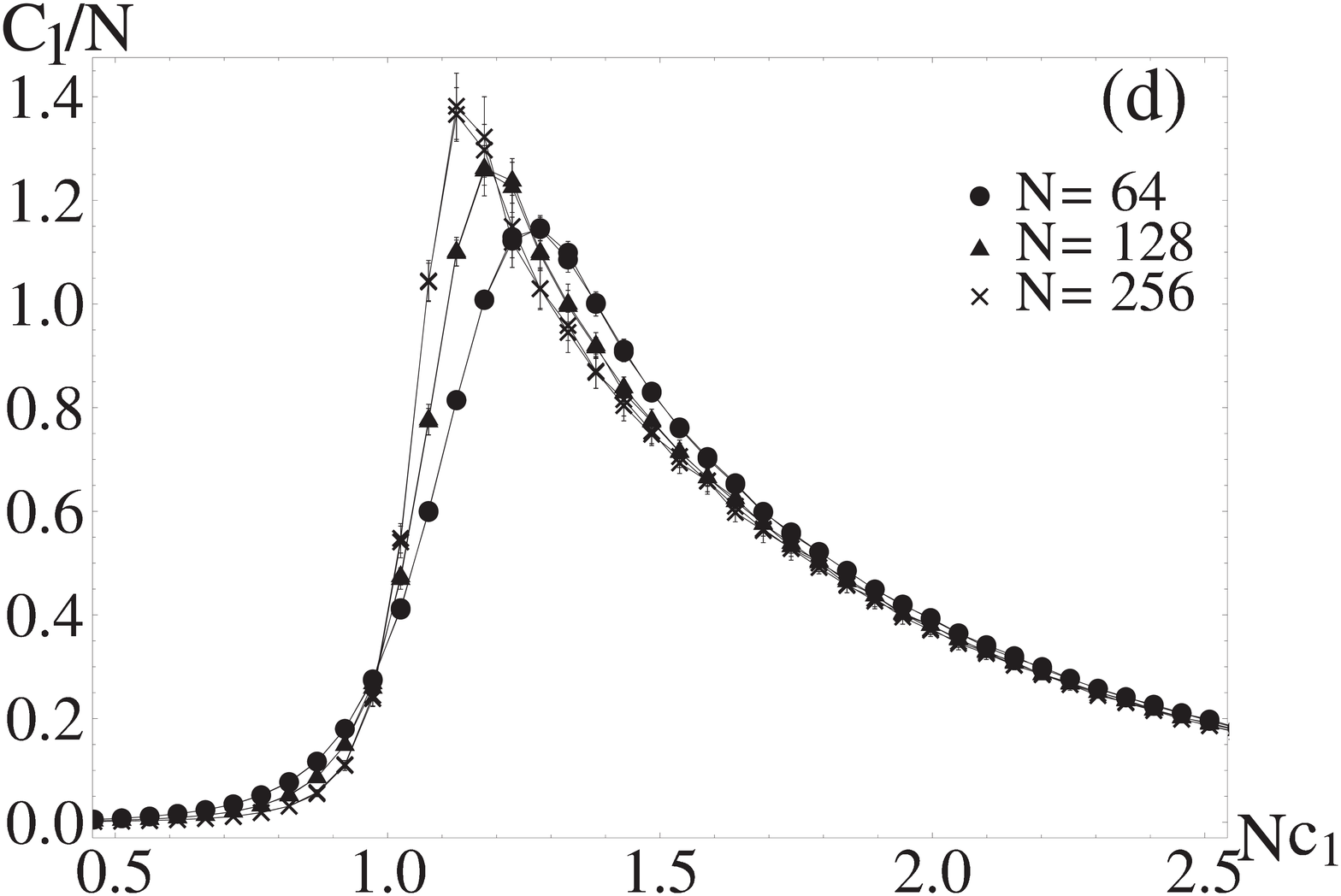}\\
\end{center}
\vspace{-0.5cm}
\caption{
(a) $U/N^2$, (b) $U_{1}/N$, (c) $C/N^2$ and (d) $C_{1}/N$ 
of Model I vs. $Nc_1$ at $c_2=10.0$ and $p=1.0$ 
($U_1$ and $C_1$ are  defined in Eq.(\ref{separateuc})).
There is a second-order transition at $c_1\simeq O(1/N)$.
In the limit of $c_2\to\infty$ and $N\to \infty$, the peak location of $C$
should approach to the value $Nc_{1c} =1.0$ of the IRI model of 
Appendix C.
}
\label{uc1higgscoulomb}
\end{figure}

In Fig.\ref{uc1higgscoulomb}
we present $U_a$ and $C_a$ of Eq.(\ref{separateuc}) for $c_2=10.0$.
There is a systematic $N$ dependence of the peak of $C_1/N$,
indicating a second-order transition at $c_1\simeq 1/N$. 
This is consistent with the 
above result (b) of IRI spin model that corresponds to $c_2=\infty$.
The critical value $c_{1c}N$ approaches to 1 as expected. \\

\clearpage

\eject

\begin{figure}[t]
\begin{center}
\hspace{-0.6cm}
\includegraphics[width=5.2cm]{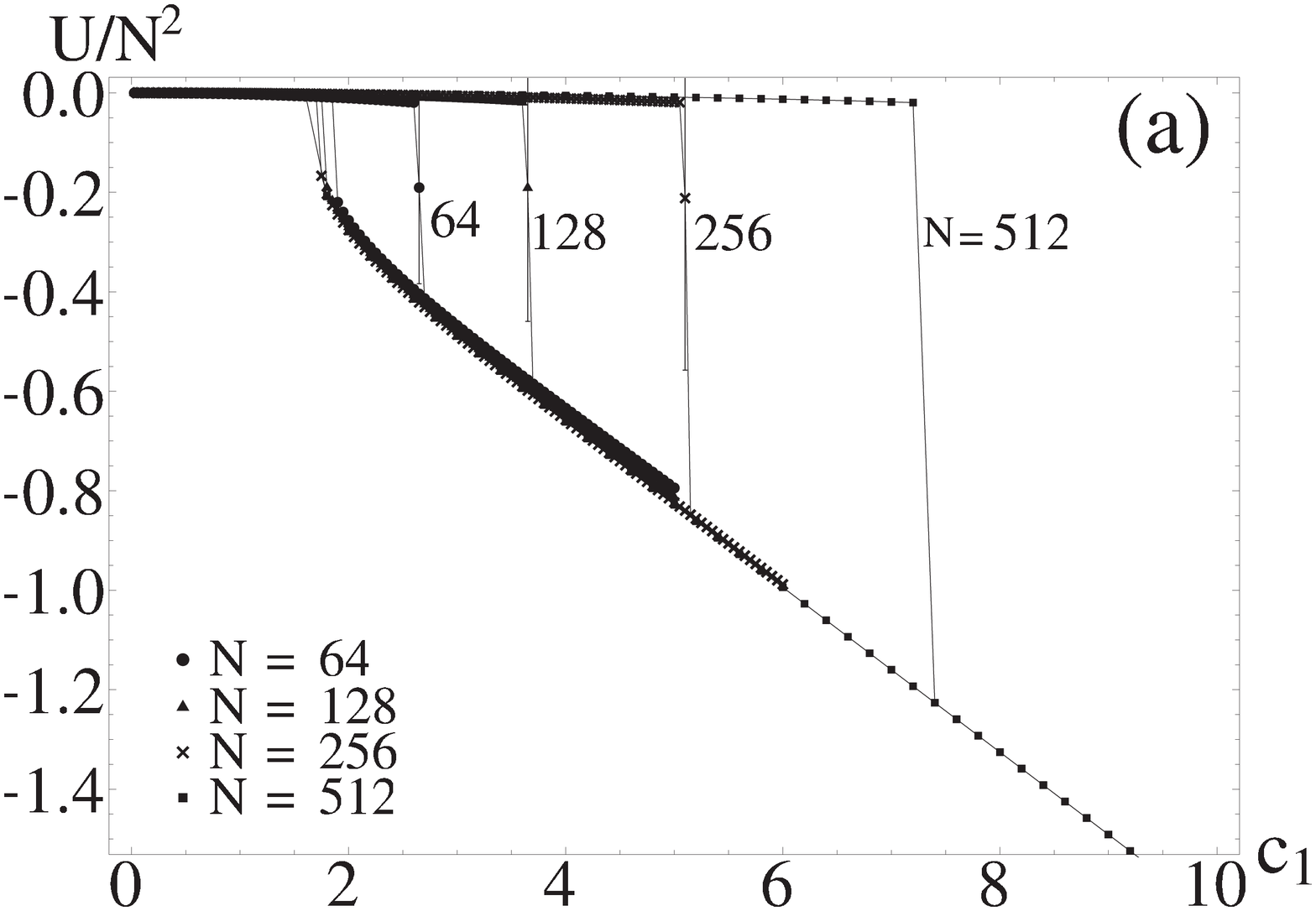}\\
\includegraphics[width=5.2cm]{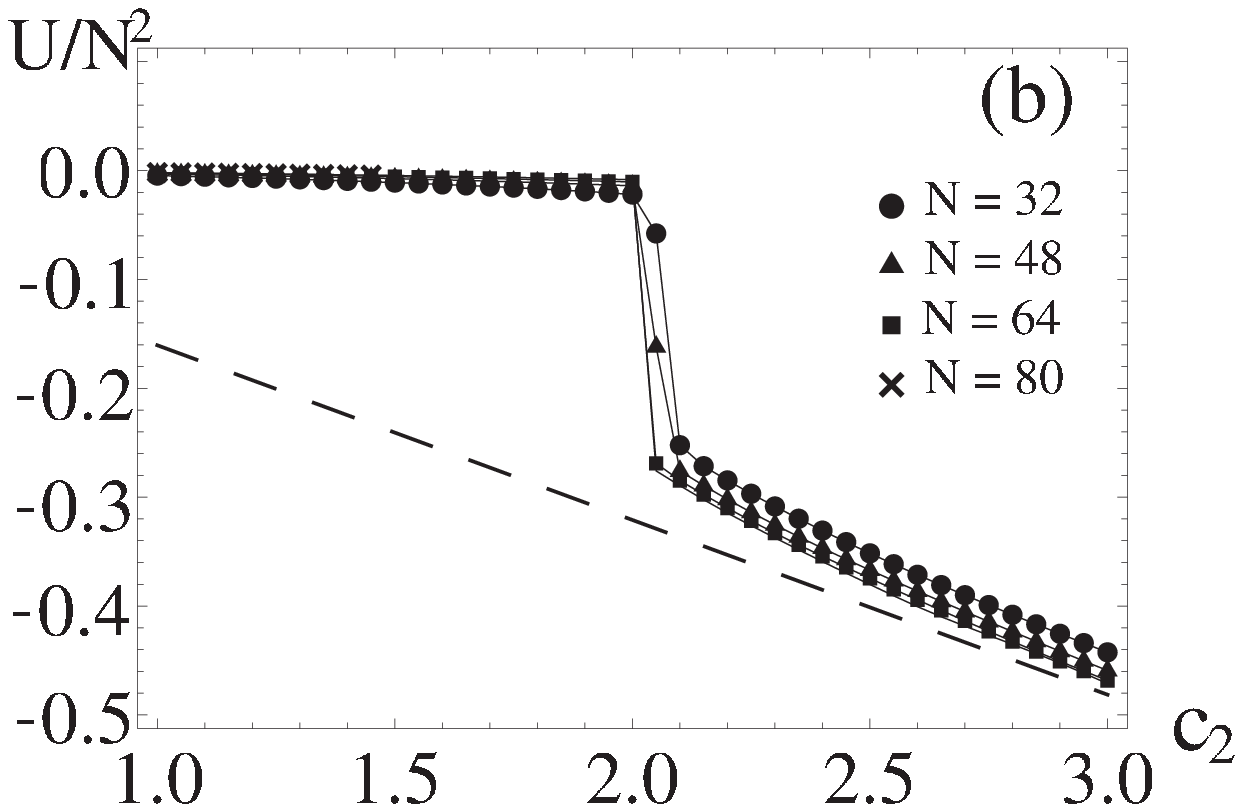}
\includegraphics[width=5.0cm]{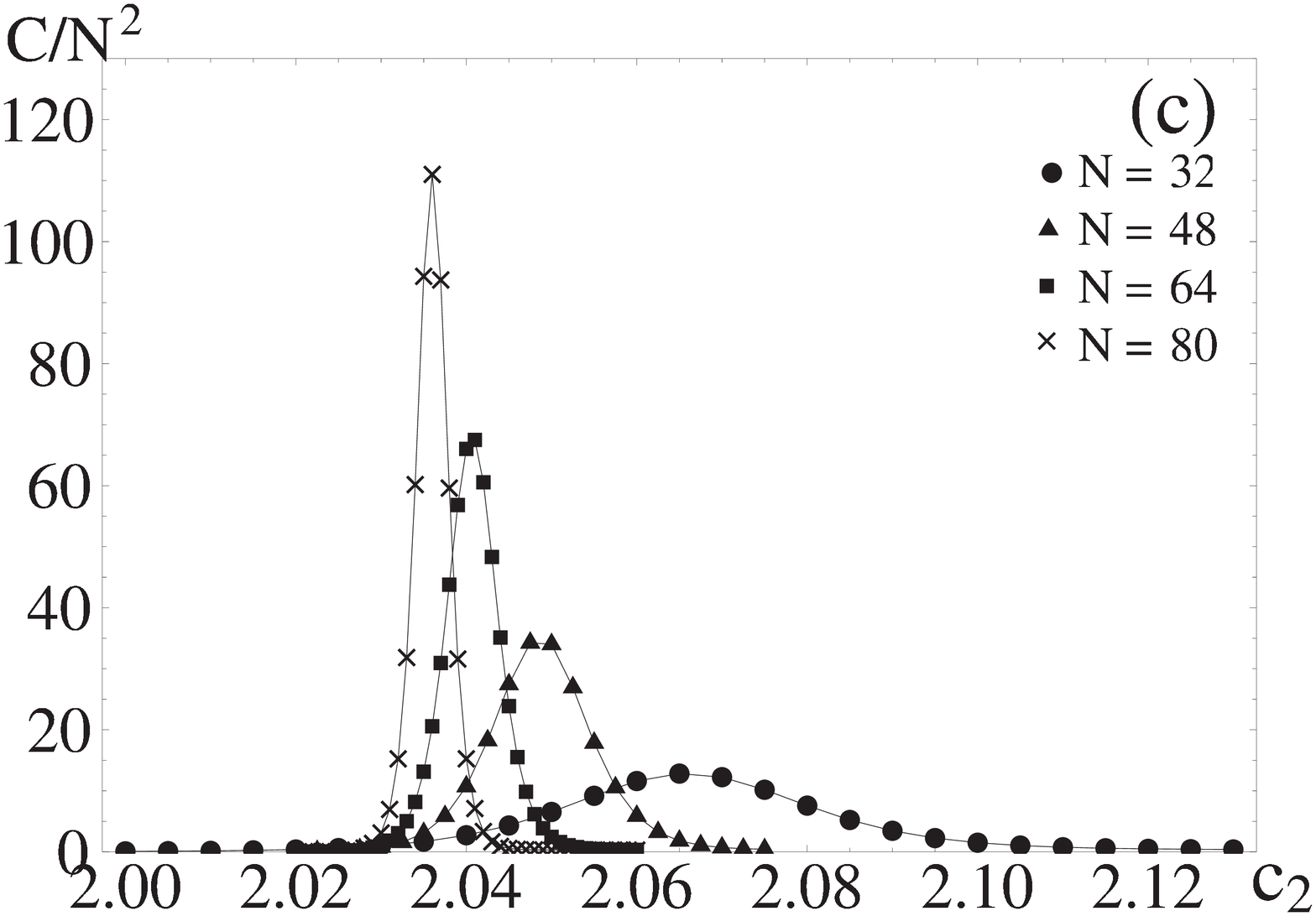}
\end{center}
\caption{
$U/N^2$ and $C/N^2$ of Model I at $c_1=0.0$ and $p=1.0$.
(a) $U/N^2$ by Metropolis updates,
(b) $U/N^2$ and (c) $C/N^2$ by multicanonical method.
The hysteresis in $U$ and the strongly $N$-dependent
development of the sharp peak in $C$ exhibit a first-order transition. 	
The range of hysteresis in $U$ is reduced significantly by
multicanonical method. The dashed line in (b) shows $U/N^2=-{}_NC_3c_2/N^3$
with $N=80$ for
the completely ordered case $J_{ij}J_{jk}K_{ki}=1$.
}
\label{uc1confinementcoulomb}
\end{figure}

\nin
(iv) Between confinement and Coulomb phases

In Fig.\ref{uc1confinementcoulomb} we present $U/N^2$ and $C/N^2$
for $c_1=0.0$. There is a sharp $N$ dependence of the peak of $C$ and
hysteresis on $U$, so there is a first-order transition.
This is in contrast to Model 0, which exhibits a second-order transition
between the confinement and Coulomb phases\cite{kemukemu}.
We note that this difference of the order of the transition at $c_1=0$ 
does not come from
the difference of the power of the $c_2$ interaction,
i.e., the quartic one $JJJJ$ and the cubic one $JJJ$.
In fact, the MFT of Appendix D 
supports this interpretation explicitly, because it predicts a first-order
transition for both cases (See Ref.\cite{kemukemu} and Appendix D). 
This difference of the transition order should reflect
the difference of connectivity, that is 
$p=1.0$ in Model I and $p=O(1/N)$ in Model III (See discussion 
of Sect.3.2 for more details).

In the Coulomb phase, the configuration of $J_{ij}$ is strongly
ordered after the transition at $c_2 \sim 2.0$.
In fact, Fig.\ref{uc1confinementcoulomb}b shows that
$U/N^2$ is near its saturated value $U/N^2=-{}_NC_3c_2/N^3$ 
(shown by the dashed line) which is given by setting $J_{ij}J_{jk}J_{ki}=1$.

We also note that the straightforward Metropolis algorithm
 (Fig.\ref{uc1confinementcoulomb}a) 
gives rise to a huge hysteresis in $U$, while a multicanonical method
(Fig.\ref{uc1confinementcoulomb}b,c) gives rise to a moderate
hysteresis. The latter is useful to locate a more accurate location
of the transition point.

\subsection{partial connections ($p < 1$)}

\begin{figure}[h]
\begin{center}
\includegraphics[width=5.8cm]{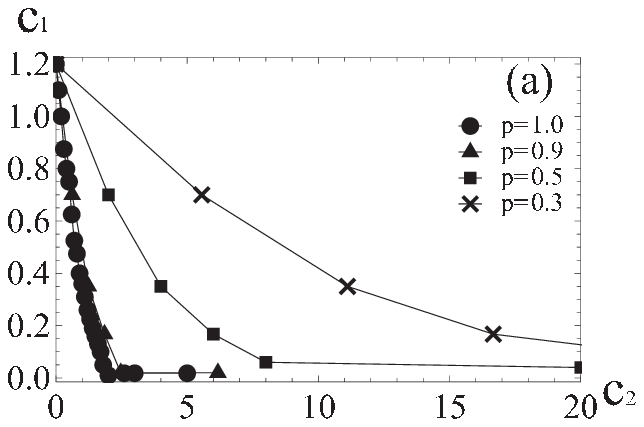}
\includegraphics[width=6cm]{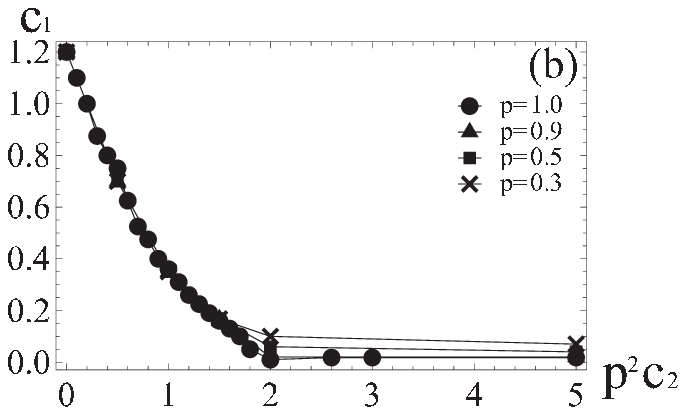}
\end{center}
\caption{
Phase diagram   of Model I
with partial connectivities, $p=0.9, 0.5, 0.3$ together with
$p=1$ ($N=64$) (a) in the $c_2$-$c_1$ plane
and (b) in the $ p^2 c_2$-$c_1$ plane. 
In (b), the curves for $0 < p^2 c_2 \lesssim 1.3$ 
show crossover, $1.3 \lesssim p^2 c_2 \lesssim 2.0$ show 
first-order transitions, and those
for $2.0 \lesssim p^2 c_2$ show second-order transitions.
The critical values in the region of  $2.0 \lesssim p^2 c_2  $
is $c_{1c}=O(N^{-1})$ for general $N$.
}
\label{phase1partial}
\end{figure}

Let us consider the phase structure for partial connections.
In Fig.\ref{phase1partial} we present the phase diagram
in the $c_2$-$c_1$ plane for $p=0.9, 0.5, 0.3$ in which
we plot the location of the peak of $C$. As in the case of $p=1.0$,
this peak exhibits  
crossover for the small $c_2$ region  ($p^2 c_2 \lesssim 1.3$) 
and  first-order transitions for 
$1.3 \lesssim p^2 c_2 \lesssim 2.0$. The curves for
$2.0 \lesssim p^2 c_2 $ are of  second-order transitions,
which have the critical value $c_{1c}\propto 1/N$ for general $N$
as explained for $p=1.0$.

In Fig.\ref{phase1partial}b we present 
the critical curves $c_{1c}$ with various $p$ 
in the $p^2 c_2$-$c_1$ plane, which
show that they have almost a scaled universal curve 
$c_1=c_{1c}(p^2 c_2)$. 
This may sound strange because 
one may expect that the effective couplings scale
as $c_1\epsilon_{ij}\to p c_1$ and $c_2\epsilon_{ij}\epsilon_{jk}
\epsilon_{ki}\to p^3c_2$. However, it is too simple and the
reason of this scaling  can be suggested  from the result
for the case of $c_2=0.0$. 
The exact study in Appendix  B 
gives rise to the location of the peak of $C$ at $c_1\simeq 1.20$, which is 
determined by the equation $c_1 \tanh c_1=1$ [See Eq.(\ref{uandcforc2zero})]
and has no $p$ and $N$ dependences, because
both $C$ and $U$ are proportional to $p$ there.
Thus one may expect for the general case of $c_2\neq 0$ that
$U\sim N  p c_1  u_1 + N^2 p^3 c_2 u_2 =p[N c_1 u_1 + N^2 p^2 c_2 u_2]$ 
with $u_1, u_2 =O(N^0p^0)$.
Then the relevant parameters may become $c_1$ and $p^2 c_2$, 
which are in fact the case as Fig.\ref{phase1partial}b shows.  

Let us comment on the confinement-Coulomb transition, which,
for $p=1.0$, is of first-order and takes place 
at $c_2\simeq 2.0$. For $p < 1.0$, it remains of first-order
and takes place at  $p^2 c_2\simeq 2.0$. We recall that 
the lattice model, Model 0, gives rise to a second-order 
confinement-Coulomb transition. Because the connectivity
of Model 0 may be estimated as $p \simeq 3N/{}_NC_2\simeq 6/N$,  
Model 0 may be viewed as  Model I 
in a special limit of dilute connectivity $p \sim 0$. And therefore
one may expect that the confinement-Coulomb transition of Model I becomes
of second-order as $p$ becomes sufficiently small, for $p < p_{c} (\neq 0)$.
It is a future problem to estimate  the possible critical value $p_c$
(The exponent $\alpha$ of $p_c = O(N^{\alpha})$ may be 0 or -1, or other
nontirivial value).

\begin{figure}[t]
\begin{center}
\includegraphics[width=5.7cm]{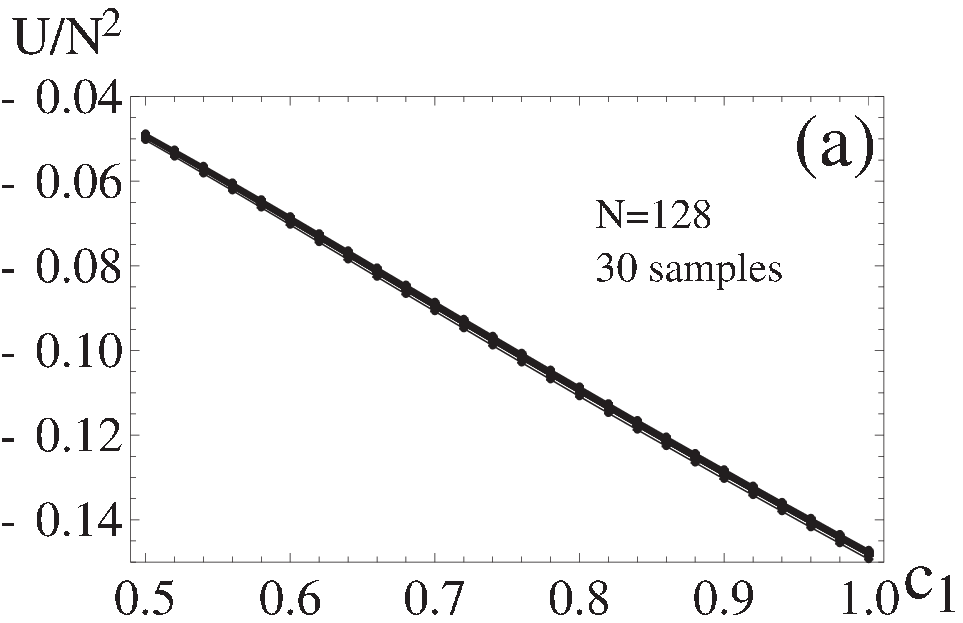}
\includegraphics[width=5.7cm]{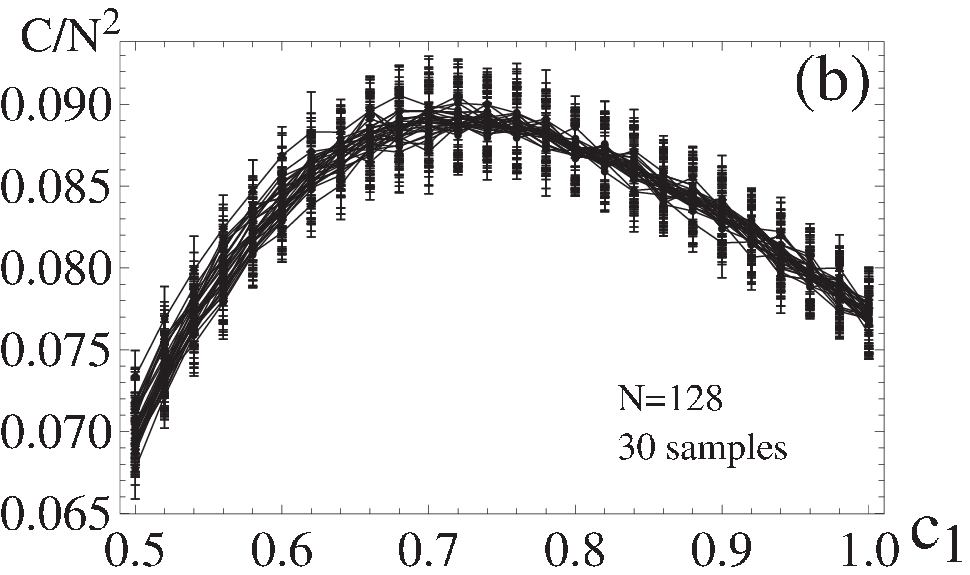}
\end{center}
\caption{
(a) $U/N^2$ and (b) $C/N^2$ of Model I at $p=0.3, c_2=5.555$ for 30 samples.
Each curve is for each sample and the error bars are errors
in thermal(MC) averages. The deviations over samples are smaller than
MC errors. 
}
\label{partialerror}
\end{figure}

We note that Fig.\ref{phase1partial} is obtained by using a set of
data $U$ and $C$ of only one sample, because we have checked that
the location of $C$ has small deviation over different samples.
For example, in Fig.\ref{partialerror} we present $U$ and $C$
along $c_2=5.555 (p^2 c_2=0.5)$ for 30 samples with p=0.3.
There are 30 curves with each curve for each sample. 
The error bars in Fig.\ref{partialerror} denote errors associated 
with MC sweeps (thermal average) of each sample.
Fig.\ref{partialerror} shows that the deviations of $U, C$ over 
samples are smaller than these errors by factor $\sim 2$.
So we judge that the result of one sample is reliable for $0.3 \leq p
\leq 1.0$.

\clearpage
\section{Model II: Quenched model with full connections}
\setcounter{equation}{0} 

\begin{figure}[b]
\begin{center}
\hspace{-0.6cm}
\includegraphics[width=5.5cm]{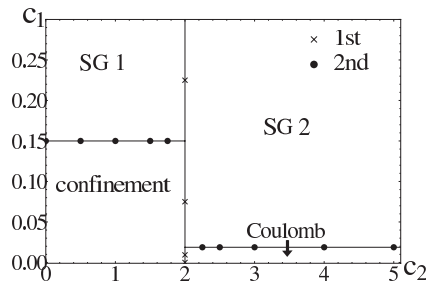}
\end{center}
\vspace{-0.6cm}
\caption{
Phase diagram of Model II in the $c_2$-$c_1$ plane for $N=64$.
There are four phases separated by boundaries with 1st and 2nd-order 
transitions as indicated. The Higgs phase of Model I is separated to 
two SG phases (SG1 and SG2). The crossovers of Model I disappear and the
1st-order transitions continue down to $c_2=0$.
The critical value  of second-order transitions
behaves as $c_{1c} =O(1/\sqrt{N})$ for $c_2 \lesssim 2.0$
and $c_{1c} =O(1/N)$ for $c_2 \gtrsim 2.0$.
}\label{uc2phase}
\end{figure}

\begin{figure}[t]
\begin{center}
\hspace{-0.6cm}
\includegraphics[width=5.3cm]{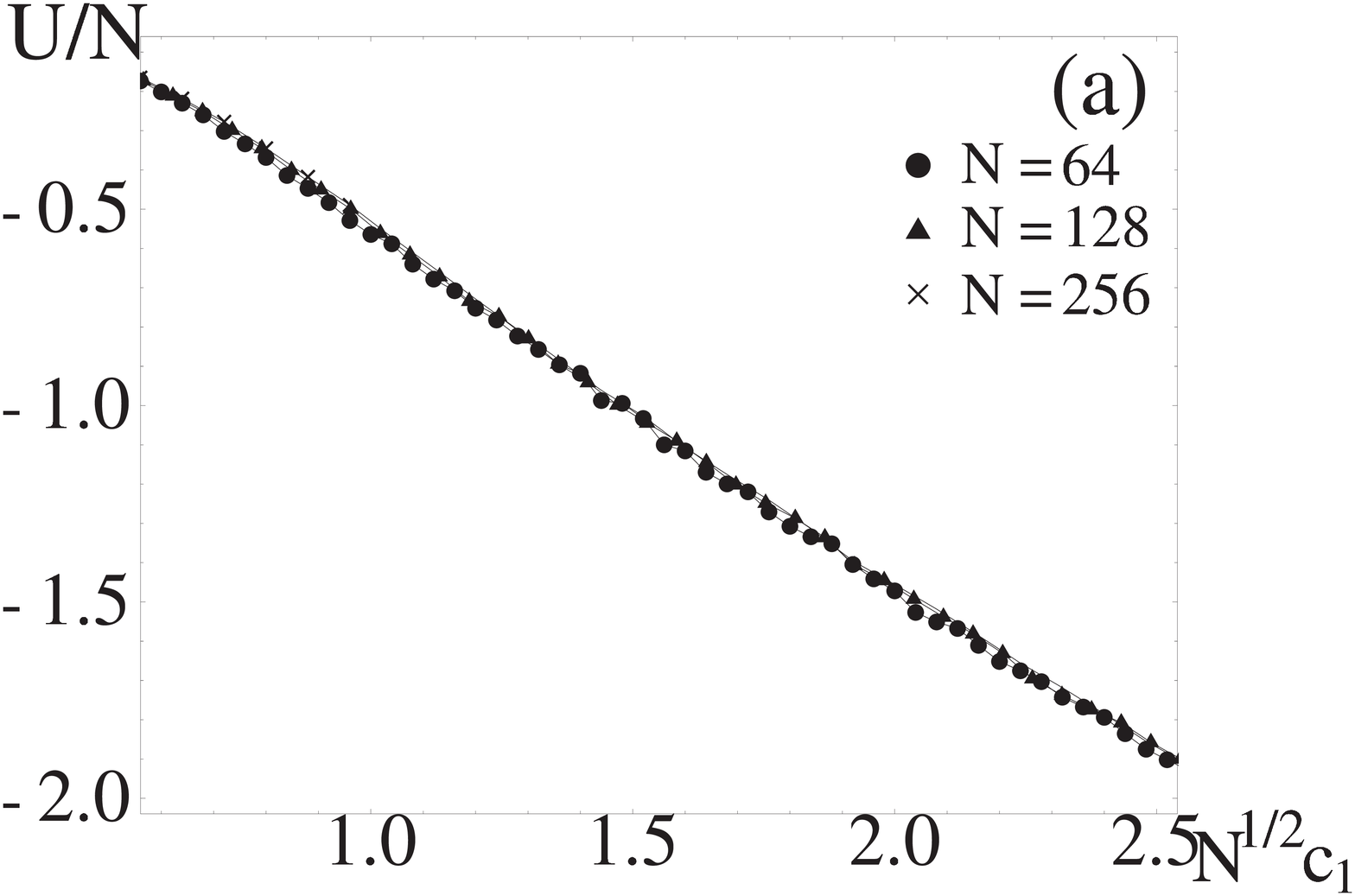}
\includegraphics[width=5.5cm]{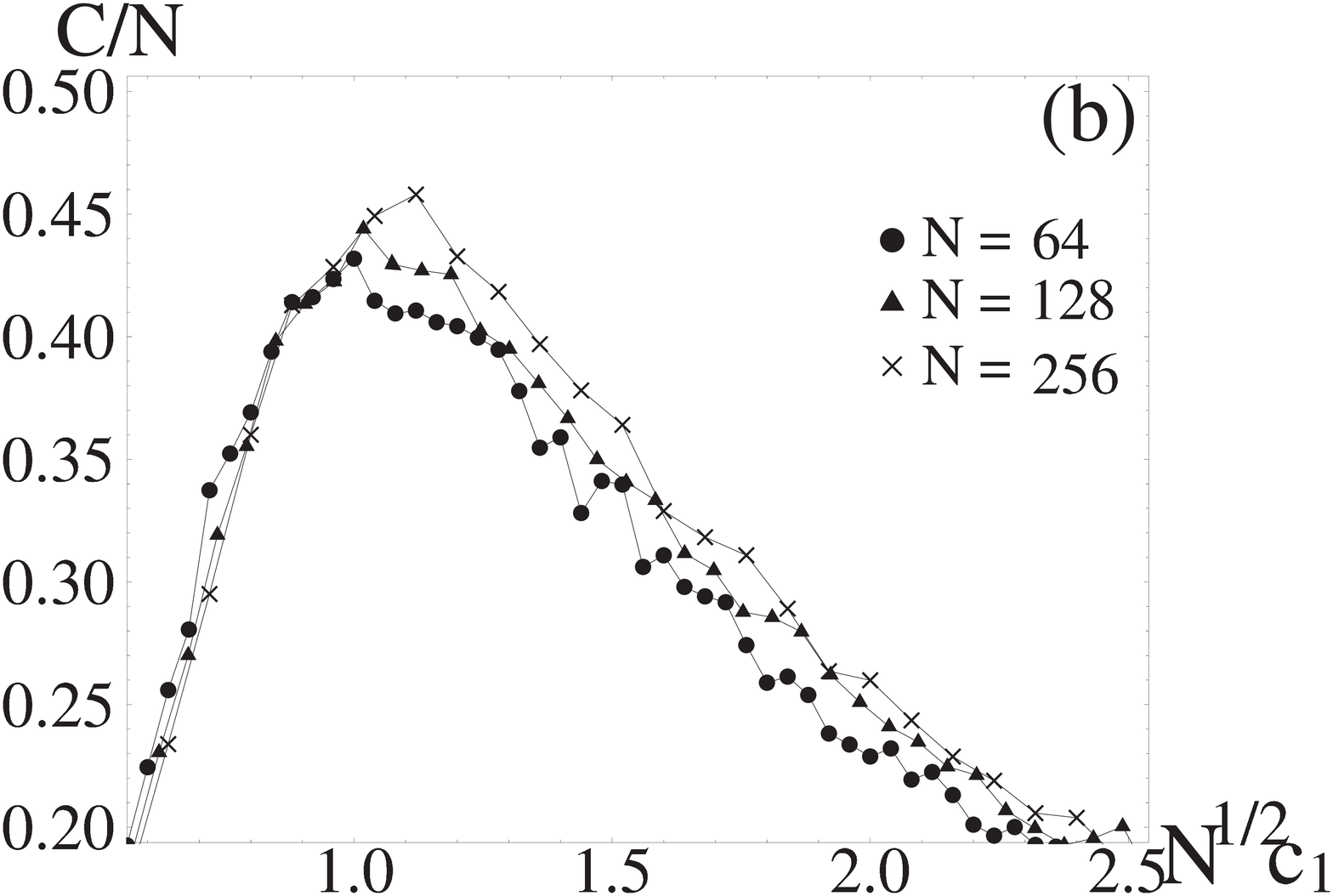}\\
\includegraphics[width=5.5cm]{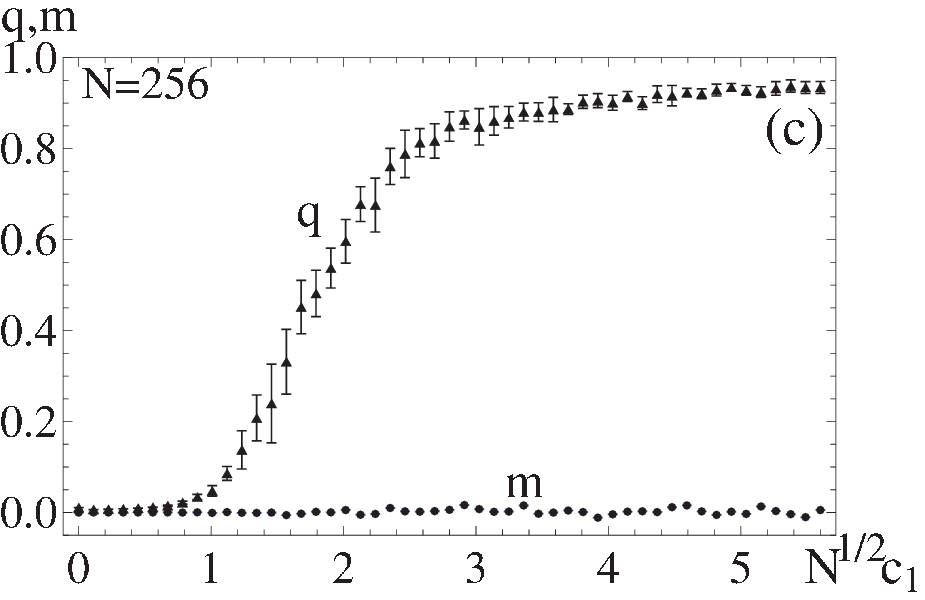}
\end{center}
\caption{
(a) $U/N$, (b)$C/N$ and (c) $q, m$ of Model II 
vs. $\sqrt{N} c_1$ for $c_2= 1.5$. It shows a second-order transition into
a SG1 phase. $m$ vanishes as expected.
}\label{uc2confinementsg}
\end{figure}

In this section we study the phase structure of Model II.
In Fig.\ref{uc2phase} we first present its 
phase diagram in the $c_2$-$c_1$ plane.
Here we recall that the configurations of synaptic variables $J_{ij}$
are completely determined by $P(J)$ of (\ref{averageo2}). From the
analysis of Sect.2.1 for $c_1=0.0$, $P(J)$ describes 
a first-order phase transition at $c_2 \simeq 2.0$ for $p=1.0$.
This transition at $c_2 \simeq 2.0$ survives
in Model II for all $c_1$. 

Fig.\ref{uc2phase} shows that there are other two 
phase transition lines, both of which is of second order.
One is in the region $c_2 \lesssim 2.0$  at 
$c_{1c}\simeq 0.15 $ for $N=64$ and separates 
the confinement phase
$c_1 < c_{1c}$ and the another phase $c_1 > c_{1c}$. We call
this phase $c_1 > c_{1c}$ a SG phase (we call it SG1 phase) 
as we shall see that the SG order parameter $q$ is nonvanishing there.
Also we shall see that the value of $c_{1c}$ scales as $c_{1c} \propto 1/\sqrt{N}$ as $N$ increases.

The other transition line is in the region $c_2 \gtrsim 2.0$ at 
$c_{1c}\simeq 0.02 $ for $N=64$ and separates 
the Coulomb phase
$c_1 < c_{1c}$ and another SG phase for $c_1 > c_{1c}$
(we call it SG2 phase).
The value of $c_{1c}$ scales as $c_{1c} \propto 1/N$ as $N$ increases,
which is similar to Model I.

Let us see each transition in details.\\

\nin
(i) confinement-SG1 transition

In Fig.\ref{uc2confinementsg} we present
$U$, $C$, $m$ and $q$ vs. $\sqrt{N} c_1$ for $c_2= 1.5$.
The $N$ dependence of the peak of $C$ indicates a second-order transition.
The behavior of $m$ and $q$ show that the phase of higher $c_1$
is the SG phase.

In Model I, the exact treatment for $c_2=0$ in Appendix B exhibits 
a crossover as $c_1$ varies. The reason that Model II exhibits a 
second-order transition line in this region ($0 \leq c_2 \leq c_{2c}$)
instead
is traced back to our treatment of $J_{ij}$ as quenched variables. 
In Appendix E we make use of the resemblance of
Model II at $c_2=0$ and the Sherrington-Kirkpatrick model
\cite{skmodel} of SG, and present a plausible argument
that Model II for $c_2=0$ 
has a second-order transition at $c_1 \propto 1/\sqrt{N}$.

This result $c_{1c} \propto 1/\sqrt{N}$ for $c_2 \lesssim 2.0 $
can be also understood as a compromise of the two results;
(a) $c_{1c} = O(N^{-1})$ of annealed Model I for 
$c_2 \gtrsim 2.0 $ where  $J_{ij}$ are almost ordered, 
and (b)  $c_{1c} = O(N^{0})$ of Model I for 
$c_2 \lesssim 2.0 $ where  $J_{ij}$ are random.
In fact, configuration of $J_{ij}$
in each sample of the quenched Model II  for $c_2 \lesssim 2.0$
is almost fixed, but the spatial average of $J_{ij}J_{jk}J_{ki}$ is much less
than its saturated value 1.
Therefore the effect of $J_{ij}$ may be smaller than the complete order
in the case (a) but larger than the complete randomness in the case (b). \\

\nin
(ii) Coulomb-SG2 transition

\begin{figure}[t]
\begin{center}
\begin{minipage}{0.49\hsize}
\hspace{0.3cm}\includegraphics[width=5.0cm]{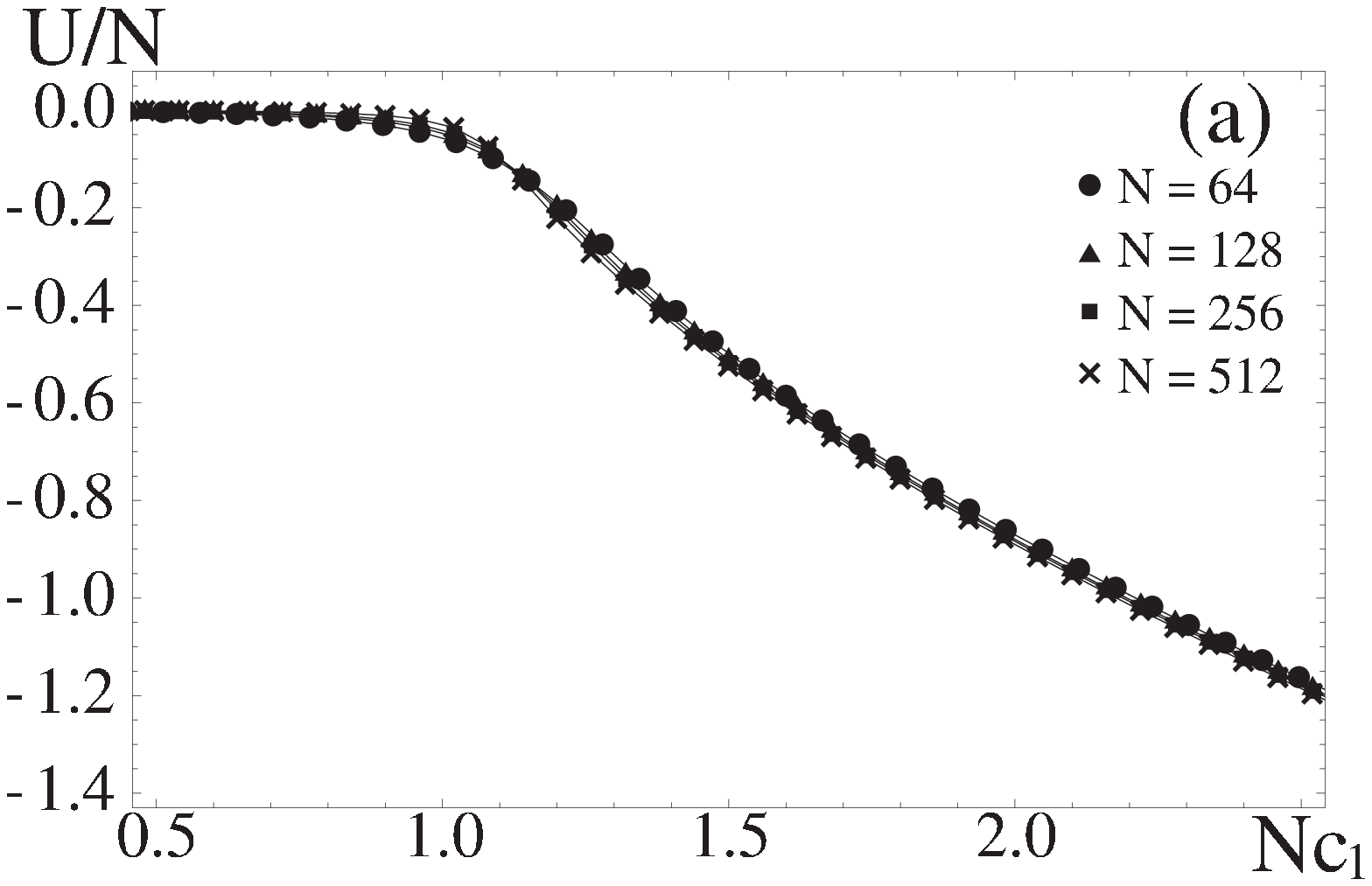}
\end{minipage}
\begin{minipage}{0.49\hsize}
\includegraphics[width=5.0cm]{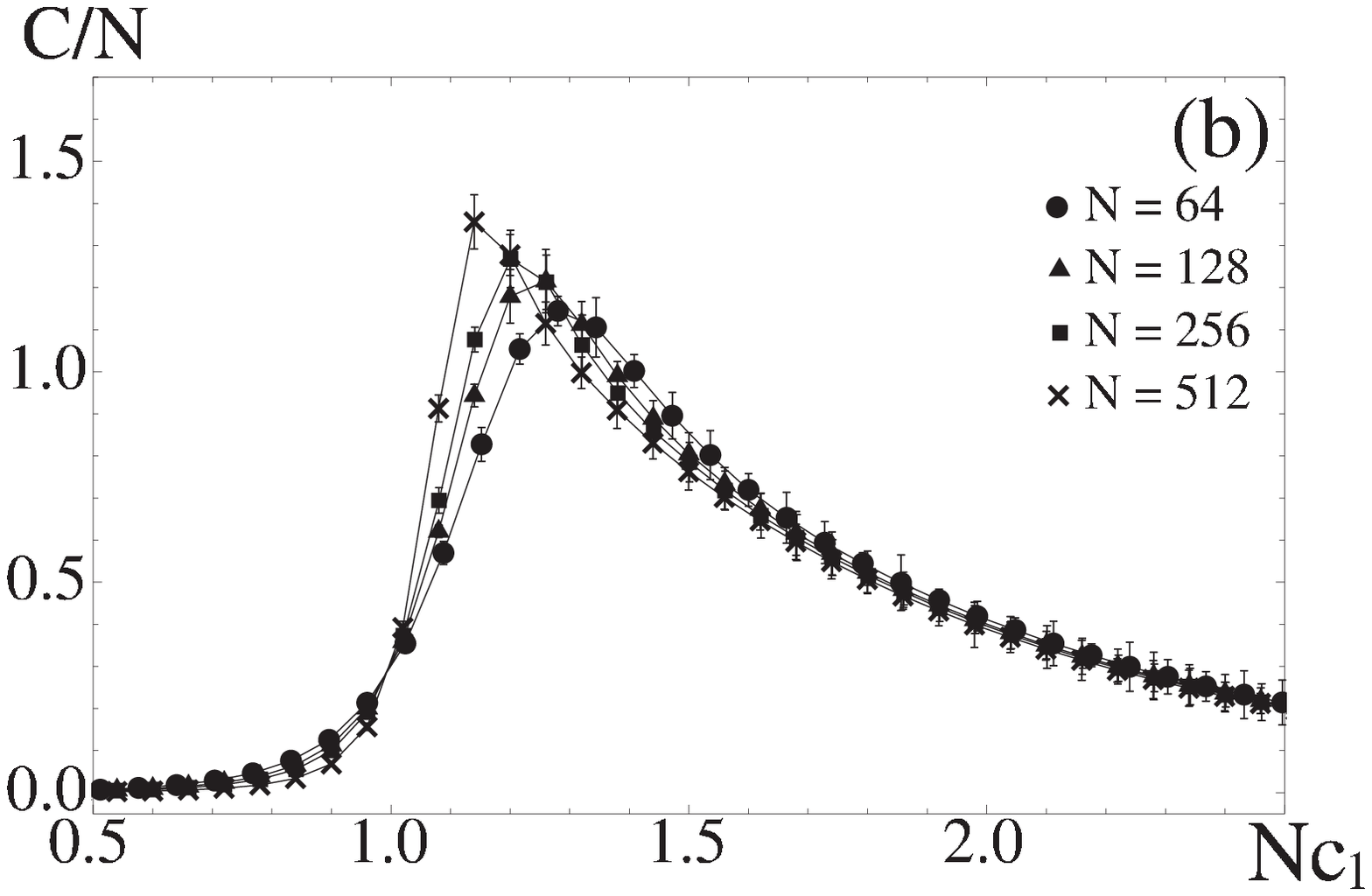}
\end{minipage}
\begin{minipage}{0.49\hsize}
\hspace{0.3cm}
\includegraphics[width=5.0cm]{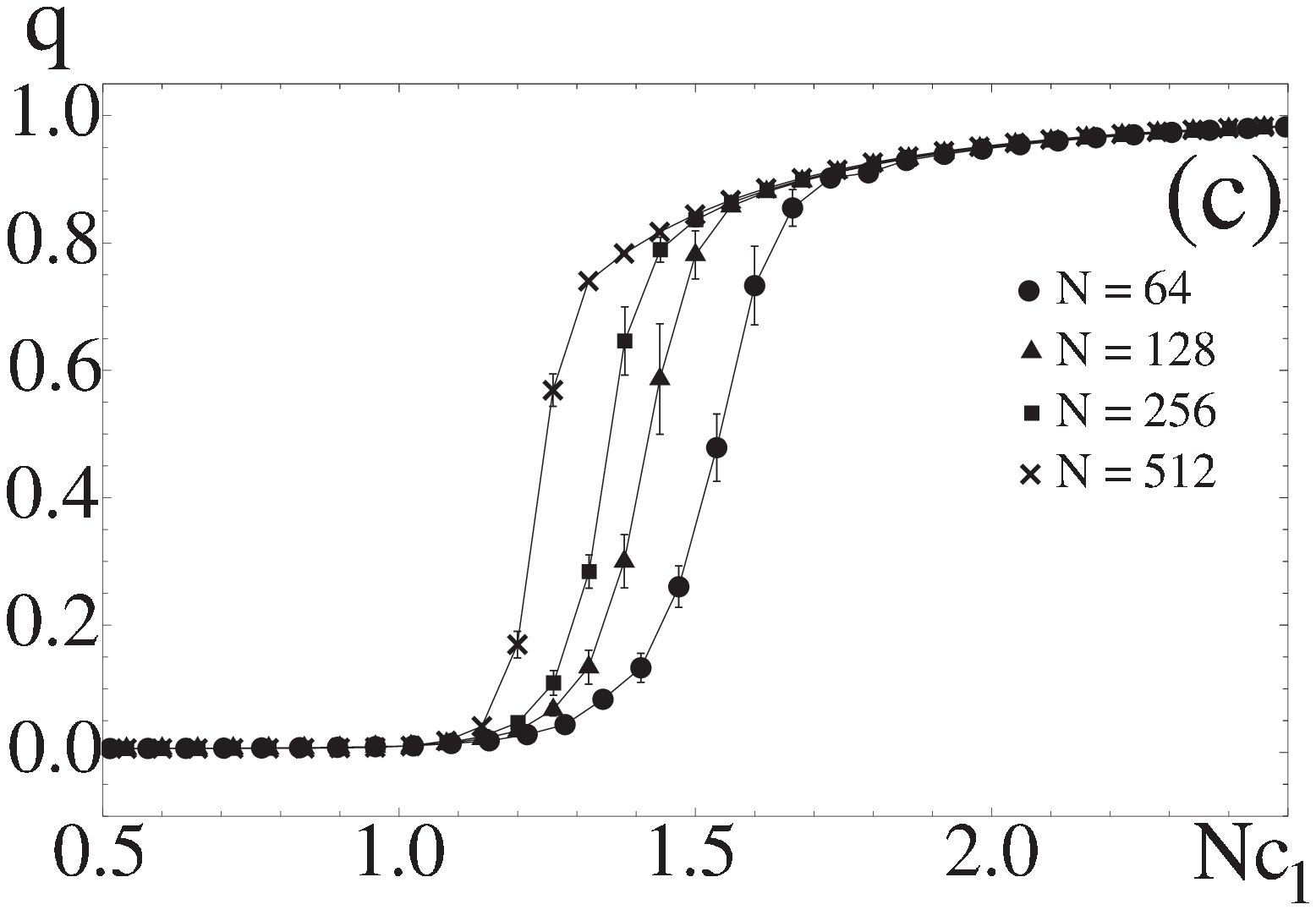}
\end{minipage}
\begin{minipage}{0.49\hsize}
\includegraphics[width=5.0cm]{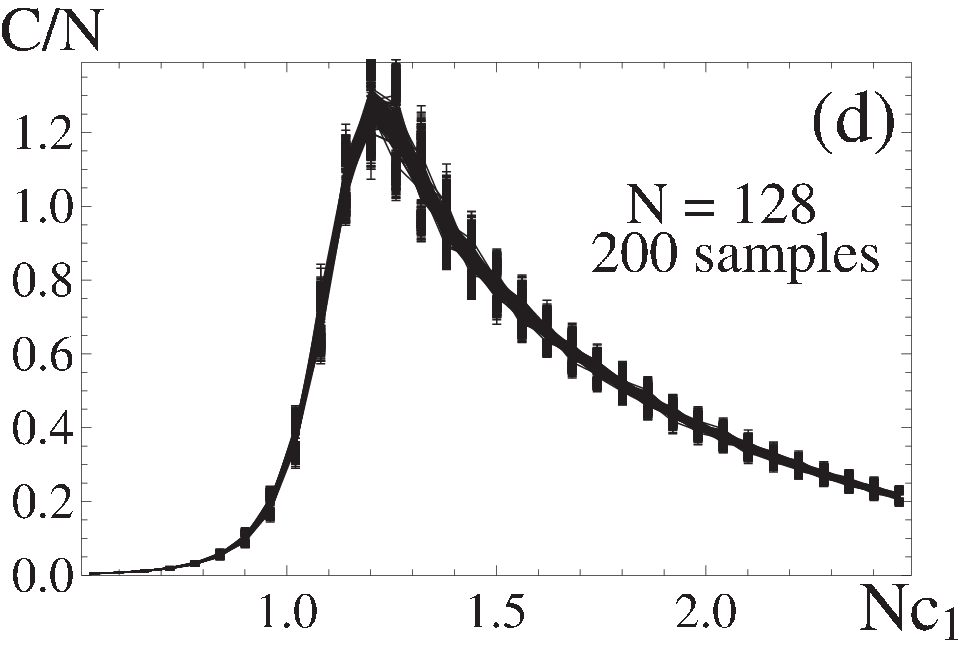}
\end{minipage}
\end{center}
\vspace{-0.5cm}
\caption{
(a)$U/N$, (b)$C/N$, (c)$q$, (d) $C/N$ of Model II 
vs. $N c_1 $ for $c_2= 2.5$. They exhibit a second-order transition at
$Nc_{1}\simeq 1.15$. (d) shows 200 curves of $C/N$ for 200 samples 
of quenched variables $J_{ij}$ separately.
Error bars in (d) are errors in MC sweeps (thermal average),
which are larger than the deviations over samples.
}\label{uc2coulombhiggs}
\end{figure}

In Fig.\ref{uc2coulombhiggs} we present
$U/N$, $C/N$ and $q$ vs. $N c_1$ for $c_2= 2.5$.
The $N$ dependence of the peak of $C$ indicates a second-order transition
at $N c_{1}\simeq 1.15$.
The behavior of $q$ shows that the phase of higher $c_1$
is the SG phase. 

Fig.\ref{uc2coulombhiggs}d shows
$C/N$ of 200 samples (different configurations of $J_{ij}$); 
each curve is for each sample.
It shows that the deviations over samples are smaller than
typical errors in thermal average over different $S_i$.
Therefore we judge that 200  samples are sufficient to obtain the location of
specific heat along a fixed $c_2$ semiquantitatively.\\

\nin
(iii) Transition across the $c_2 \simeq 2.0$ line

\begin{figure}[h]
\begin{center}
\begin{minipage}{0.49\hsize}
\includegraphics[width=5.2cm]{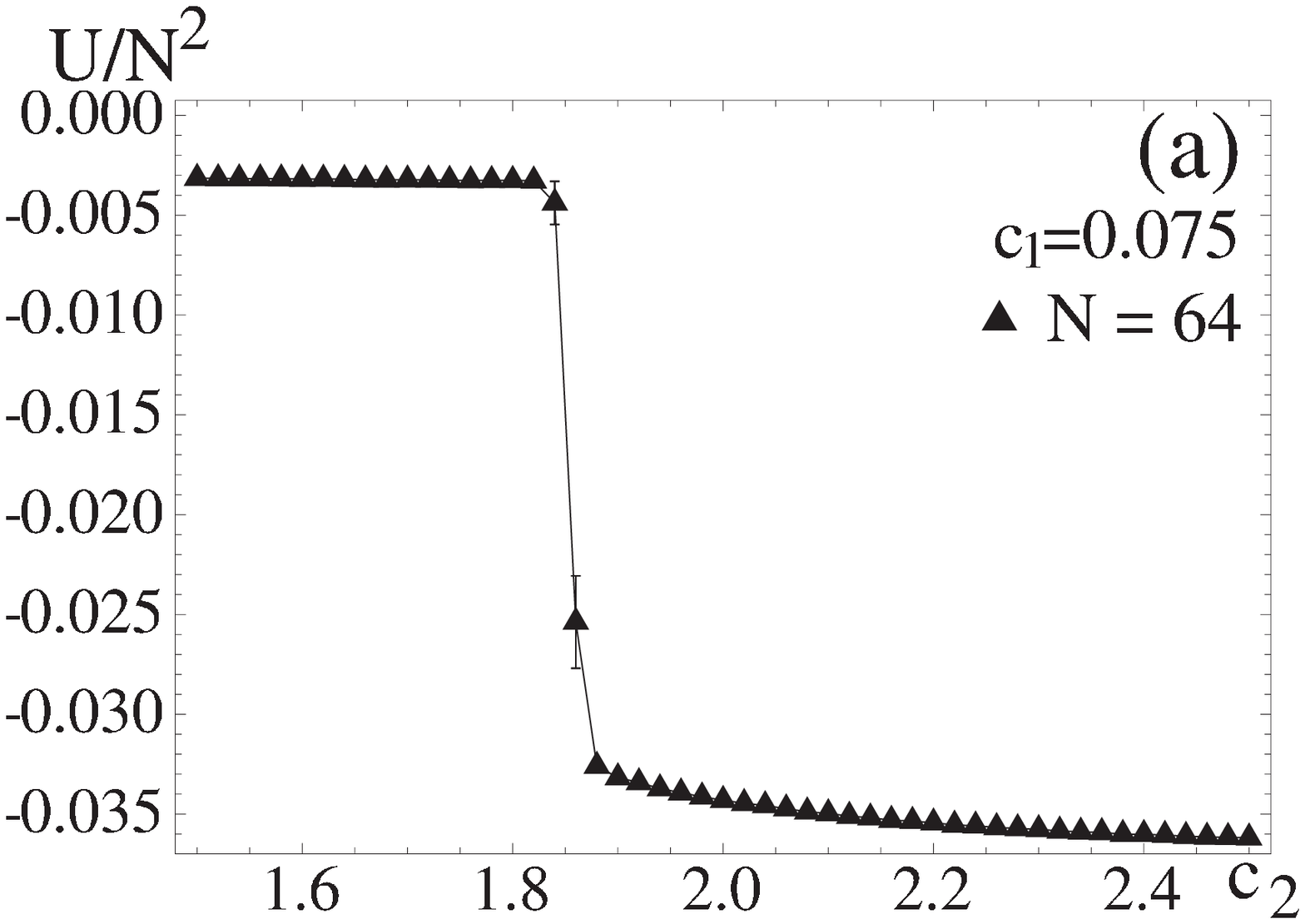}
\end{minipage}
\begin{minipage}{0.49\hsize}
\includegraphics[width=4.8cm]{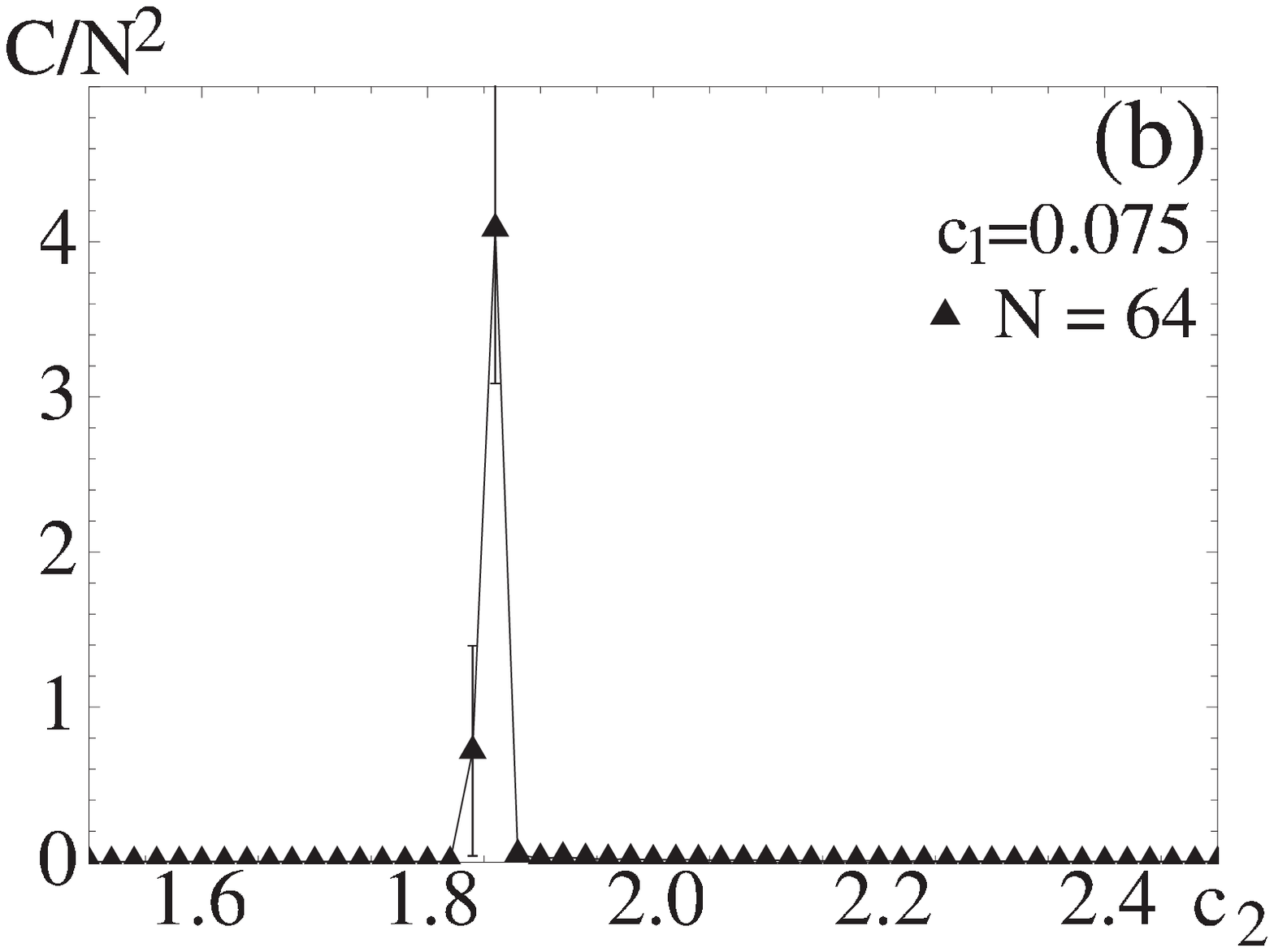}
\end{minipage}
\begin{minipage}{0.49\hsize}
\hspace{0.2cm}
\includegraphics[width=4.8cm]{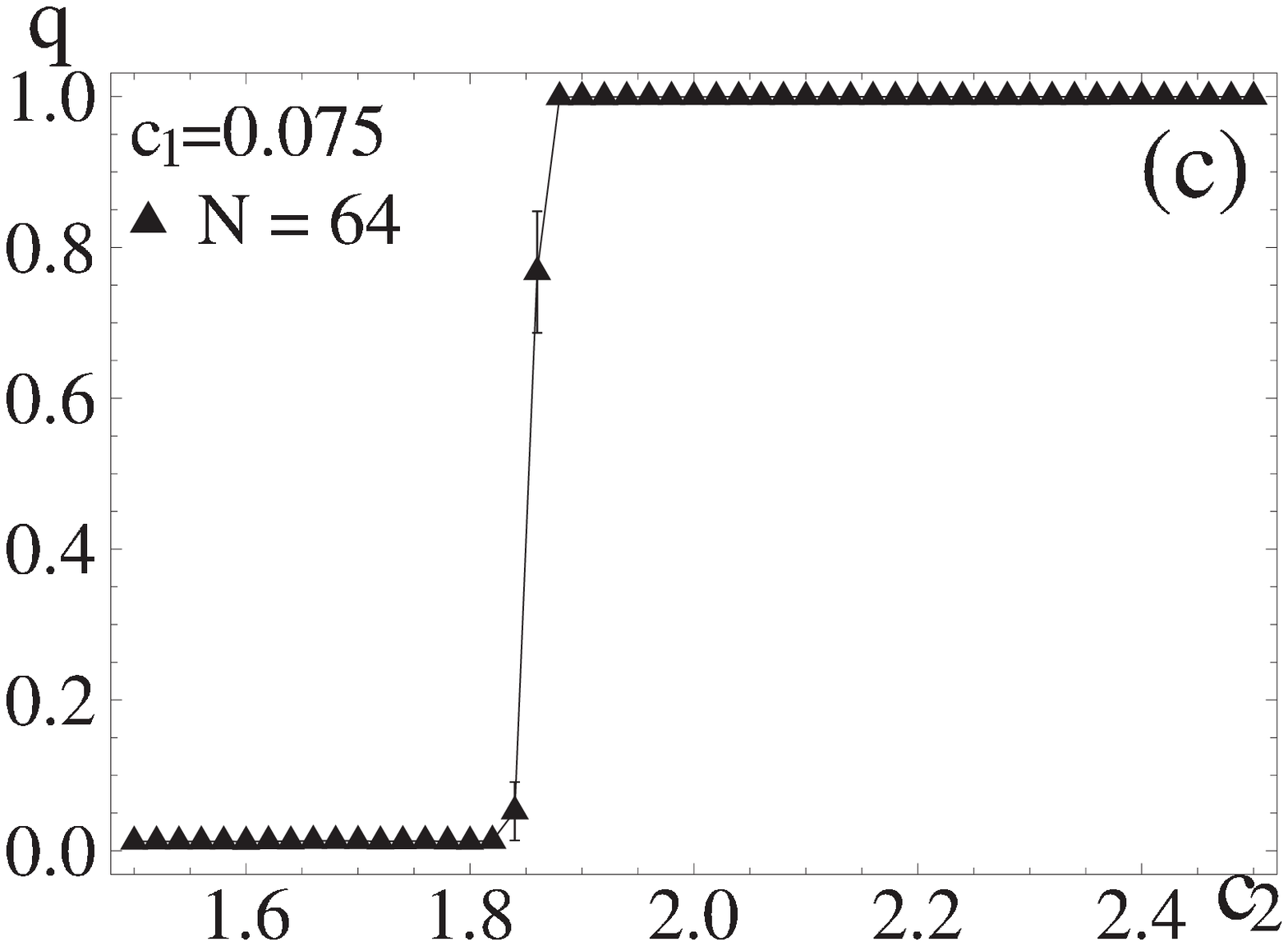}
\end{minipage}
\begin{minipage}{0.49\hsize}
\hspace{-0.4cm}
\includegraphics[width=5.2cm]{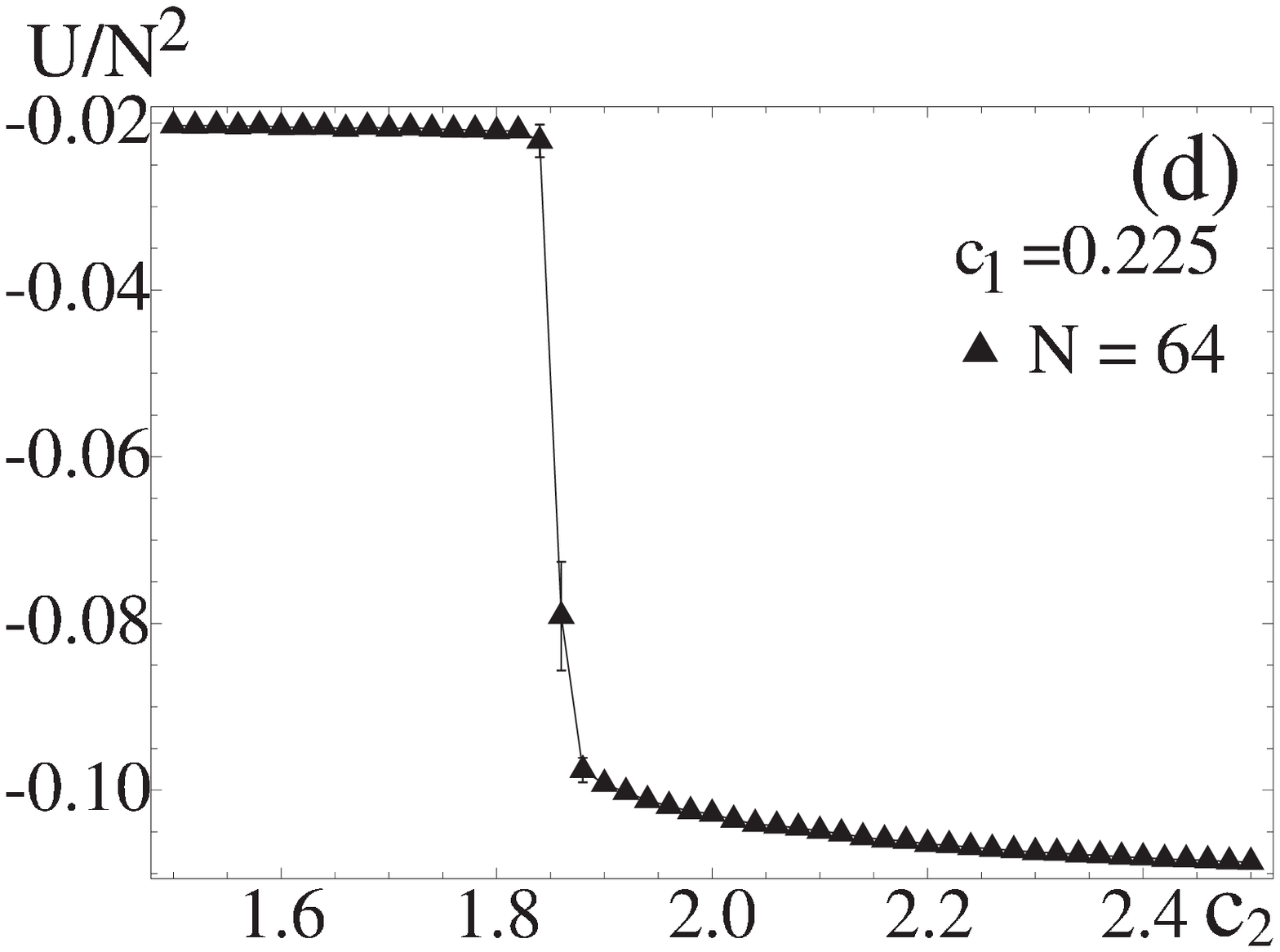}
\end{minipage}
\begin{minipage}{0.49\hsize}
\hspace{0.2cm}\includegraphics[width=4.8cm]{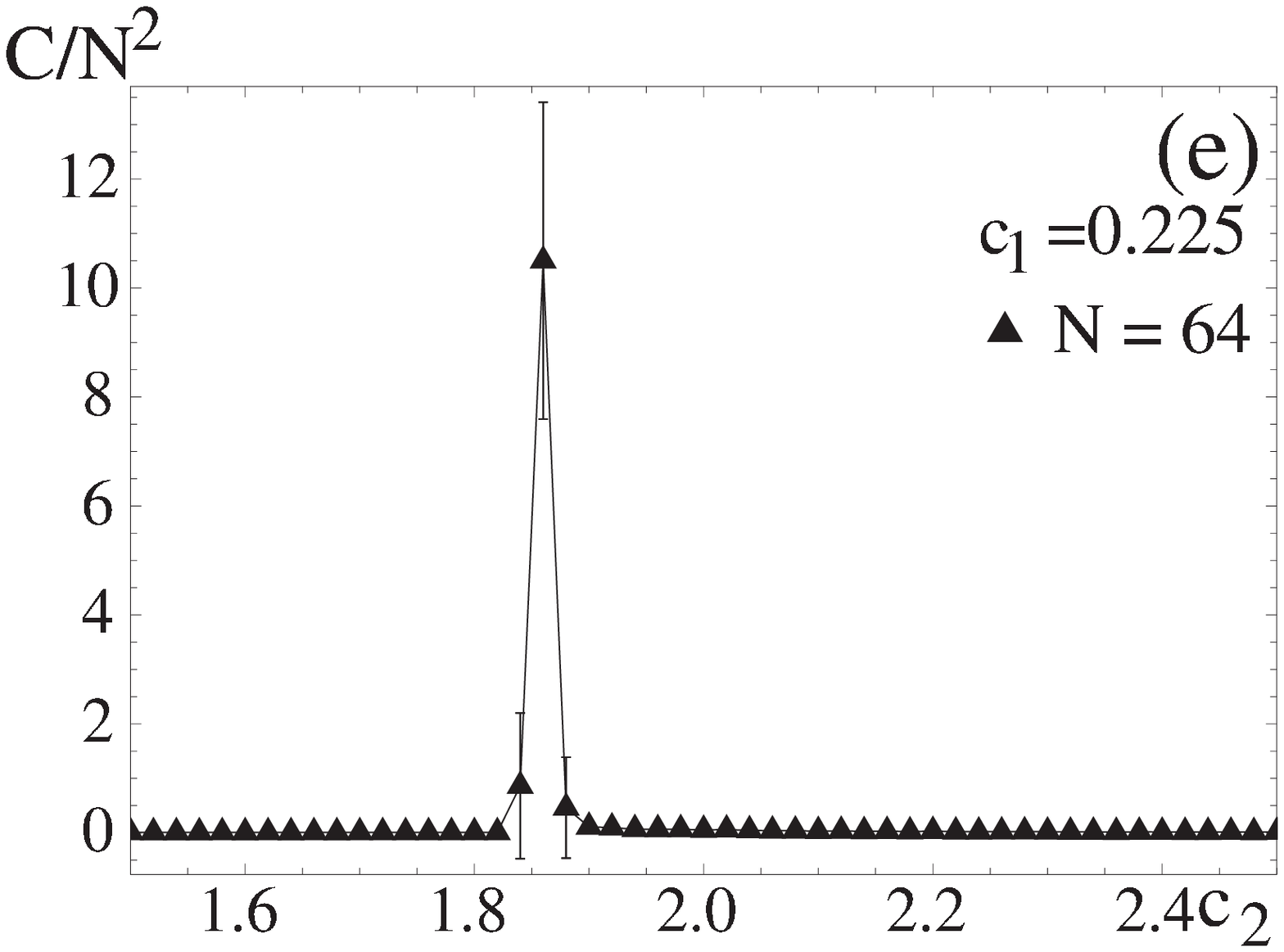}
\end{minipage}
\begin{minipage}{0.49\hsize}
\includegraphics[width=4.8cm]{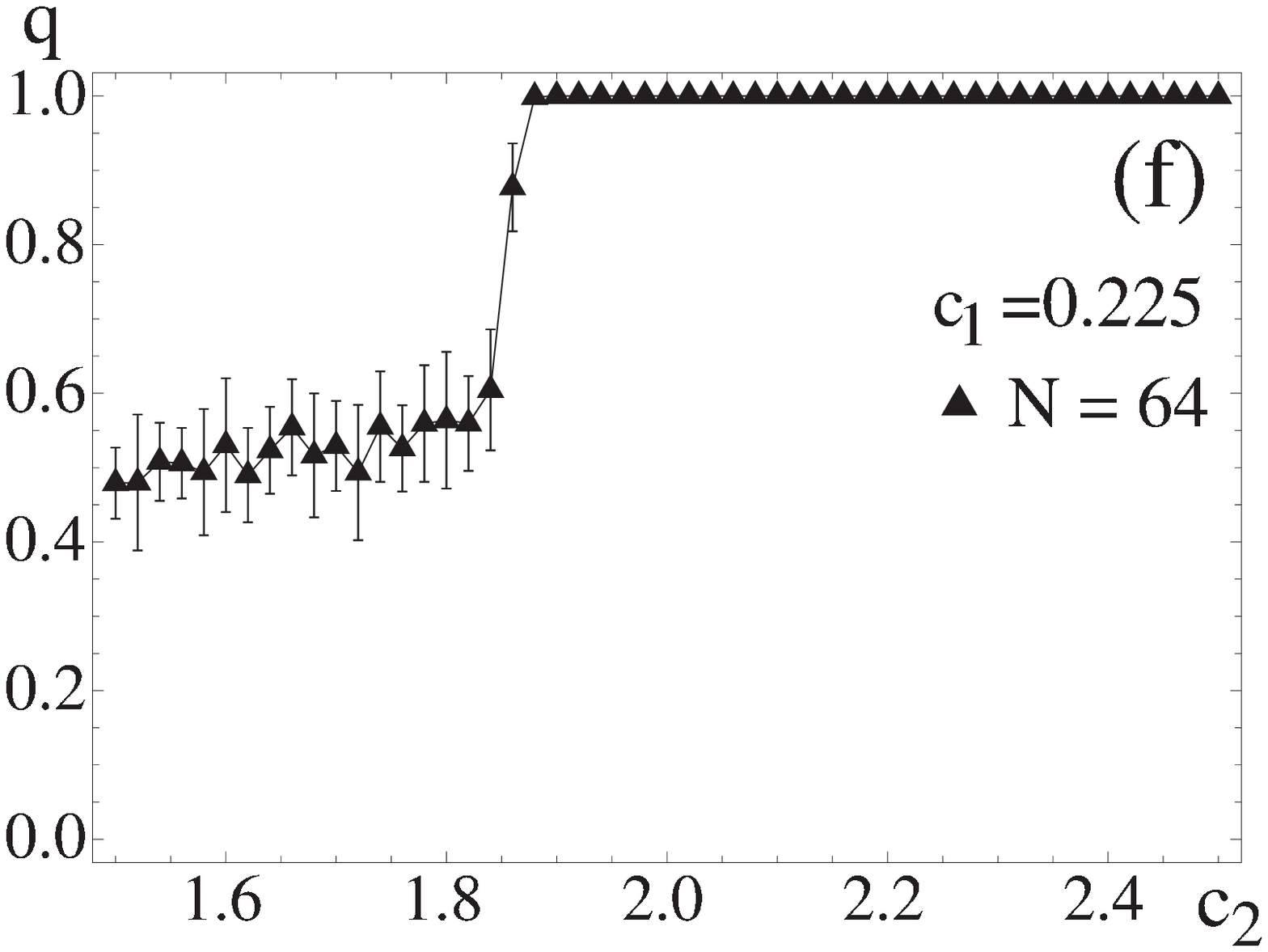}
\end{minipage}
\begin{minipage}{0.49\hsize}
\hspace{0.2cm}
\includegraphics[width=4.8cm]{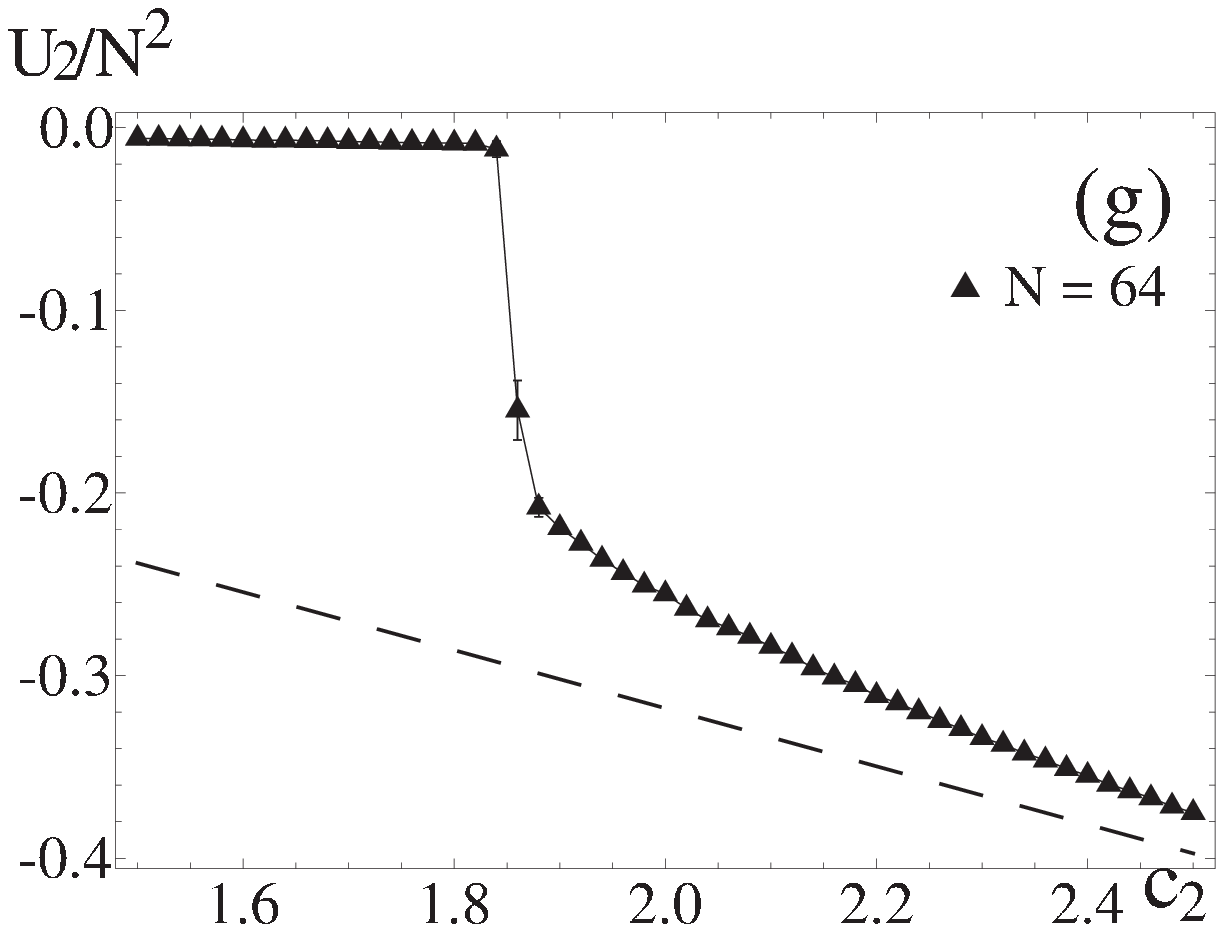}
\end{minipage}
\begin{minipage}{0.49\hsize}
\includegraphics[width=4.7cm]{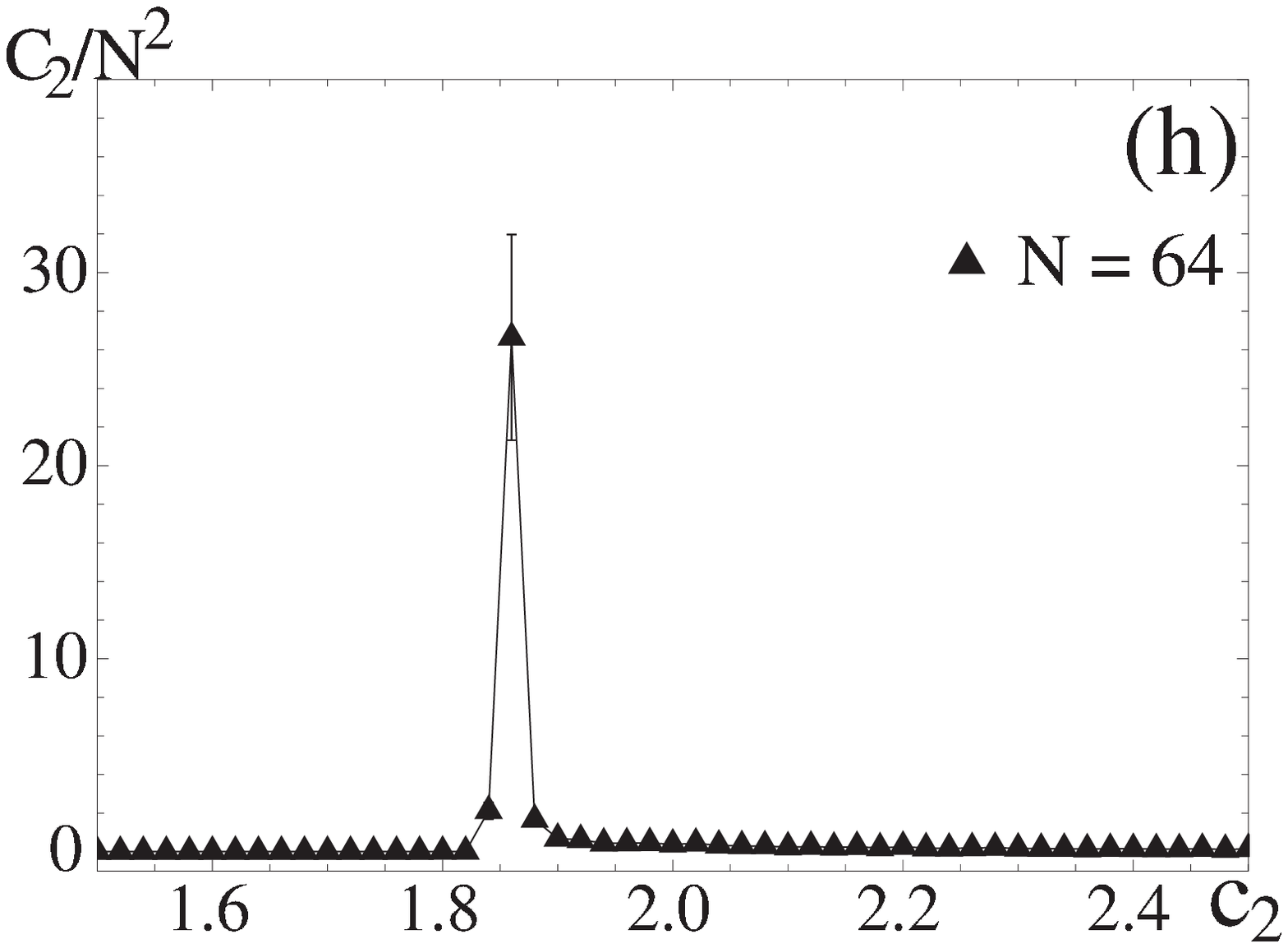}
\end{minipage}
\end{center}
\caption{
Metropolis calculations of 
$U/N^2$, $C/N^2$ and $q$ of Model II at (a-c) $c_1=0.075$
for the confinement-SG2 transition and  
at (d-f) $c_1=0.225$ for the SG1-SG2 transition. Both of them
are of first-order.   
(g) $U_2/N^2$ and (h) $C_2/N^2$ are defined in 
Eq.(\ref{U2C2}) and should be compared with 
Fig.\ref{uc1confinementcoulomb}b,c of multicanonical calculations.
The dashed line in (g) is $U_2/N^2=-c_2{}_NC_3/N$ for $N=64$
corresponding to the saturated value $J_{ij}J_{jk}J_{ki}=1$.
}\label{fig1011}
\end{figure}

As explained, this first-order transition curve reflects $P(J)$ of 
(\ref{averageo2}) as already shown in Fig.\ref{uc1confinementcoulomb} 
for Model I at $c_1=0$.
Explicit calculation of $U$ and $C$ across this transition for fixed
$c_1$ is time-consuming because one needs multicanonical method 
due to  large hysteresis of Metropolis updates as 
Fig.\ref{uc1confinementcoulomb}a shows. In place of such a calculation, 
in Fig.\ref{fig1011}
we present $U/N^2$,  $C/N^2$  of (\ref{ucmodel2}) and $q$, which are 
calculated with Metropolis updates by selecting runs along the 
lower-energy branch of the hysteresis curve 
in Fig.\ref{uc1confinementcoulomb}a. 
These runs are realized by a cold start such as $S_i=J_{ij}=1$
and the results are reliable qualitatively because the transition point
$c_{2c}\simeq 1.85 (N=64)$ determined by this method is not far from the 
true value $c_{2c}\simeq 2.04$
given in Fig.\ref{uc1confinementcoulomb}b,c ($\sim 10\%$ deviation).
In Fig.\ref{fig1011}g,h we also present $U_2$ and $C_2$
for the $c_2$-term defined by
\be
E_2 &\equiv& -\frac{c_2}{N} \sum_{i < j < k}J_{ij}J_{jk}J_{ki},\nn
U_2 &\equiv& -\la E_2\ra,\ C_2\equiv\la E_2^2\ra-\la E_2\ra^2,
\label{U2C2}
\ee
and calculated by this Metropolis updates.
These definitions  are equivalent to $U, C$ of Model I at 
$c_1=0$, and therefore they  should be compared with
Fig.\ref{uc1confinementcoulomb}b,c.
Actually they have no significant differences.

Fig.\ref{fig1011}f shows that $q$ has a jump $\Delta q \simeq 0.4$
across the SG1-SG2 phase transition, and $q\simeq 1$ in SG2 phase.
It shows the difference of two SG phases clearly.

\clearpage
\begin{figure}[b]
\begin{center}
\hspace{-0.6cm}
\includegraphics[width=5.5cm]{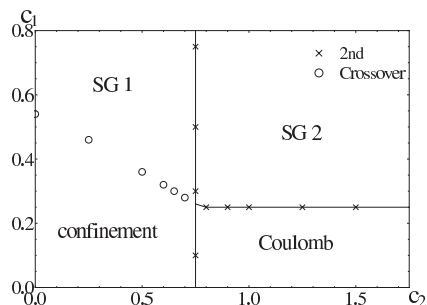}\\
\end{center}
\vspace{-0.7cm}
\caption{
Phase structure of Model III, the quenched 3D lattice model, 
in the $c_2$-$c_1$ plane. The boundaries are determined by the 
location of the specific heat $C$ for $N=12^3$. The second-order 
 confinement-SG1 transition in Model II becomes crossover.
}\label{latticephase}
\end{figure}

\section{Model III: Quenched lattice model}
\setcounter{equation}{0} 


In this section we study the phase structure of Model III.
In Fig.\ref{latticephase} we present 
the phase diagram in the $c_2$-$c_1$ plane.
The overall phase structure is similar to that of Model II, but
the second-order transition between 
the confinement and the SG1 phases of Model II ($c_2 \lesssim 2.0$) 
becomes a crossover. 
This may be accounted for by the fact that the connectivity among $S_i$ 
in Model III is restricted to the nearest-neighbor neurons and much weaker   
than in Model II. 
The argument of obtaining the second-order transition
for Model II by referring to the Sherrington-Kirkpatrick model in Appendix E
fails due to the scarce connectivity of Model III, which does not validate
the saddle-point estimation of Ref.\cite{skmodel}.
Therefore, it is harder to obtain an ordered  phase of $S_i$
in Model III compared with Model II.
  
Furthermore, the critical value $c_{1c}$ for
$c_2 \gtrsim 2.0 $ depends on  $N$ weakly, but
it is  almost constant in contrast with $1/N$-dependence
of Model I and  Model II. This is also due to the scarce connectivity and 
consistent with the previous result
for annealed 3D model in which $c_{1c}$ for
$c_2 \gtrsim 2.0 $ is $O(N^0)$ (Note there is no extra factor $N^{-1}$
in the $c_2$-term in Eq.(\ref{energy0}) and Eq.(\ref{annealpj})). 

Let us see each phase transition and crossover.\\

\vspace{-0.3cm}
(i) Crossover between the confinement and SG1 phases 

\begin{figure}[t]
\begin{center}
\begin{minipage}{0.49\hsize}
\includegraphics[width=5.2cm]{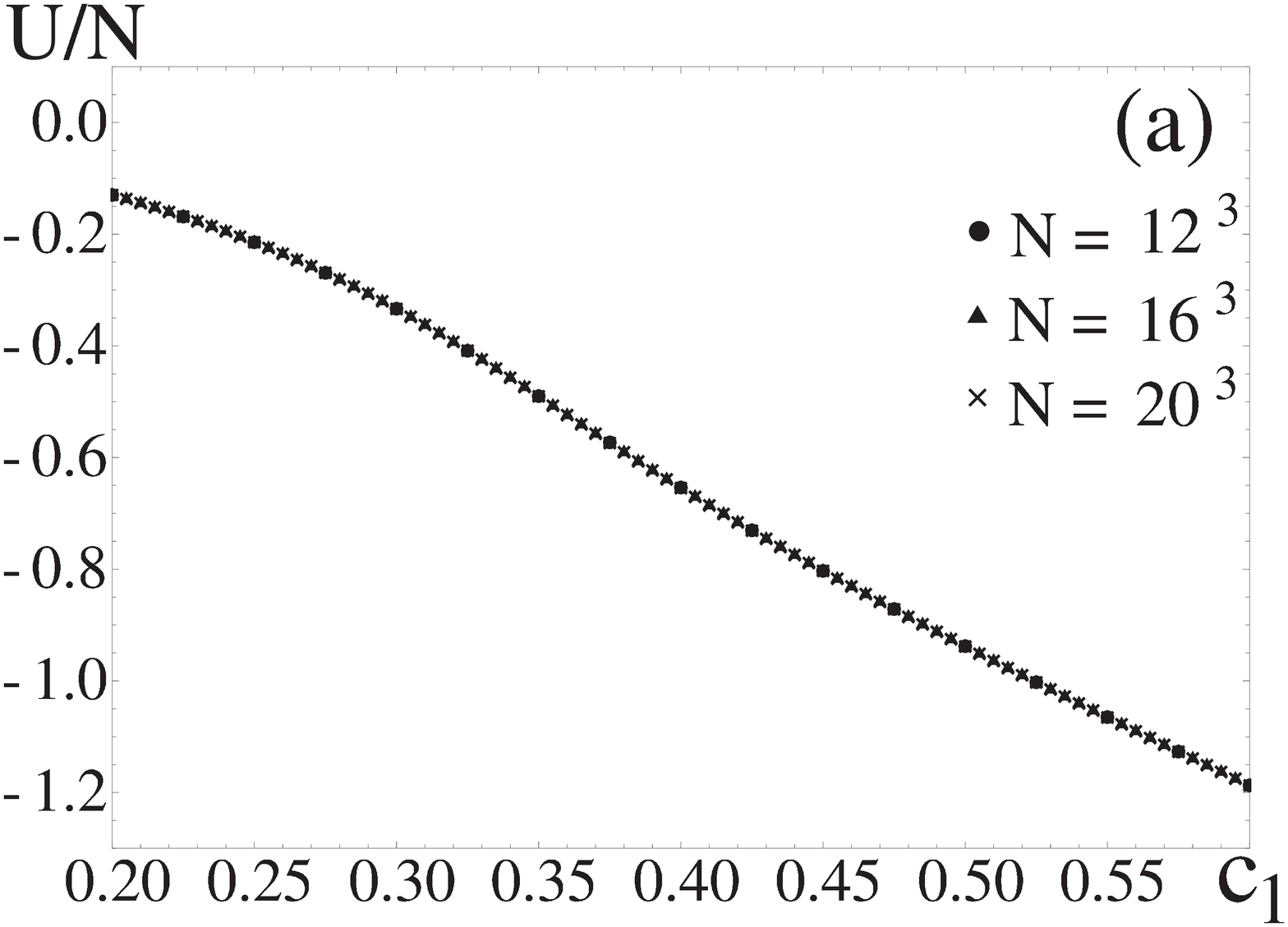}
\end{minipage}
\begin{minipage}{0.49\hsize}
\includegraphics[width=5cm]{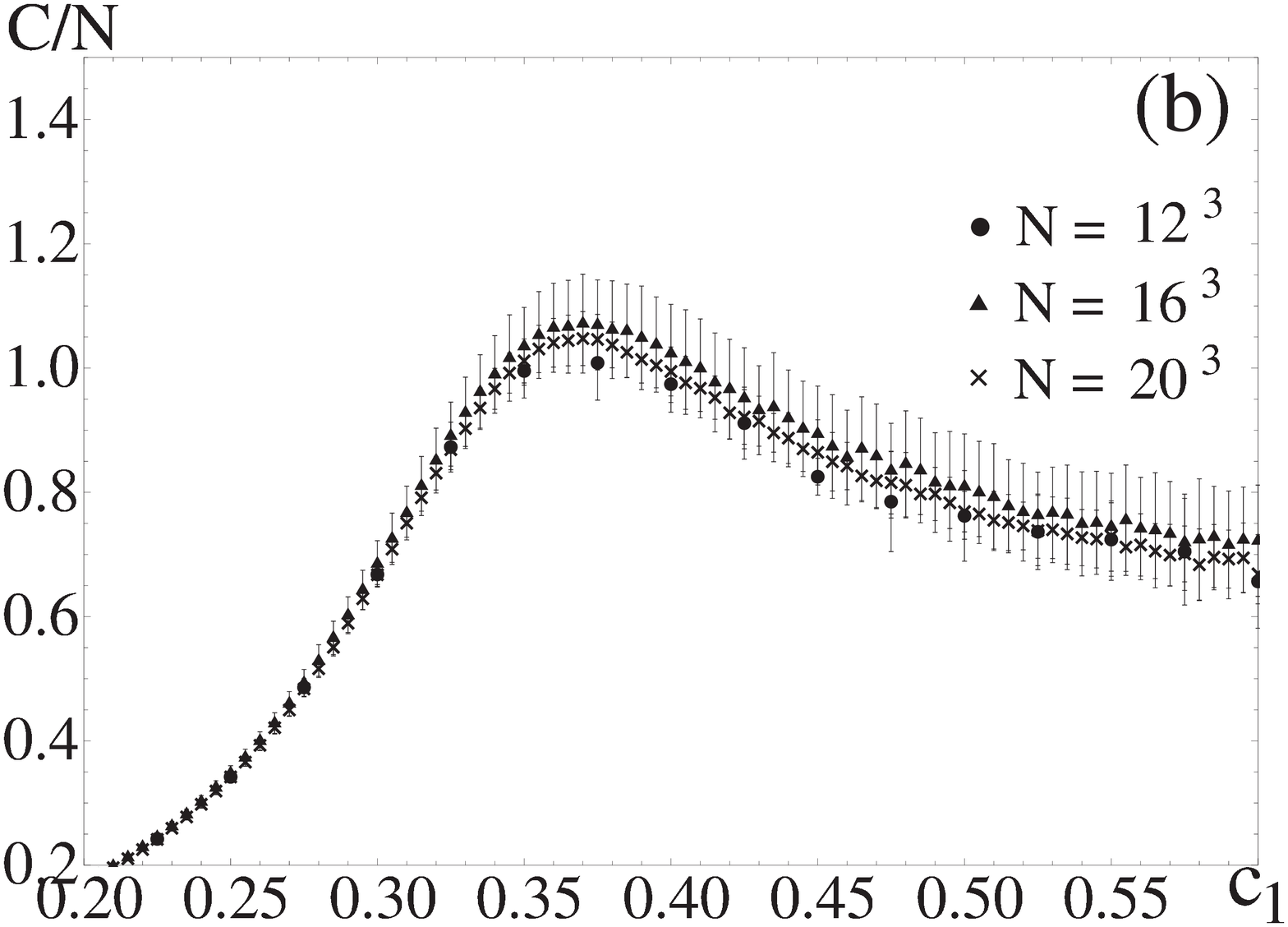}
\end{minipage}
\includegraphics[width=5cm]{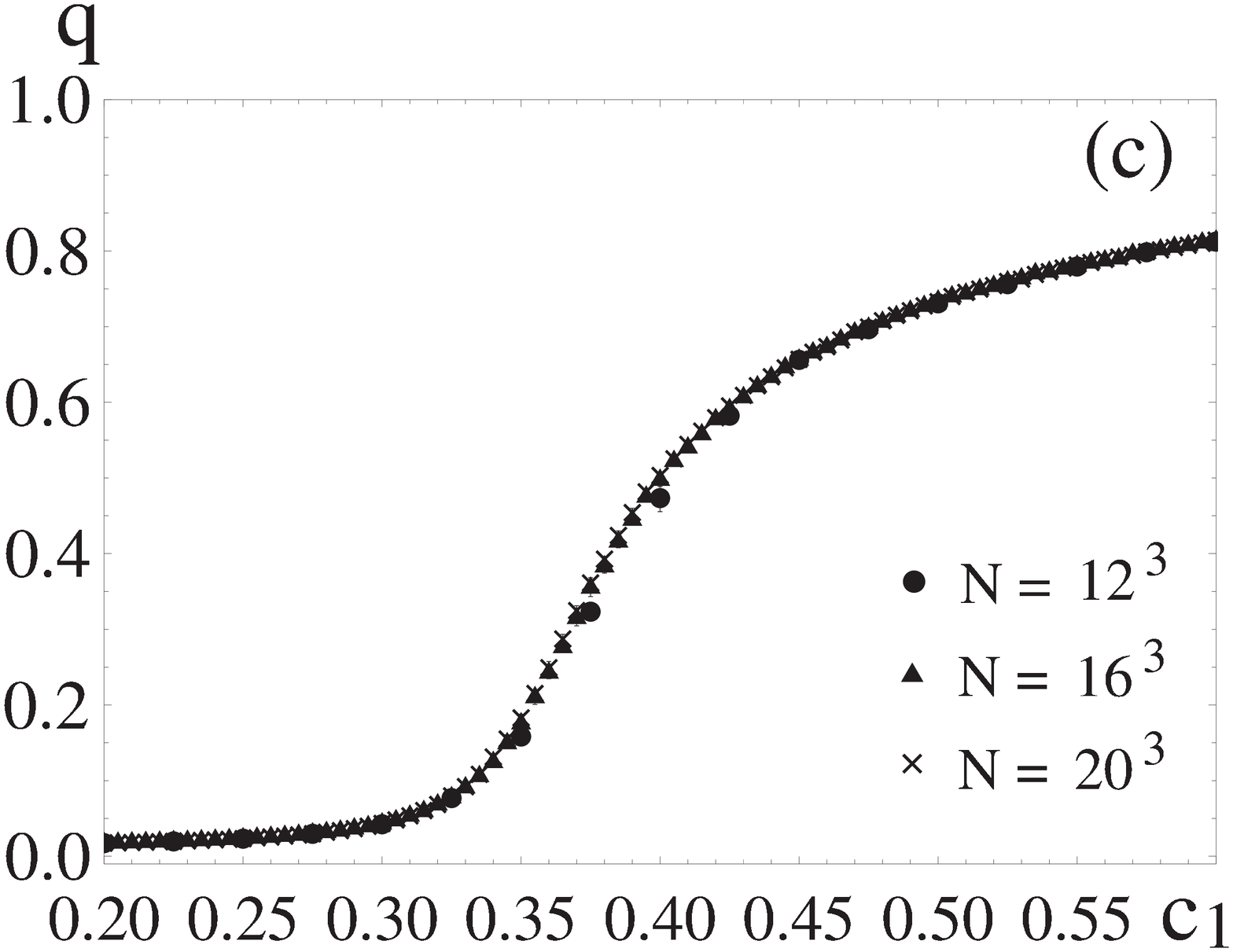}
\end{center}
\vspace{-0.7cm}
\caption{
(a) $U/N$, (b) $C/N$, (c) $q$  of Model III vs. $c_1$ 
for $c_2= 0.5$. There are no
sharp transitions but a crossover between the 
confinement and SG1 phases at $\sqrt{N} c_1\simeq 0.35 \sim 0.40$.
}\label{latticeconfinementspinglass}
\end{figure}

In Fig.\ref{latticeconfinementspinglass} we present
$U/N$, $C/N$ and $q$ vs. $c_1$ for $c_2= 0.5$.
$C/N$ shows a crossover between confinement and SG1 phases 
because it has almost no $N$ dependence. \\

\vspace{-0.3cm}
(ii) Coulomb-SG2 transition 

\begin{figure}[h]
\vspace{-0.7cm}
\begin{center}
\begin{minipage}{0.49\hsize}
\vspace{-0.5cm}
\hspace{0.6cm}
\includegraphics[width=4.5cm]{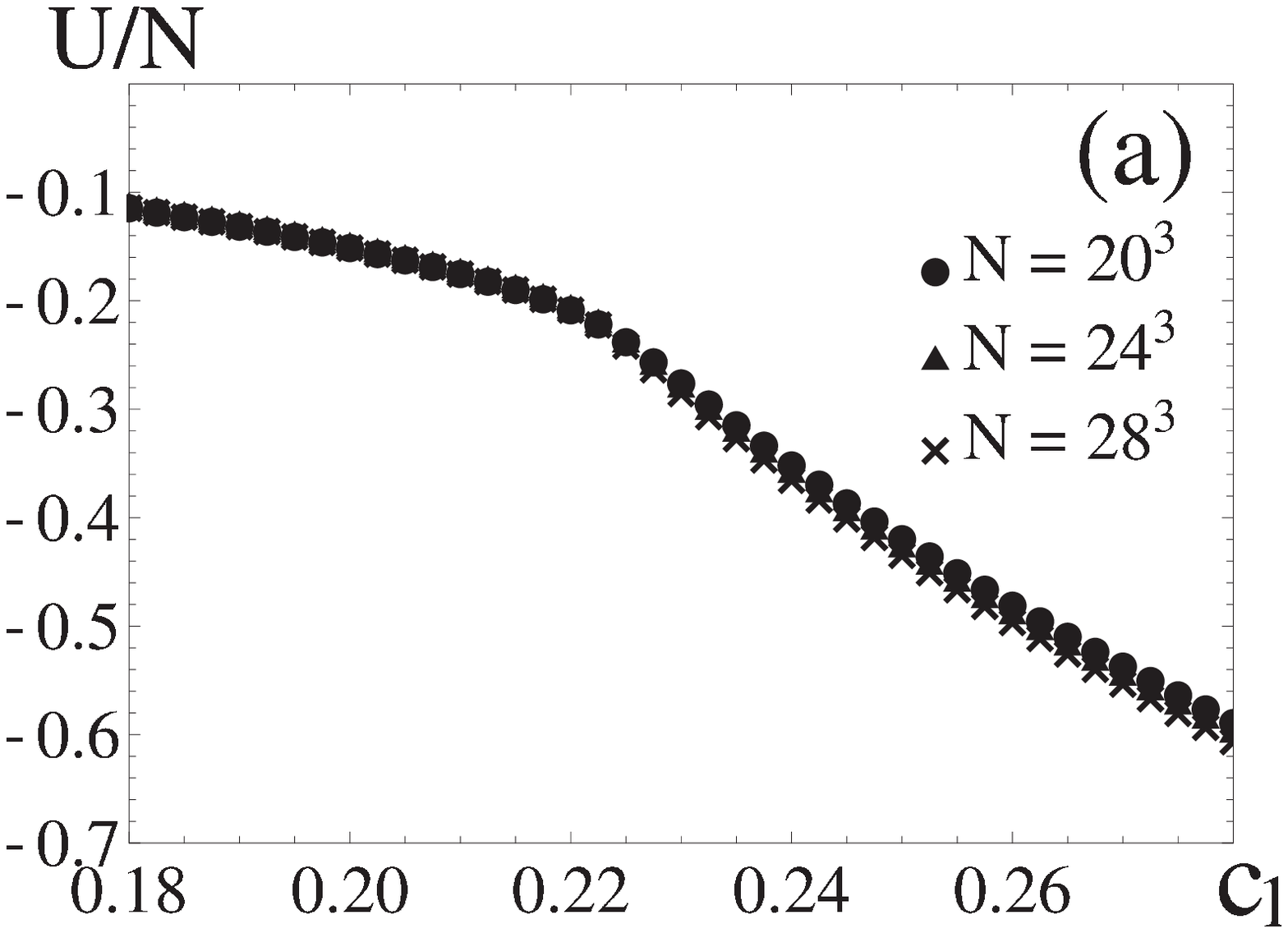}
\end{minipage}
\begin{minipage}{0.49\hsize}
\includegraphics[width=4.5cm]{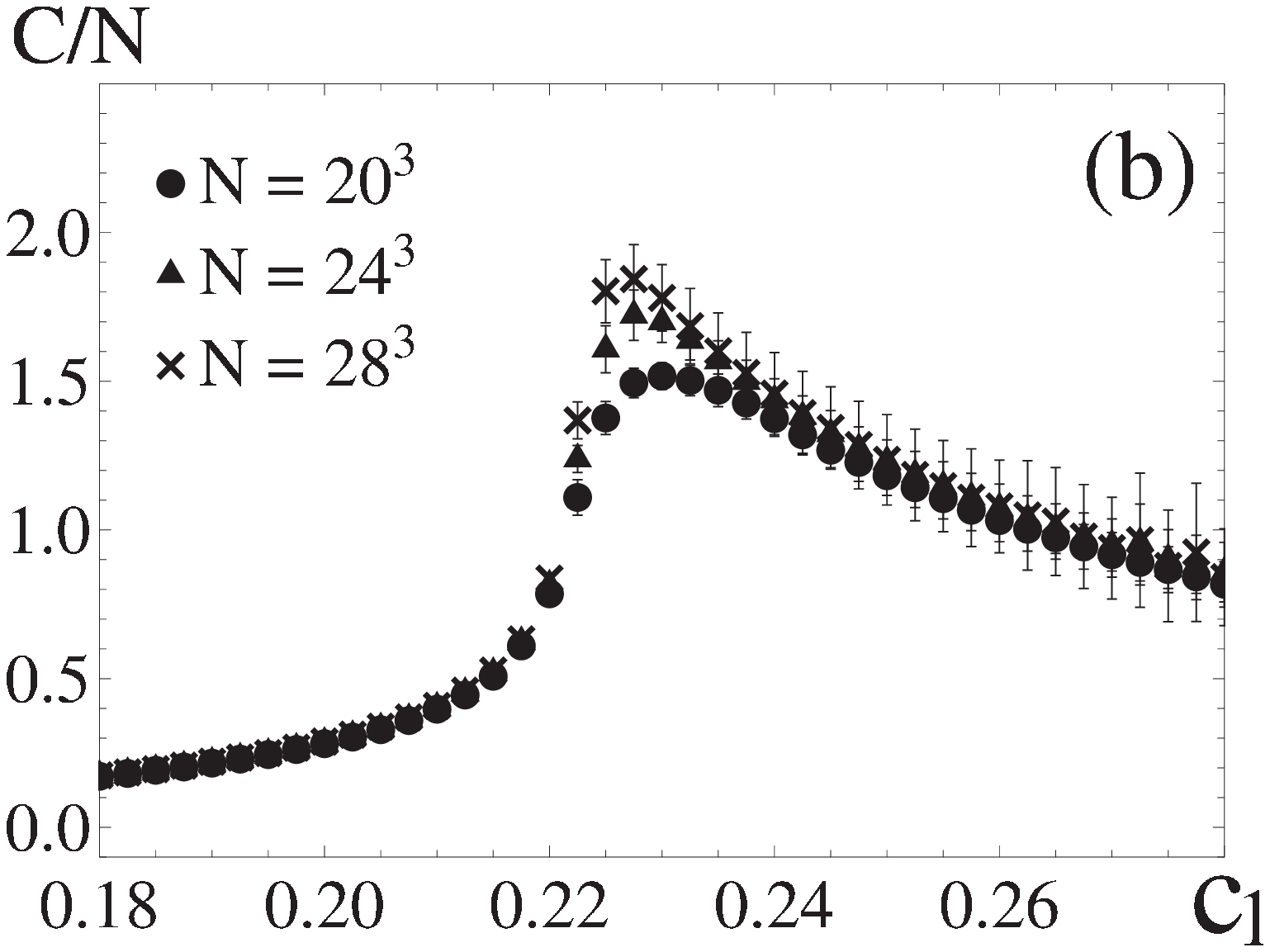}\\
\end{minipage}
\begin{minipage}{0.49\hsize}
\vspace{-0.5cm}
\includegraphics[width=4.5cm]{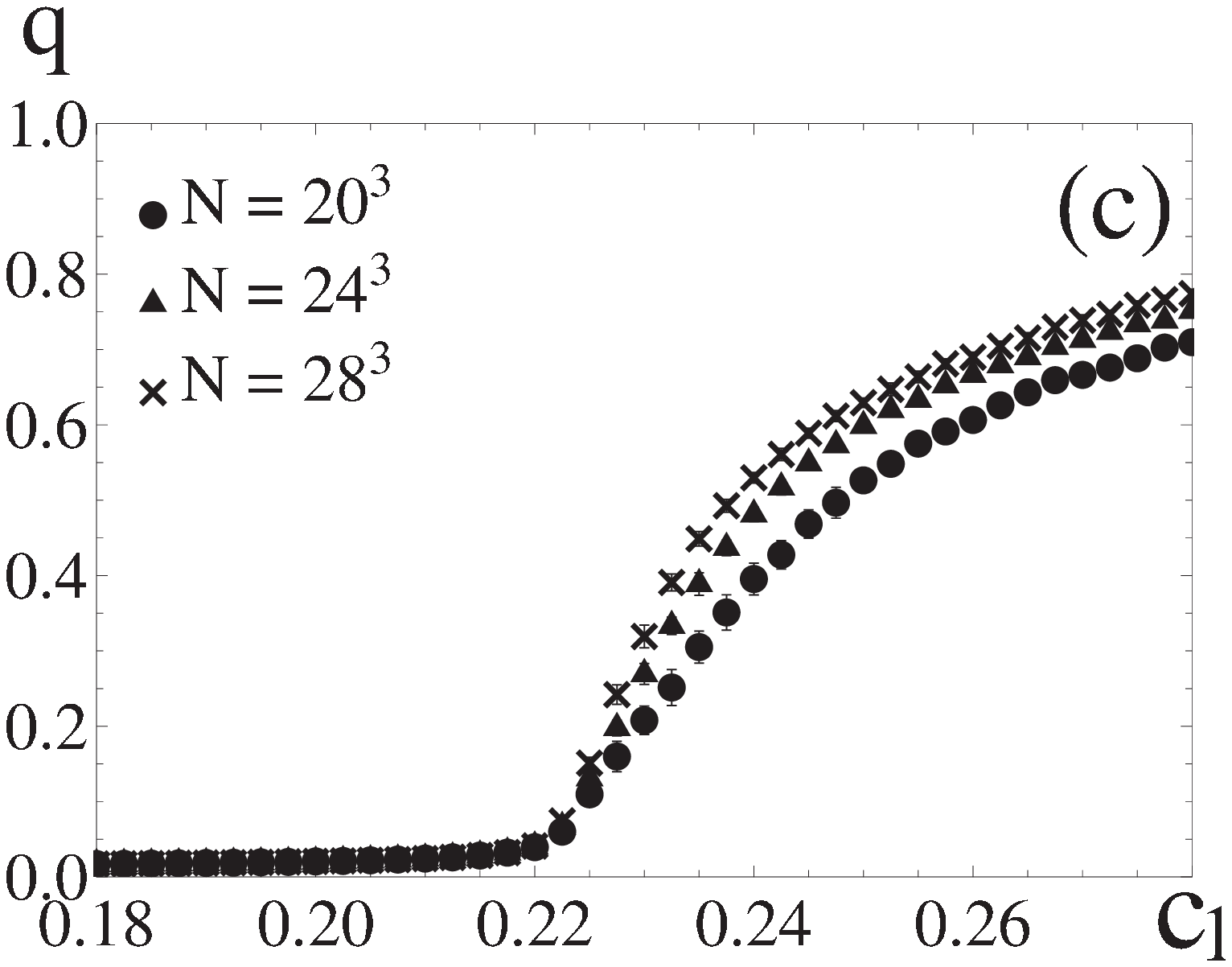}
\end{minipage}
\end{center}
\vspace{-0.6cm}
\caption{
(a) $U/N$, (b) $C/N$, and (c) $q$ of Model III vs. $c_1$ for $c_2= 1.0$.
There is a second-order transition between the Coulomb and SG2 phases 
at $c_1\simeq 0.225$.
}\label{latticecoulombhiggs}
\end{figure}

In Fig.\ref{latticecoulombhiggs}
we present $U/N$, $C/N$  and $q$ vs. $c_1$ for $c_2= 1.0$.
$C/N$ shows a second-order transition between Coulomb and SG2 phases 
because its peak develops systematically as $N$ increases.\\

\clearpage

\clearpage
\vspace{-0.3cm}
(iii) Transition across the line $c_2 \simeq 0.75$. 

\begin{figure}[h]
\begin{center}
\vspace{-0.2cm}
\hspace{-0.8cm}
\includegraphics[width=5.3cm]{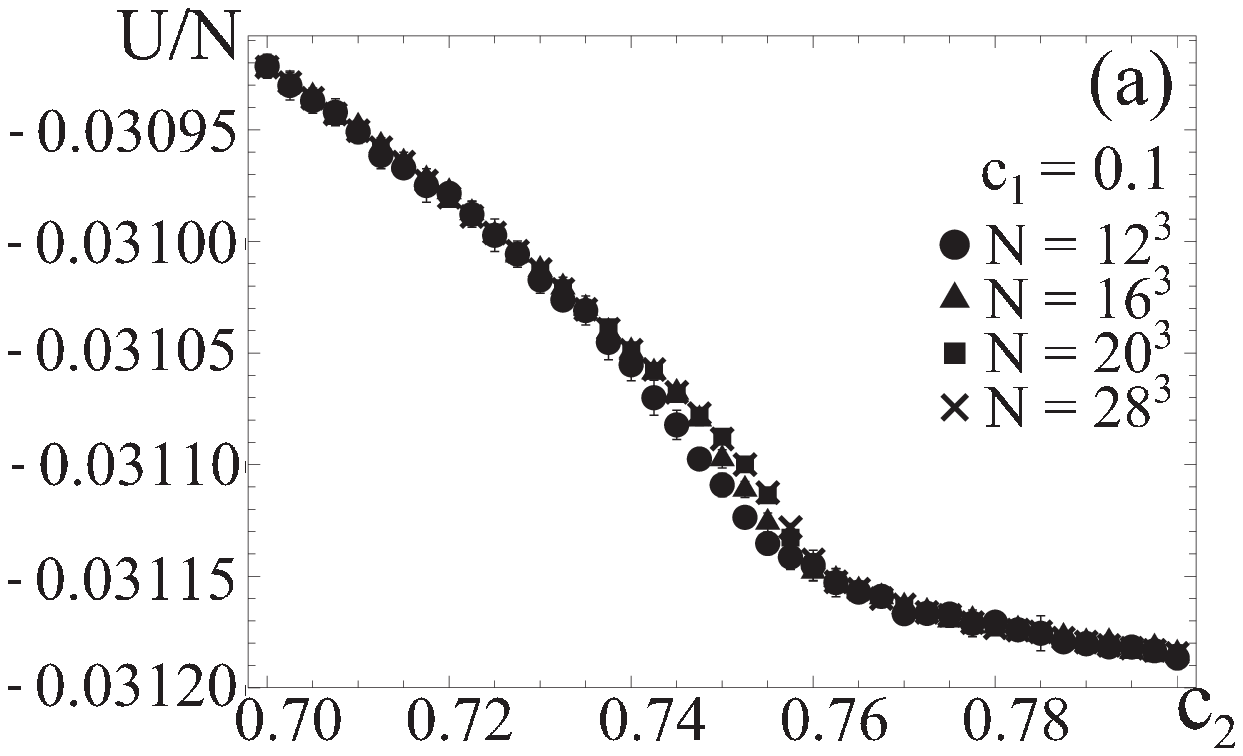}
\includegraphics[width=5.0cm]{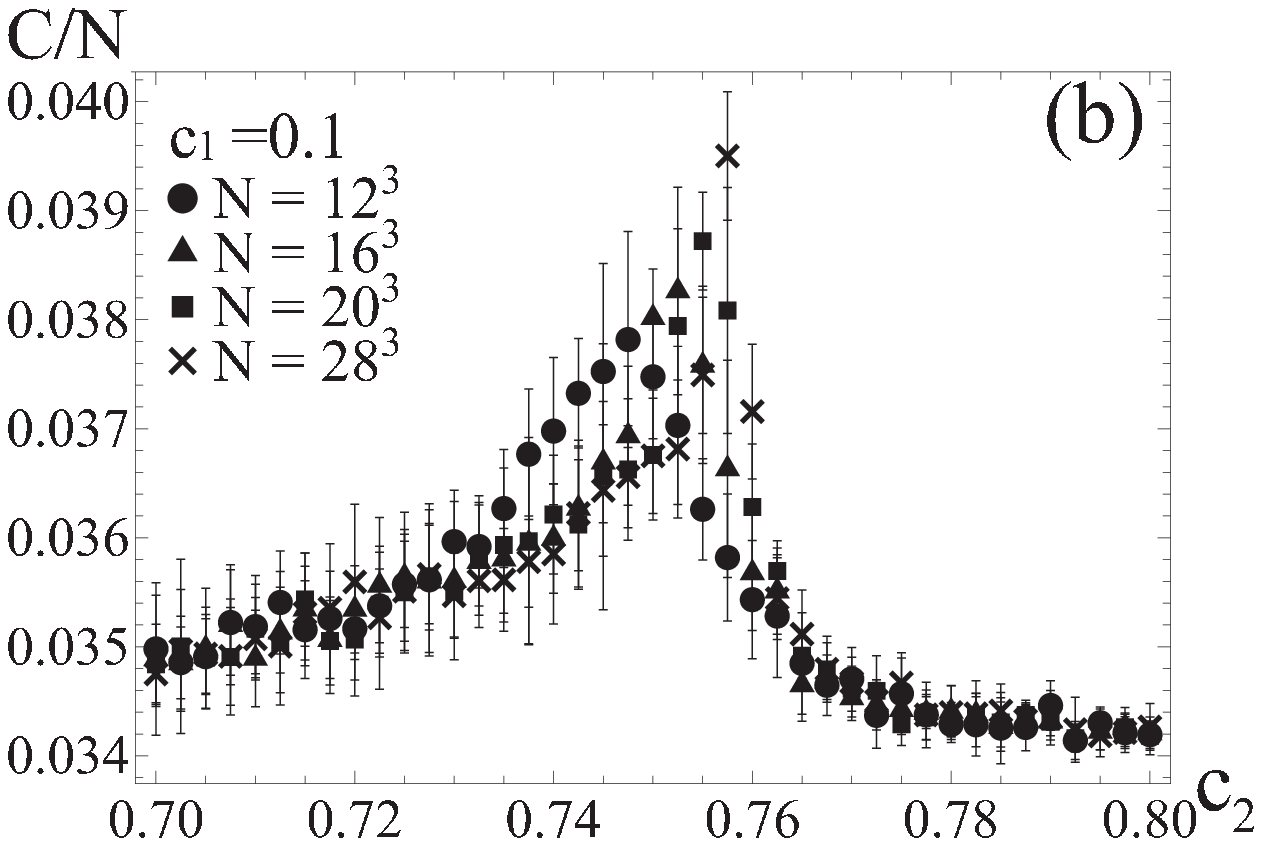}\\
\hspace{-0.6cm}
\includegraphics[width=4.6cm]{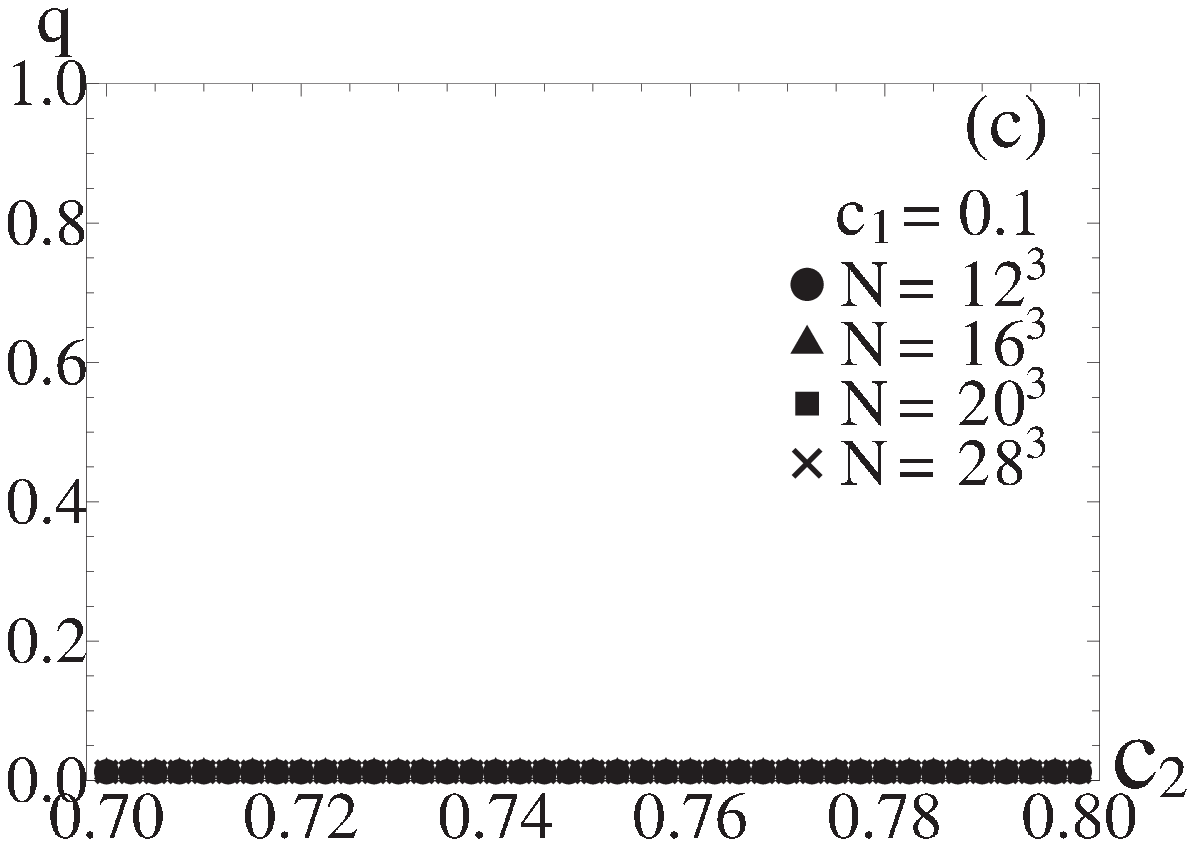}
\hspace{0.2cm}
\includegraphics[width=4.6cm]{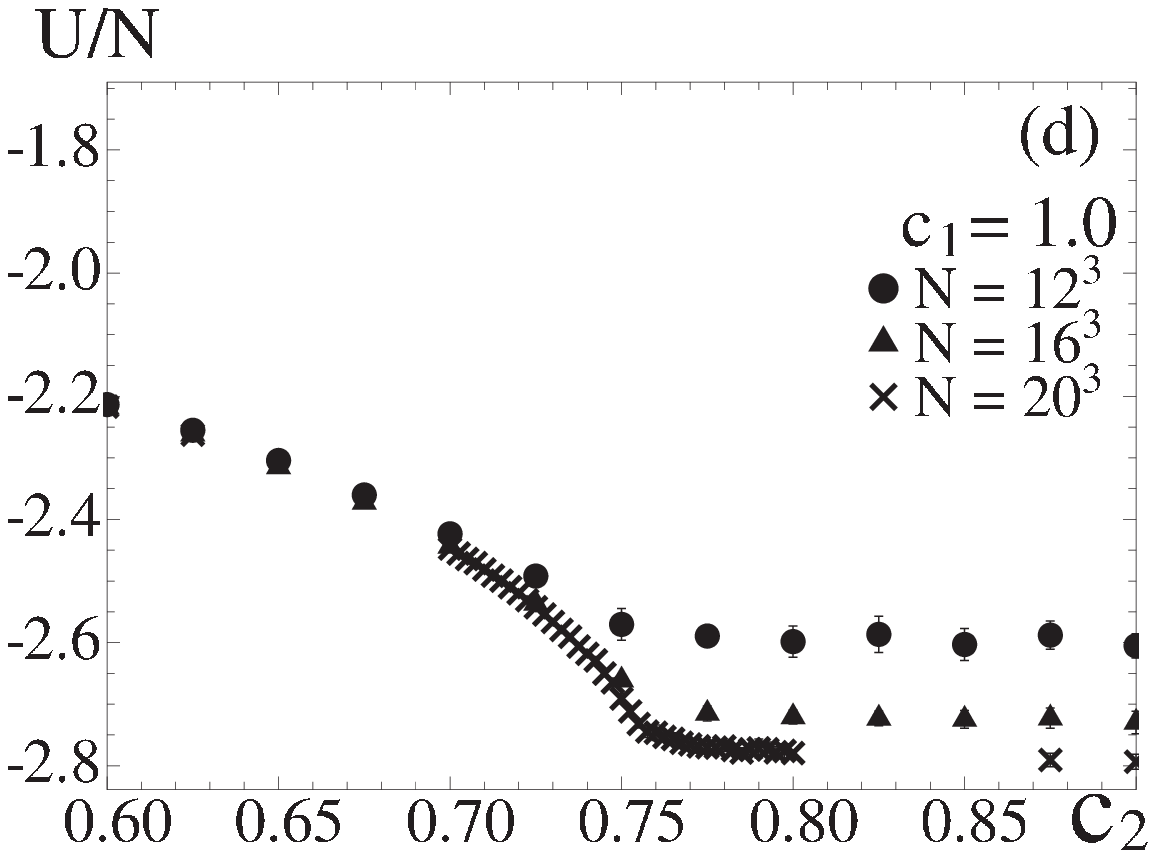}\\
\hspace{-0.5cm}
\includegraphics[width=4.5cm]{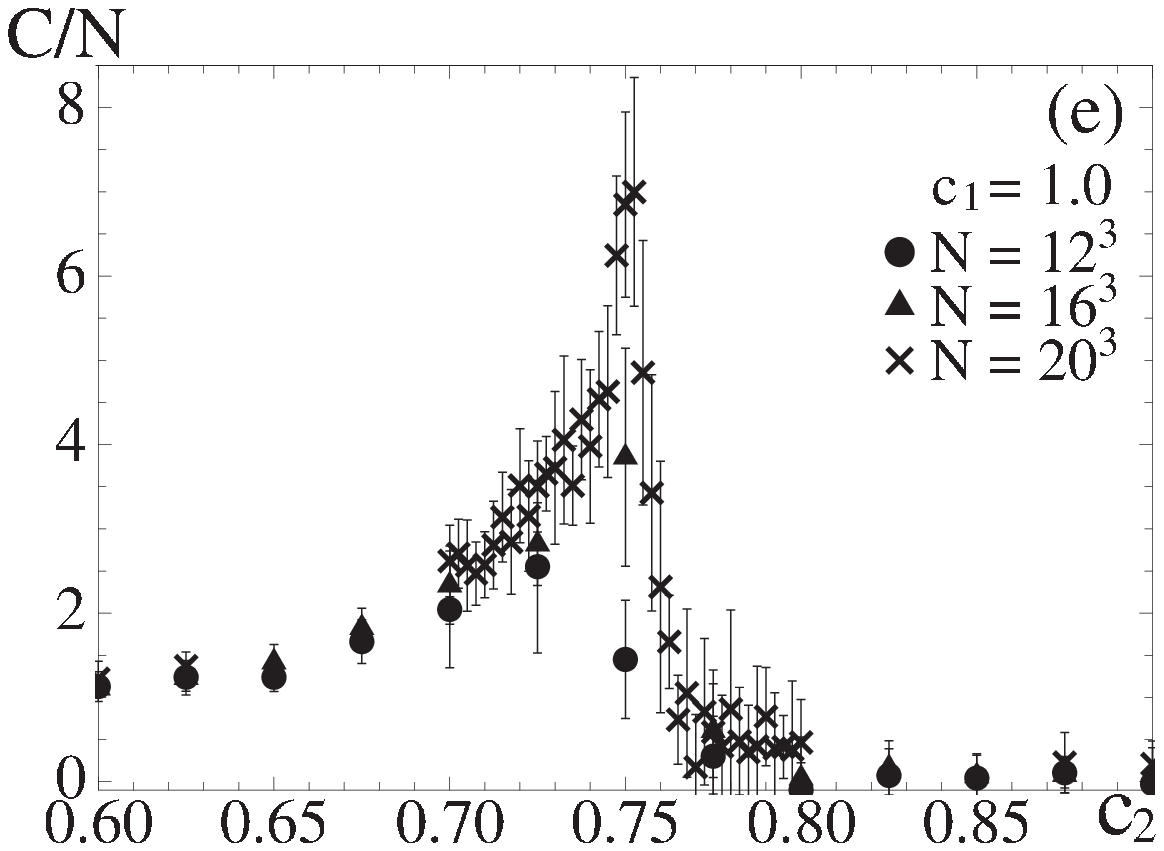}
\hspace{0.5cm}
\includegraphics[width=4.6cm]{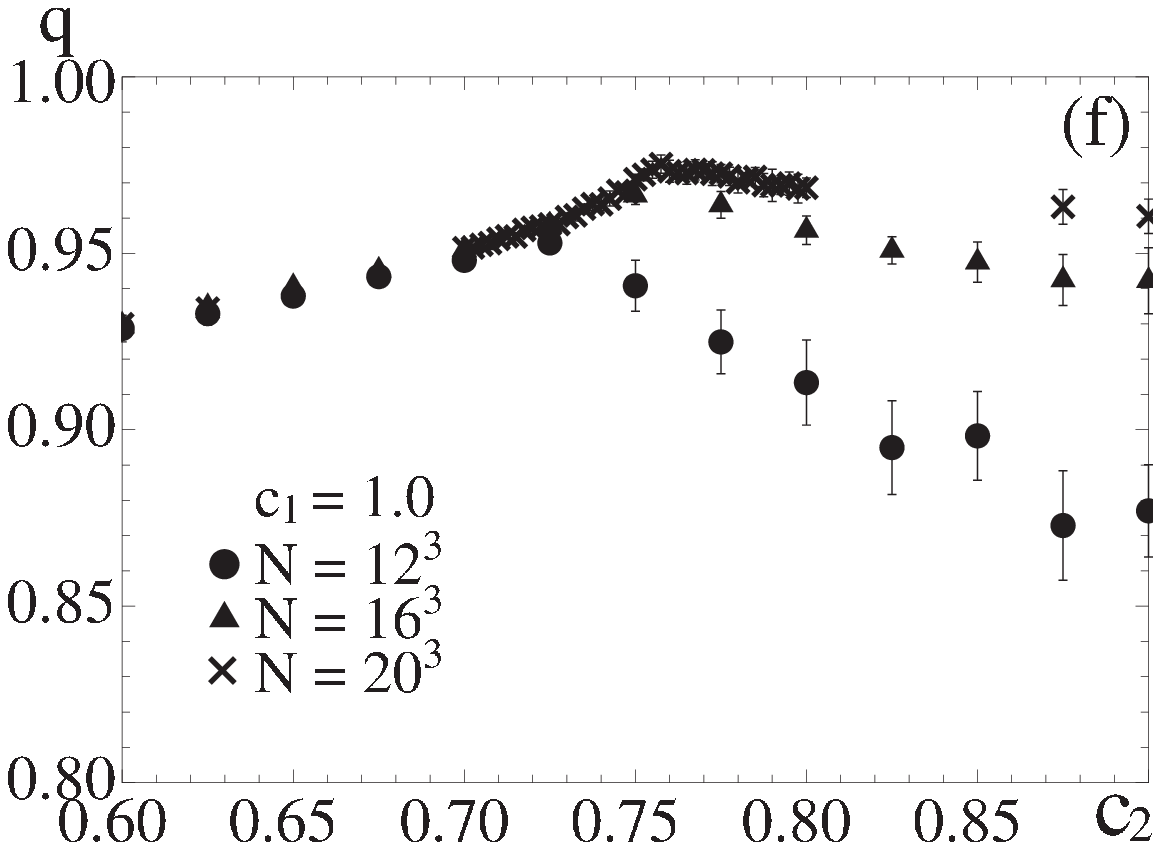}
\end{center}
\vspace{-0.1cm}
\caption{
$U/N, C/N$ and $q$ of Model III at (a-c) $c_1=0.1$ and (d-f) $c_1=1.0$. 
They show a second-order transition reflecting $P(J)$ of (\ref{annealpj}).
}\label{fig14}
\end{figure}

Because the system is quenched one, this second-order transition 
reflects the $c_2$-term of the energy. It has been studied 
in Model 0 at $c_1=0$\cite{kemukemu}. In this case, 
after the duality transformation,
this pure-gauge system becomes equivalent to the Ising spin model 
in three-dimensions,
which is well known to exhibit a second-order transition. 
In Fig.\ref{fig14} we present $U$ and $C$ at $c_1$=0.1 and 1.0.
They exhibit a second-order transition as expected.\\

\section{Conclusions and Discussions}
\setcounter{equation}{0} 

In this paper we have studied three versions of the Z(2) gauge neural network,
Models I, II, and III, and compared them each other and with 
the annealed 3D lattice model (Model 0). 
The effect of reverberating signals is, in short, to
enhance the order and stability of synaptic connections $J_{ij}$.
For example, as $c_2$ is increased along the line of $c_1=0$, 
the confinement phase for $c_2 < c_{2c}$ is converted to
the Coulomb phase for $c_2 > c_{2c}$ 
(See Figs.\ref{phase1}, \ref{uc2phase}, \ref{latticephase}).

Concerning to the phase structure, the obtained phases and 
the order of transitions   are summerized in Table 3.
These results are consistent each other; one may
interpret them in a coherent manner considering
how each term of $U$ and $C$, critical values $c_{1c}$ and $c_{2c}$, and 
the order of transition depend on the total number of sites $N$  and 
their connectivity $p$ as discussed in Sect.3-5.\\

{\small
\begin{center}
\begin{tabular}{|p{1.3cm}|p{2.2cm}|p{2cm}|p{2cm}|} 
\hline Model
& Higgs-confinement & confinement-Coulomb & Coulomb-Higgs 
\\ \hline
\ 0&\ CO-1st &\ 2nd &\ 2nd   
\\ \hline
\ I&\ CO-1st &\ 1st &\ 2nd   
\\ \hline
\end{tabular}\\
\vspace{0.1cm}
\begin{tabular}{|p{1.3cm}|p{2.2cm}|p{2cm}|p{2cm}|} 
\hline Model
& Across $c_2=c_{2c}$ line
& SG1-confinement & SG2-Coulomb  
\\ \hline
\ II&\ 1st($c_{2c}\simeq 2.0$)&\ 2nd &\ 2nd   
\\ \hline
\ III& 2nd($c_{2c}\simeq 0.75$) &\ CO &\ 2nd   
\\ \hline
\end{tabular}\\
\end{center}
Table3. Orders of phase transitions for various models.
CO implies crossover. 
The upper table is for the annealed models, Models 0 and I, and
the lower table is for the quenched models, Models II and III, where
SG1 is the phase at $c_2 < c_{2c}$ and SG2 is at $c_2 > c_{2c}$.}\\

For the annealed model, Model I, the obtained phases
are same as three phases of Model 0, but the order of 
confinement-Coulomb transition becomes 1st order
instead of 2nd order. As discussed in Sec.3.1, this reflects
the difference of connectivity.

For the quenched models, Models II and III, the Higgs phase of the
annealed models is better classified as the SG phase.
Actually,
the quenched transition at the critical value $c_2=c_{2c}$, 
which is independent of $c_1$, partitions  the Higgs phase
into two separate SG phases, SG1 ($c_2 < c_{2c}$) and 
SG2 ($c_2 > c_{2c}$).
These two phases are both characterized by nonvanishing SG
order parameter $q$, but are distinguished by disorder (SG1)
and order (SG2) of gauge variables $J_{ij}$
as explained by using Fig.\ref{fig1011}f,g.
This is another example of the effect of reverberating signals.

There we introduced $U_2=\la E_2\ra$ and $C_2$ defined in (\ref{U2C2}).
We note that this $U_2$ may be viewed as an example of Wilson loop.
In the usual lattice gauge theory {\it without matter fields}, 
which has only a plaquette interaction $JJJJ$,
the confinement phase and the Coulomb phase are distinguished
by the behavior of the Wilson loop\cite{wilson} as
\be
W[C]&\equiv& \la \prod_{C}J_{x\mu} \ra \sim
\left\{
\begin{array}{ll}
\exp(-\alpha S)& {\rm confinement\ phase}\nn
\exp(-\alpha' P)& {\rm Coulomb\ phase},\\
\end{array}
\right.
\nn 
\ee
where the product is taken along a closed loop $C$ on the lattice,
and 
$S$ is the minimum area having its edge $C$, and $P$ is the perimeter of $C$.



To examine the critical properties of the present models, 
it is necessary to study their scaling properties such as 
critical exponents of their second-order transitions by applying  
finite-size scaling argument to MC results, 
although such a study is  beyond the scope of the present paper. 
Concerning to this point, we recall a work by 
Hashizume and Suzuki\cite{suzuki2}. They studied the 3D lattice 
model, which is  equivalent to the 
present model\cite{suzuki1}, Model 0 and Model III at $c_1=0$, 
by a kind of MFT and correlation identities, and  obtained approximately
the transition temperature, scaling functions, and critical exponents, etc. 
Such an analytical and simple method may
give us some hints to calculate approximate critical exponents and related
quantities for other cases of the models studied in the present paper.

As general subjects for future investigations of the Z(2) gauge neural 
network, following points may be listed up as interesting extension of 
the models themselves. \\

- In this paper, we restricted ourselves to the region of $c_1, c_2 \geq 0$.
We chose this region because the $c_1$-term with $c_1 > 0$ 
may be regarded as  a rescaled energy of the Hopfield model and 
the  $c_2$-term of reverberating signals corresponds 
to the energy of magnetic field for $c_2 > 0$\cite{wilson}.
Study beyond this 
region may lead us to some new phases and transitions among 
them\cite{negativec2}.\\

-  We put the constraint $|J_{ij}|=1$ for the synaptic strength
for simplicity. Even if one uses other distribution of $J_{ij}$ 
with same mean and covariance in place of $|J_{ij}|=1$,
the global phase structure should be unchanged as long as one uses
the same energy as argued
 in Ref.\cite{wilsonkogut}. 
However, 
modification of the energy together with 
relaxing $J_{ij}$ to $0 \leq |J_{ij}| < \infty$ 
 will  serve as a model to investigate spontaneous
distribution of $|J_{ij}|$\cite{suzuki}. 
This is an interesting possibility because
some parts of the human brain has a log-normal distribution
of $|J_{ij}|$ which is a key structure to explain 
some activities of the human brain\cite{lognormal}.\\

- One may consider nontirivial  structure of connectivity 
$\epsilon_{ij}$ such as a small-world network\cite{smallworld}, etc.
This is interesting because the actual network structure of 
some parts of the human brain are known to be small-world type. \\  

- It is of interest to 
study the asymmetric case with two independent gauge variables,
$J_{ij}$ and $J_{ji}$ for a pair $i < j$\cite{symmetric}.  
This case  is expected to describe
 some interesting effects such as spontaneous oscillations in 
 time-development of the system.
 

\setcounter{section}{0} 
\renewcommand{\theequation}{\Alph{section}.\arabic{equation}}

\appendix
\section{Elitzur's theorem for quenched systems}

In this appendix we derive Elitzur's theorem for quenched systems.
Let us start by a brief derivation of the theorem for the annealed model,
Model I.
The average $\la O(S,J) \ra$ of (\ref{averageo1}) is written in the form,
\begin{eqnarray}
\langle O(S,J)\rangle&=&\sum_{S}\sum_{J}O(S,J)\exp(-E(S,J)).
\label{osj1}
\end{eqnarray}
Here it is sufficient to consider the average over each sample with 
definite $\epsilon_{ij}$, because the final average is just the sum 
(\ref{averageo12}) of such average.
By regarding the gauge transformation (\ref{z2gaugetrsf2})
as a change of variables $S_i\to S'_i,\ J_{ij}\to J'_{ij}$,
$O(S,J)$ of (\ref{osj1}) is rewritten as
\be
\la O(S,J)\ra&=&\sum_{S'}\sum_{J'}O(S',J')\exp(A(S',J'))\nn
&=&\sum_{S}\sum_{J}O(S',J')\exp(A(S,J))
= \la O(S',J')\ra,
\label{oo'}
\ee
where we used 
\be
\hspace{-0.5cm}
A(S',J')=A(S,J),\quad \sum_{S'_i}=\sum_{S_i},\quad
\sum_{J'_{ij}}=\sum_{J_{ij}}.
\ee
Let us restrict $O(S,J)$ to those satisfying
\be
O(S',J')=G(V)O(S,J).
\label{ogo}
\ee
Then (\ref{oo'}) claims that
\be
\la O(S,J)\ra =G(V)\la O(S,J)\ra.
\label{o=go}
\ee
If $O(S,J)$ is a gauge-invariant quantity, then $G(V)=1$,
and (\ref{o=go}) poses no restrictions to $\la O(S,J)\ra$.
If $O(S,J)$ is a {\it gauge-variant} quantity, then
$G(V)\neq 1$ and the following theorem is derived,
\be  
\hspace{-1cm}{\rm If\ }O(S,J)\ {\rm is\ gauge\ variant,\ then}\
\la O(S,J)\ra =0. 
\label{elitzur}
\ee

For a general $O(S,J)$ that does not satisfy (\ref{ogo}),
it may be expressed as a sum
\be
O(S,J)&=&\sum_\ell O_\ell(S,J),\nn
O_\ell(S',J')&=&G_\ell(V)O_\ell(S,J).
\ee
Then it is straightforward to derive the theorem (\ref{elitzur}).

Let us consider the quenched model, Model II for example.
The average is given by (\ref{averageo2}),
\be
\la O(S,J)\ra &=&\sum_J \sum_S O(S,J) \frac{\exp(-E_1(S,J))}
{Z_1(J)}P(J),\nn
Z_1(J)&=&\sum_{S}\exp\left(-E_1(S,J)\right).
\ee
We repeat the same change of variables (\ref{z2gaugetrsf2}) 
and note the gauge invariance,
\be
E_1(S',J')=E_1(S,J),\ Z_1(J')=Z_1(J),\ P(J')=P(J),
\ee
to get
\be
\la O(S,J)\ra = \la O(S',J')\ra.
\ee
Then we follow the same steps as for the annealed model 
to arrive at the theorem (\ref{elitzur}).

\section{Exact solution for $c_2=0$}

\renewcommand{\thefigure}{\arabic{figure}}
\setcounter{figure}{17}

\begin{figure}[b]
\begin{center}
\includegraphics[width=4.3cm]{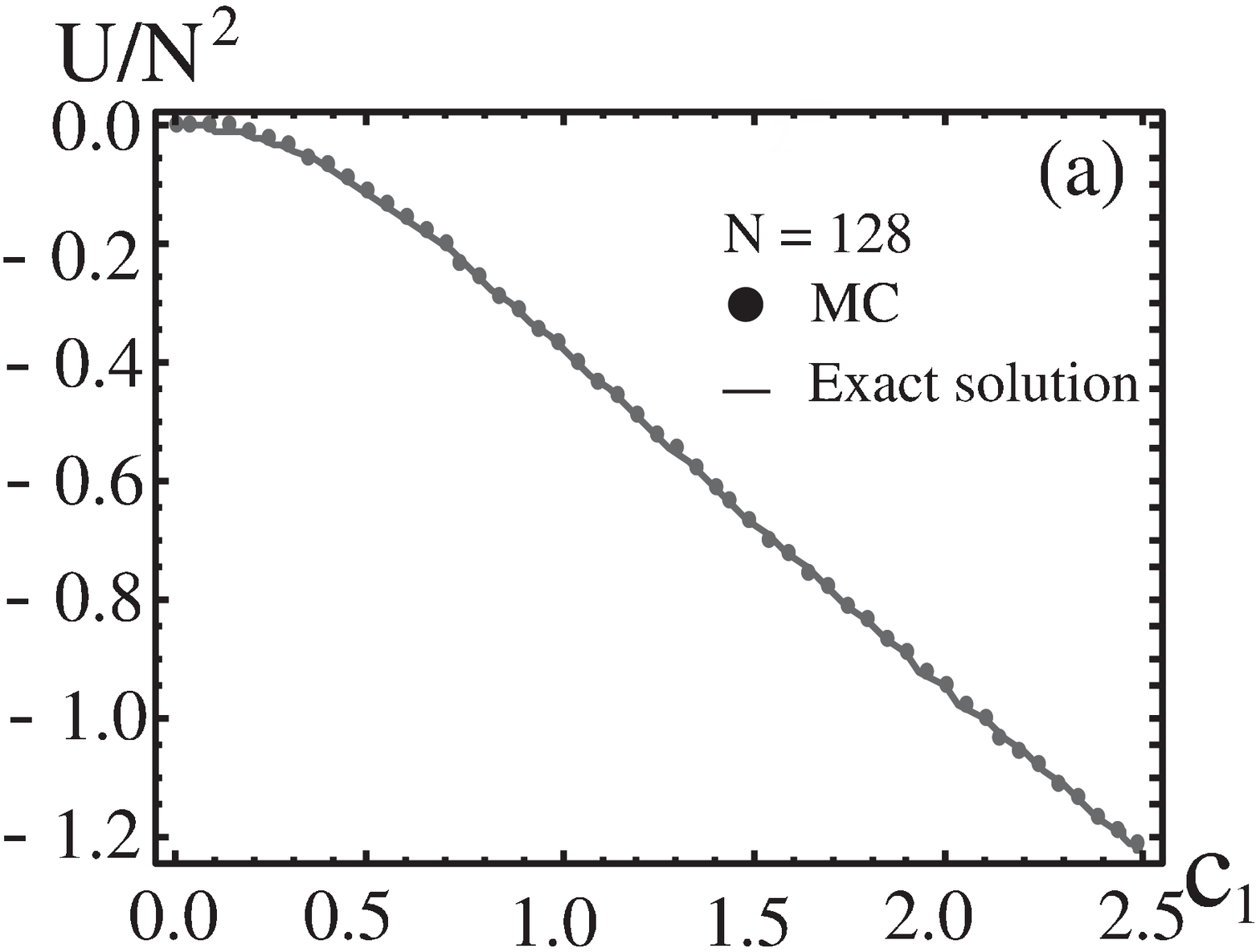}
\includegraphics[width=4.3cm]{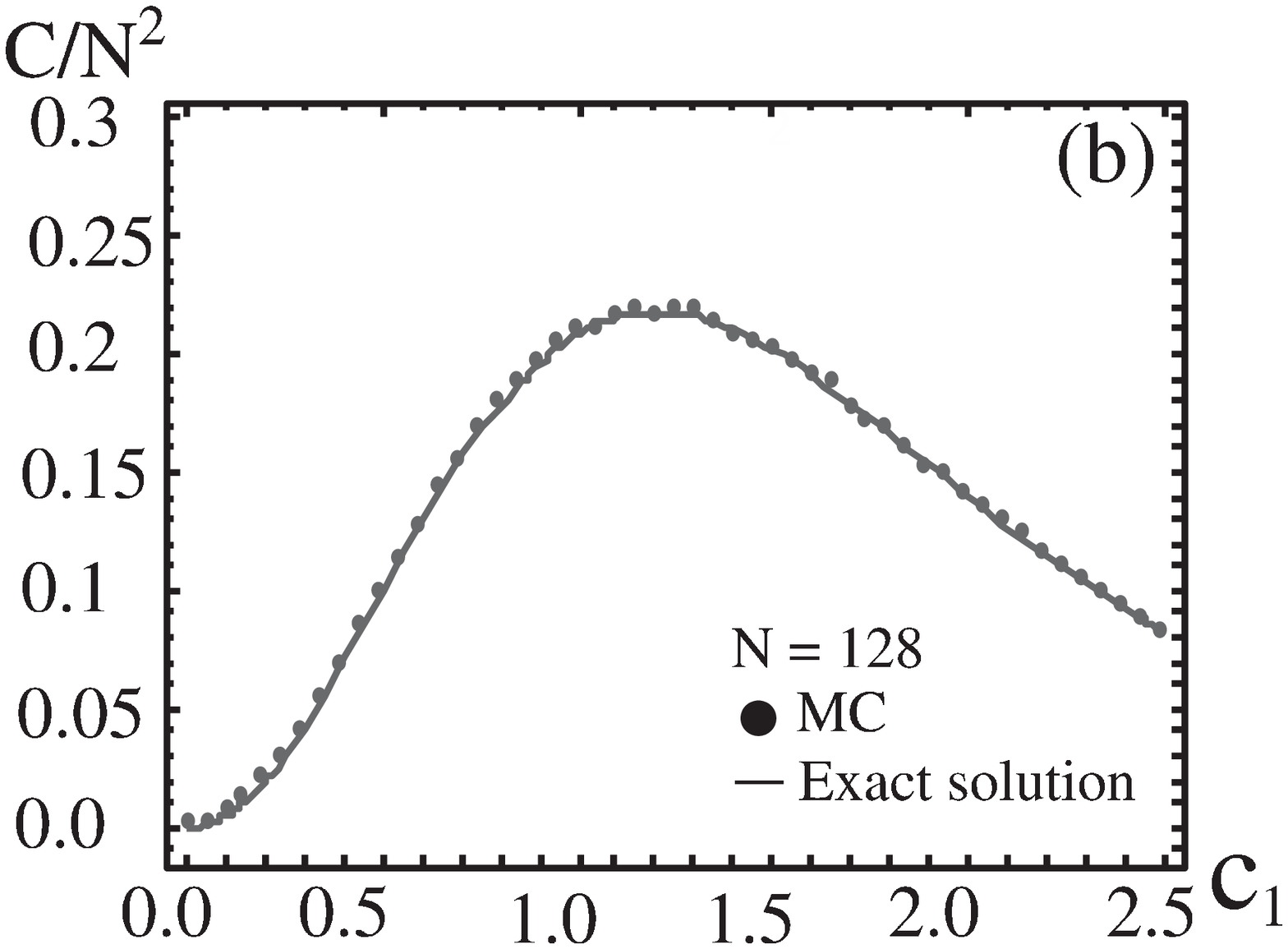}
\end{center}
\caption{The MC results of (a) $U$ and $(b) C$ of Model I at $c_2=0$
for $p=1.0$ and $N=128$. They
agree with the analytic expressions of (\ref{uandcforc2zero}).}
\label{figuandcforc2zero}
\end{figure}

In this Appendix, we study the exact solution of Model I for $c_2=0$.
The partition function for a  sample with  a definite $\epsilon_{ij}$ 
is calculated as
\begin{eqnarray}
Z_{c_2=0}(\epsilon) 
&=& \sum_{S} \sum_{J} \exp (-E_{c_2=0}(\epsilon)) = \sum_{S} \sum_{J}
              \exp ( c_1 \sum_{i<j}\epsilon_{ij}J_{ij}S_iS_j) \nonumber \\
          &=& \sum_{S} \prod_{i < j}\sum_{J_{ij}=\pm1} 
              \exp( c_1 \epsilon_{ij}J_{ij}S_iS_j) \nonumber \\
          &=& \sum_{S} \prod_{i < j}
              \left[\exp( c_1 \epsilon_{ij}S_iS_j)
              + \exp(- c_1 \epsilon_{ij}S_iS_j) \right] \nonumber \\
          &=& \sum_{S} \prod_{i < j}
              \left[2\delta_{\epsilon_{ij},1}\cosh c_1
              +2\delta_{\epsilon_{ij},0}\right]\nn
              &=& 2^N\prod_{i < j}
              \left[2\delta_{\epsilon_{ij},1}\cosh c_1
              +2\delta_{\epsilon_{ij},0}\right].
\ee
Then the partition function $Z_{c_2=0}$ averaged over  samples is given by
\be 
\hspace{-0.5cm}
Z_{c_2=0}&=& \frac{1}{N_\epsilon}\sum_\epsilon \delta_{p(\epsilon),p}
Z_{c_2=0}(\epsilon)= 2^N \left(2\cosh c_1\right)^{\frac{N(N-1)}{2}p}2^{\frac{N(N-1)}{2}(1-p)},
\end{eqnarray}
where we used the fact that the number of links of $\epsilon_{ij}=1(0)$
in a sample is ${}_NC_2 p\ [{}_NC_2 (1-p)]$.
Then the internal energy $U$ and the specific heat $C$ are
calculated as
\begin{eqnarray}
U &=& \langle E \rangle_p =
 -c_1\frac{d}{dc_1}\ln Z_{c_2=0} = -\frac{N(N-1)}{2}p c_1 \tanh c_1,\nn
C &=& \frac{dU}{dT} =
           -c_1^2 \frac{d}{dc_1}\left(\frac{U}{c_1}\right) 
          = \frac{N(N-1)}{2}\frac{p\; c_1^2 }{\cosh^2c_1}.
\label{uandcforc2zero}          
\end{eqnarray}
We note that $U$ and $C$ are proportional to  $p$ 
 as it should be.
We have checked that the MC results agree with these results as
shown in Fig.\ref{figuandcforc2zero}. 

\section{Infinite-Range Ising spin model}
Let us study the IRI spin model (\ref{irm}).
The partition function is rewritten as
\be
Z_{\rm IRI}&=&\sum_{S}\exp(c_1\sum_{i<j}S_iS_j)
=\sum_{S}\exp(\frac{c_1}{2}\sum_{i,j}S_iS_j-\frac{c_1}{2}N)\nn
&=&\exp(-\frac{c_1}{2}N)(2\pi c_1)^{-\frac{1}{2}}
\sum_{S}\int_{-\infty}^{\infty} d\chi \exp(-\frac{1}{2c_1}\chi^2+\chi
\sum_{i}S_i)\nn
&=&\exp(-\frac{c_1}{2}N)(2\pi c_1)^{-\frac{1}{2}}
\int_{-\infty}^{\infty} d\chi
\exp\left(-\frac{1}{2c_1}\chi^2+N\ln(2\cosh\chi)\right)\nn
&=&\exp(-N F_{\rm IR} + O(N^0)),\nn
F_{\rm IRI}&=& \frac{1}{2c_1 N}\chi_0^2-\ln(2\cosh\chi_0),
\ee
where we assumed that $c_1=O(N^{-1})$ and used the saddle-point
evaluation for large $N$. $\chi_0$ is the solution of the saddle-point
equation,
\be
-\frac{\chi}{N c_1}+\tanh\chi=0.
\ee
$\chi_0$ exhibits a second-order transition at $c_{1c}=1/N$,
\be
\chi_0  \left\{
\begin{array}{ll}
=0, & N c_1 < 1,\\
\neq 0, & N c_1> 1.
\end{array}
\right.
\ee\\

\section{Mean field theory for Model I with $p=1.0$}

In this Appendix we study MFT of Model I with $p=1.0$ based on
the Feynman's method\cite{feynman}.
It is formulated as a variational principle
for the Helmholtz free energy $F$ by using the variational (trial) energy $E_0$
as follows;
\be
Z &=& \sum_{S,J} \exp(-\beta E) \equiv \exp(-\beta F),\nn
Z_0 &=& \sum_{S,J}\exp(-\beta E_0) \equiv \exp(-\beta F_0),\nn
\la O \ra_0 &\equiv& Z_0^{-1} \sum_{S,J} O \exp(-\beta E_0), \nn 
F &\le& F_v \equiv F_0 + \la E -E_0 \ra_0.
\end{eqnarray}
We adjust the variational parameters
contained in $E_0$ optimally so that $F_v$ is minimized.

For $E_0$ we use 
\begin{eqnarray}
E_0 = - W \sum_{i<j}J_{ij} - h \sum_i S_i,
\label{variationalenergy}
\end{eqnarray}
where $W$ and $h$ are real variational parameters.
Then we have  
\begin{eqnarray}
f_v &\equiv& \frac{F_v}{N}=-\frac{N_l}{N} \ln( 2\cosh \beta W) - 
\ln( 2\cosh \beta h) \nn
&&-c_1\frac{N_l}{N}m^2 M  
 -c_2 \frac{{}_NC_3}{N} M^3  + \frac{N_l}{N}WM + hm,\nonumber\\
m &\equiv& \langle S_i 	\rangle_0 =  \tanh  h, \
M\equiv \langle J_{ij} 	\rangle_0 = \tanh  W.
\label{fv}
\end{eqnarray}

The minimization of $f_v$ yields
the three phases characterized as follows;
\be
 \begin{tabular}{|c|c|c|} 
\hline
   phase    & $M$ & $m $  
\\ \hline
Higgs       & $\neq 0$  & $\neq 0$ 
\\ \hline
Coulomb     & $ \neq 0$  & $0$  \\ \hline
Confinement & $0$   & $0$  \\
\hline
\end{tabular}
\label{tab:phases}
\ee

\begin{figure}[h]
\begin{center}
\hspace{-0.6cm}
\includegraphics[width=5.5cm]{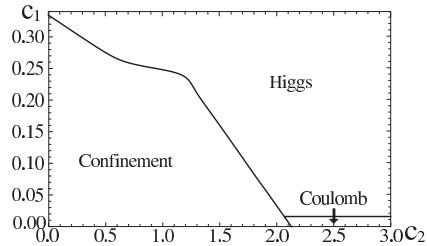}\\
\end{center}
\caption{
MFT result for the phase structure of Model I with $p=1.0$  
in the $c_2$-$c_1$ plane ($N=64$). There is a first-order transition
curve starting at $c_{1c}\simeq 0.334 (c_2=0)$ and ending at 
$c_{1c}=0 (c_2\simeq 2.13)$, and  
a second-order one along $c_{1c}\simeq 1/N (c_2 \gtrsim 2.0)$ .
}
\label{mft}
\end{figure}

The phase boundaries are shown in Fig.\ref{mft}.
The discontinuity of order parameters  of each transitions are  as follows;
\be
 \begin{tabular}{|c|c|c|c|} 
\hline phase boundary& order
     & $\Delta M $ & $\Delta m$  
\\ \hline
Confinement-Coulomb       & 1st &$\neq 0$  & $0$ 
\\ \hline
Higgs-Coulomb     & 2nd&$ 0$  & $ 0$  \\ \hline
Higgs-Confinement & 1st&$\neq 0$  & $\neq 0$  \\
\hline
\end{tabular}
\label{tab:phases2}
\ee

For the Higgs-Coulomb transition, the critical value of $c_1$ is estimated as
\be
c_{1c}=\frac{N}{2N_l}\frac{1}{M }\simeq \frac{1}{N},
\ee
for large $N$ and large $c_2$ at which $M\simeq 1$.

\section{Comparison of Model II at $c_2=0$ and the 
Sherrington-Kirkpatrick model}

In this Appendix we study a possible phase transition
of Model II at $c_2=0$ by using the known result of
Sherrington-Kirkpatrick (SK) model\cite{skmodel}.

The energy of the SK model is given by
\be
\hspace{-0.5cm}
E_{\rm SK} &=& -\sum_{i < j}J_{ij}S_iS_j,\ S_i=\pm1,\ J_{ij}\in (-\infty,\infty).\ee
The quenched variable 
$J_{ij}$ is a real number(we use the same symbol with our $J_{ij}=\pm1$),
and distributes by the Gaussian weight,
\be
P_{\rm SK}(J)&=&\prod_{i<j}\frac{N^{1/4}}{\sqrt{2\pi}\tilde{J}}
\exp\left(-\frac{(J_{ij}-\tilde{J}_0/N)^2}{2\tilde{J}^2/N}\right).
\label{psk}
\ee
Then the replica-symmetric solution for large $N$ 
(which is accepted as correct ones for small
$J_0$) gives rise to a phase-diagram in the 
$\tilde{J_0}/\tilde{J}-(1/\tilde{J})$ plane in which
there is a horizontal second-order phase transition line
along $1/\tilde{J}=1$ for $\tilde{J_0}/\tilde{J} \leq 1$
separating the SG phase $m=0, q \neq 0$ ($1/\tilde{J} < 1$)
and the paramagnetic phase $m=0, q=0$ (1/$\tilde{J} > 1$).
From the point $1/\tilde{J}=1, \tilde{J_0}/\tilde{J}=1$, two
transition curves spring out to border the ferromagnetic phase $m\neq 0,
q\simeq m^2$ in the larger $\tilde{J_0}/\tilde{J}$ region (Note that Z(2) gauge symmetry is violated for $\tilde{J}_0 \neq 0$).

Let us turn to Model II at $c_2=0$ and deform it by
replacing $J_{ij}=\pm1$ to a real variable with an optimally
determined distribution of the form of 
$P_{\rm SK}$ of (\ref{psk}). 
We choose $P_{\rm SK}$ to generate 
the same mean value and variance as $P(J)$
of (\ref{averageo2}), i.e.,
$\la J_{ij} \ra_{P_{\rm SK}}=0, \la J_{ij}^2 \ra_{P_{\rm SK}}=1$.
This treatment of Z(2) variable by real Gaussian variable 
may preserve universal
critical properties of the system\cite{wilsonkogut}.
This determines the optimal $P_{\rm SK}$ as 
\be
P_{\rm SK}(J)&=&\prod_{i<j}\frac{N^{1/4}}{\sqrt{2\pi}\tilde{J}}
\exp\left(-\frac{J_{ij}^2}{2}\right).
\label{psk2}
\ee
To adjust $E_{\rm SK}$ to $E_{1}$ of (\ref{averageo2}), 
we replace $J_{ij}$ of the SK model
by $c_1 J_{ij}$. Then $P_{\rm SK}(J)$ of (\ref{psk})
becomes the same as Eq.(\ref{psk2}) by choosing $\tilde{J}_0=0$ and
\be
\frac{c_1^2}{\tilde{J}^2/N}=1.
\ee
Then, the established transition point of SK model for $\tilde{J}_0=0$,
$\tilde{J}=1$ predicts the location of second-order transition
of Model II at $c_2=0$ as
\be
c_{1c} =\frac{1}{\sqrt{N}}.
\ee
This gives an estimate $c_{1c}=0.125$ for $N=64$, which should be
compared with the MC result of Fig.\ref{uc2phase}, $c_{1c} \simeq 0.15$.
Inclusion of the $c_2$-term makes summation over $J_{ij}$ variables
difficult analytically.



\begin{thebibliography}{99}
\bibitem{c3}
In Ref.\cite{kemukemu},
we have included the $c_3$-term with $SJJJS$.
In this paper we cite the result for $c_3=0$
for simplicity. It may also help to enhance the contrast with the case 
with full connections $(p=1.0)$ of Model I.
\bibitem{reverberating}
We note that the $c_2$-term is relevant and 
generated after renormalization of the 
$c_1$-term as one can see the relation, 
$S_x J_{x\nu}S_{x+\nu}\times S_{x+\nu}J_{x+\nu,\mu}
S_{x+\mu+\nu}\times S_{x+\mu+\nu}J_{x+\mu,\nu}S_{x+\mu}
\times S_{x+\mu}J_{x\mu}S_x = J_{x\nu}J_{x+\nu,\mu}
J_{x+\mu,\nu}J_{x\mu}$ because 
$S_x^2=1$ for Models 0 and III, and
$S_iJ_{ij}S_j\times S_jJ_{jk}S_k\times
S_kJ_{ki}S_i = J_{ij}J_{jk}J_{ki}$ for Models I and II.
\bibitem{mftmodel0}
We note that, instead of a second-order one, 
the MFT incorrectly predicts a first-order transition 
for the confinement-Coulomb transition, 
because the number of dimensions (three) is not high enough for
MFT\cite{kemukemu}.
\bibitem{symmetric}
We note that,
even for the general case of two independent variables $J_{ij}$ and $J_{ji}$, 
the Hopfield energy $c_1\sum_{i < j}S_iJ_{ij}S_j$ itself
is independent of the asymmetric part $(J_{ij}-J_{ji})/2$.
On the other hands, the $c_2-$term reflects this asymmetry.
\bibitem{negativec2}
In Ref.\cite{kemukemu} some results for $c_2 < 0$ are obtained.
Concerning to this point, we note that there hold
the relations, $Z_{\rm 0,III}(c_1,c_2)=Z_{\rm 0,III}(-c_1,c_2)$, 
$Z_{\rm I,II}(c_1,c_2)=Z_{\rm I,II}(-c_1,-c_2)$, which can be derived
by the change of variables $J_{ij}\to -J_{ij}$.
\end{thebibliography}
\end{document}